\documentclass[useAMS,usenatbib]{mnras}

\usepackage{times}
\usepackage{graphicx}
\usepackage{subfigure}
\usepackage{amsmath}
\usepackage{booktabs}

\newcommand{\gsim}{\mathrel{\hbox{\rlap{\lower.55ex \hbox {$\sim$}}
                   \kern-.3em \raise.4ex \hbox{$>$}}}}
\newcommand{\lsim}{\mathrel{\hbox{\rlap{\lower.55ex \hbox {$\sim$}}
                   \kern-.3em \raise.4ex \hbox{$<$}}}}

\title[IMF variation with redshift \& metallicity]{Variation of the low-mass end of the stellar initial mass function with redshift and metallicity}
\author[M. R. Bate]{Matthew R. Bate$^{1}$\thanks{E-mail:
M.R.Bate@exeter.ac.uk}\\
$^{1}$ Department of Physics and Astronomy, University of Exeter, Stocker
Road, Exeter EX4 4QL, United Kingdom
}

\date{Accepted 2025 January 7. Received 2024 November 19; in original form 2024 September 9}
\begin{document}
\maketitle
\begin{abstract}
We report the stellar mass functions obtained from 20 radiation hydrodynamical simulations of star cluster formation in 500~M$_\odot$ molecular clouds with metallicities of 3, 1, 1/10 and 1/100 of the solar value, with the clouds subjected to levels of the cosmic microwave background radiation that are appropriate for star formation at redshifts $z=0, 3, 5, 7,$ and 10.  The calculations include a thermochemical model of the diffuse interstellar medium and treat dust and gas temperatures separately.  We find that the stellar mass distributions obtained become increasingly bottom light as the redshift and/or metallicity are increased.  Mass functions that are similar to a typical Galactic initial mass function are obtained for present-day star formation ($z=0$) independent of metallicity, and also for the lowest-metallicity (1/100 solar) at all redshifts up to $z=10$, but for higher metallicities there is a larger deficit of brown dwarfs and low-mass stars as the metallicity and redshift are increased.  These effects are a result of metal-rich gas being unable to cool to as lower temperatures at higher redshift due to the warmer cosmic microwave background radiation.  Based on the numerical results we provide a parameterisation that may be used to vary the stellar initial mass function with redshift and metallicity; this could be used in simulations of galaxy formation.  For example, a bottom-light mass function reduces the mass-to-light ratio compared to a typical Galactic stellar initial mass function, which may reduce the estimated masses of high-redshift galaxies.

\end{abstract}

\section{Introduction}
\label{introduction}

For more than a decade, hydrodynamical simulations of the gravitational collapse and fragmentation of molecular clouds, that include either radiation transport or radiative heating from young stars, have been able to model star formation in groups of hundreds of stars, enough to examine the statistical properties of low-mass stars.  
Some of these simulations have been quite successful in reproducing many of the observed statistical properties of young low-mass stars in nearby Galactic star-forming regions, such as their stellar mass distributions (\citealt*{Bate2012, Bate2019, KruKleMcK2012}; \citealt{Guszejnov_etal2021, MatFed2021}; \citealt*{MatFedSet2022}; \citealt{Grudic_etal2022, Chon_etal2024}), and the dependence of multiplicity on primary stellar mass \citep{Bate2012, Bate2019, KruKleMcK2012, MatFed2021, MatFedSet2022, Guszejnov_etal2022, Grudic_etal2022, Chon_etal2024}.  The calculations of \cite{Bate2012, Bate2019}  even resolve many protostellar discs, and the disc mass and radius distributions from these calculations are found to be in good agreement \citep{Bate2018, ElsBat2021, Elsender_etal_2023} with those of observed, very young (Class 0), protostellar systems \citep[e.g.,][]{Tychoniec_etal2018,Tobin_etal2020}.  Other recent papers have also explored how the statistical properties of discs may vary with magnetic field strength \citep{Lebreuilly_etal2024}, and how best to compare simulated and observed discs \citep{Tung_etal2024}.

Over the same time period, many studies have explored the impacts of various initial conditions or physical processes on stellar properties, including molecular cloud density \citep{Bate2009b, JonBat2018a,TanKruFed2022}, protostellar outflows (\citealt{KruKleMcK2012, MatFed2021}; \citealt*{TanKruFed2022}), episodic outflow feedback \citep{Rohde_etal2021}, variations in turbulence (\citealt*{LomWhiHub2015,NamFedKru2021}; \citealt{MatFedSet2022}), and magnetic fields (\citealt{Myersetal2013, Myers_etal2014, Krumholz_etal2016, Cunningham_etal2018}), including non-ideal magnetohydrodynamic (MHD) processes \citep{WurBatPri2019}.

One particular question that has been investigated in some detail is how stellar properties are predicted to vary with the metallicity of the molecular gas.  Early radiation hydrodynamical calculations simply scaled the dust opacity of the molecular gas linearly with metallicity \citep{Myers_etal2011, Bate2014} and found that the resulting stellar mass distribution was invariant, even when changing the metallicity from 1/100 to 3 times the solar value (Z$_\odot$).  

More recently, the effects of varying the metallicity of the gas have been treated in more detail.  \cite{BatKet2015} combined radiative transfer with a thermochemical model of the diffuse interstellar medium (ISM), in which gas and dust temperatures are treated separately, and prescriptions are included to account for heating and cooling mechanisms that are important for low-density molecular gas (heating from the interstellar radiation field, cosmic rays, and molecular hydrogen formation; cooling via atomic and molecular line emission).  The method also includes simple chemical models for hydrogen (i.e., the fractions in atomic and molecular forms) and carbon (i.e., the fractions in the forms of C$^+$, neutral carbon, and CO).  The \cite{BatKet2015} ISM model is similar in general to that used by \citet{GloCla2012c} to study molecular cloud evolution, although greatly simplified.

Using this more detailed radiative transfer / thermochemical method, \cite{Bate2019} again investigated the dependence of low-mass stellar properties on metallicity.  However, the overall result was the same: it was found that the low-mass stellar mass distribution did not vary greatly if the metallicity was between 1/100 to 3 times the solar value (Z$_\odot$).  When considering the effects of metallicity on the thermodynamics of collapsing and fragmenting molecular gas, there are two competing effects.  Low-metallicity gas is typically warmer than high-metallicity gas, as expected due to the reduced ability of low-metallicity gas to cool.  Since the Jeans mass depends on gas temperature, $T_{\rm gas}$, as $M_{\rm J} \propto T_{\rm gas}^{3/2}$, this may be expected to reduce gravitational fragmentation and, therefore, perhaps to increase stellar masses.  However, a competing effect is that at lower-metallicities, collapsing molecular gas remains optically thin (and able to cool) to higher densities.  This means that the so-called `opacity limit for fragmentation' \citep{Hoyle1953, LowLyn1976, Rees1976, BoyWhi2005, Whitworth_etal2024} is reached later and since the Jeans mass also depends on density, $\rho$, as $M_{\rm J} \propto \rho^{-1/2}$ this increases the amount of fragmentation on small-scales.  For present-day star formation, in environments similar to those found in Galactic star-forming regions near the Sun, these competing effects (less cooling on large scales, but more cooling on small scales) apparently cancel out so as to give a relatively invariant low-mass stellar initial mass function (IMF).  

Note that \cite{ChoOmuSch2021} performed star formation calculations, including a chemical model, with metallicities ranging from $10^{-6} - 0.1$~Z$_\odot$ and found that the characteristic (median) stellar mass decreased with decreasing metallicity from 0.1 to $10^{-3}$~Z$_\odot$, before increasing again from $10^{-4}-10^{-6}$~Z$_\odot$.  However, they did not include the effects of radiation transport or radiation from protostars.  When heating from protostars was included \citep{Chon_etal2024}, no significant variation in the median stellar mass was found for metallicities $10^{-3}-1$~Z$_\odot$.

Although the stellar mass distribution is relatively invariant to metallicity, the enhanced small-scale fragmentation at low metallicity does have several important effects on the stars \citep{Bate2019}.  First, the frequency of close binary stars is higher at lower metallicity, due to both the increased small-scale fragmentation (including cloud and disc fragmentation), and the fact that first hydrostatic cores \citep[pressure supported objects $\sim 5$~au in radius;][]{Larson1969} cool more quickly and collapse to form stellar cores more quickly (thus, reducing the time that first hydrostatic cores have to merge, rather than form binary protostars).  Such an  anti-correlation between close binary fraction and metallicity has been observed for solar-type stars (\citealt{Badenes_etal2018}; \citealt*{ElBRix2019,MoeKraBad2019}).  Second, in the calculations there is also an increase in the fraction of protostellar mergers at low metallicities.  Third, the protostellar discs produced in the calculations are found to have smaller characteristic radii of protostellar discs with decreasing metallicity and the discs and orbits of pairs of protostars tend to be less well aligned at lower metallicity \citep{ElsBat2021}.

Recently, \cite{Bate2023} presented the results from similar low-mass star formation calculations, but instead of taking the star formation to be in a present-day Galactic environment, they began to explore the dependence of low-mass star formation on the age of the Universe.  Specifically, calculations that were identical to those of \cite{Bate2019} were performed, except that the cosmic microwave background (CMB) radiation that comes into the ISM model was modified to be that at a redshift, $z=5$.  Three calculations with metallicities of 1/100, 1/10, and 1 Z$_\odot$ were performed.  For these calculations, the stellar mass distributions for the two low-metallicity calculations were indistinguishable from the present-day mass functions.  But for the solar metallicity case there was a substantial deficit of brown dwarfs and low-mass stars, producing a bottom-light IMF.  Equivalently, the characteristic (median) stellar mass was found to increase from $M_{\rm char} \approx 0.15$~M$_\odot$ to $M_{\rm char} \approx 0.40$~M$_\odot$.  Thus, the stellar IMF is found to be insenstive to the metallicity for present-day star formation ($z=0$), but metallicity dependent for ($z=5$).  

The reason that the $z=5$ solar metallicity calculation becomes bottom-light compared to the $z=0$ calculation is due to the change in the thermal evolution of high-density molecular gas.  At high metallicities in present-day star formation, collapsing highest density gas becomes very cold due to dust extinction of the interstellar radiation field ($T_{\rm gas} \approx 6-10$~K).  However, the temperature of the CMB scales proportional to $(1+z)$, so rather than the present-day value of 2.73~K, at $z=5$ the CMB temperature is 16.4~K.  This long-wavelength radiation is able to penetrate deep within molecular clouds, essentially providing a temperature floor and keeping even the densest gas much warmer at $z=5$ compared to $z=0$.  This in turn leads to less fragmentation and, thus, fewer low-mass stars.  A similar effect of the CMB radiation acting as a `temperature floor' at high redshift was reported by \cite{Chon_etal2022} who performed calculations at $z=0,5,10,20$.  They also found that the median stellar mass increased with increasing redshift at intermediate to high metallicity ($Z=0.01-0.1$~Z$_\odot$).  However, they did not include radiation transport or heating from protostars in these calculations, but they found that including such heating substantially increased the median masses in their recent present-day star formation calculations \citep{Chon_etal2024}.  Thus, it is difficult to meaningfully compare the \cite{Chon_etal2022} and \cite{Bate2023} results.

In this paper, we map out the redshift-metallicity dependence of the stellar IMF in more detail.  We report results from 20 radiation hydrodynamical calculations of stellar cluster formation in molecular clouds that have identical initial conditions to each other, except for the CMB radiation that the clouds are subjected to and the metallicity of the gas.  The calculations include the \cite{BatKet2015} combined radiative transfer and thermochemical model.  The calculations at redshift $z=0$ are extensions in time of those reported by \cite{Bate2019}, and the calculations at redshift $z=5$ and with metallicity $Z=0.01-1$~Z$_\odot$ are extensions of those published in \cite{Bate2023}.  The other 13 calculations are entirely new and have been performed specifically for this paper.  We follow the collapse of each of the molecular clouds to form a cluster of stars and then compare the mass distributions of the stars and brown dwarfs.  In Section \ref{sec:method} we provide summaries of the method and initial conditions. The results are presented in Section \ref{sec:results}, and we fit the numerical results with analytic expressions for the stellar mass functions and present a parameterisation that can be used to predict the form of the stellar initial mass function in the metallicity and redshift ranges of $Z=0.01-3$~Z$_\odot$ and $z=0-10$, respectively.   In Section \ref{sec:discussion}, we compare our results to other theoretical studies, and we discuss observational evidence for variation of the stellar mass function in the context of our results.  Finally, in Section \ref{conclusions} we present our conclusions.

\section{Method}
\label{sec:method}

The method used to perform the calculations is almost identical to
that used for the present-day (redshift $z=0$) calculations that were
presented by \cite{Bate2019} and the $z=5$ calculations of \cite{Bate2023}.
The smoothed particle hydrodynamics (SPH) code, {\sc sphNG}, 
was used.  This code originated from \citeauthor{Benz1990} 
(\citeyear{Benz1990}; \citealt{Benzetal1990}), but has been substantially
modified and extended over the past 30 years using the methods described 
in \citet{BatBonPri1995}, \citet{PriMon2007},
\citet*{WhiBatMon2005}, \citet{WhiBat2006}, \cite{BatKet2015} and 
parallelised using both OpenMP and MPI.

The code uses a binary tree to compute the gravitational forces between 
particles and a particle's nearest neighbours.  
The calculations used the standard M4 SPH kernel and the 
smoothing lengths of particles were set such that the smoothing
length of each particle $h = 1.2 (m/\rho)^{1/3}$ where $m$ is the 
SPH particle's mass \cite[see][for further details]{PriMon2007}. 
As in the earlier calculations, to reduce numerical 
shear viscosity, the \cite{MorMon1997} artificial viscosity was employed 
with $\alpha_{\rm_v}$ varying between 0.1 and 1 while $\beta_{\rm v}=2 \alpha_{\rm v}$
\citep[see also][]{PriMon2005}.

The {\sc sphNG} code employs individual time steps for each particle
\citep*{BatBonPri1995}.  The code has the choice of
two second-order integrators: the original Runge-Kutta-Fehlberg 
integrator \citep{Fehlberg1969}, and leap-frog integrator 
\citep[the later is implemented as in the {\sc Phantom} SPH code;][]{Price_etal2018}.
The calculations of \cite{Bate2019} and \cite{Bate2023} used the
Runge-Kutta-Fehlberg integrator, but the leap-frog integrator was used 
to perform the new calculations for this paper and to extend the previous $z=0,5$ 
calculations to later times for two reasons.  First, the leap-frog
integrator only requires a single force calculation per time step and, therefore, is
somewhat quicker.  Second, when the orbit of two bound sink particles 
(used to model protostars) is integrated using the Runge-Kutta-Fehlberg
integrator the orbital separation slowly increases and a very small integrator
tolerance needs to be used to integrate the orbits accurately.  This slow
increase of the separation does not occur when using the leap-frog
integrator.
Calculations of stellar cluster formation have been performed using both 
integrators and compared; the statistical results and conclusions
are not altered by the choice of integrator.  In addition, as will be seen below, 
although the Runge-Kutta-Fehlberg integrator was used to perform the 
bulk of the $z=0$ and $z=5$ calculations, and the leap-frog integrator
was used to perform the $z=3,7,10$ calculations, the trends in the stellar
properties that we obtain from the calculations are consistent regardless of
which integrator was used.

\begin{figure}
\vspace{-0.5cm}
\centering
    \includegraphics[width=9cm]{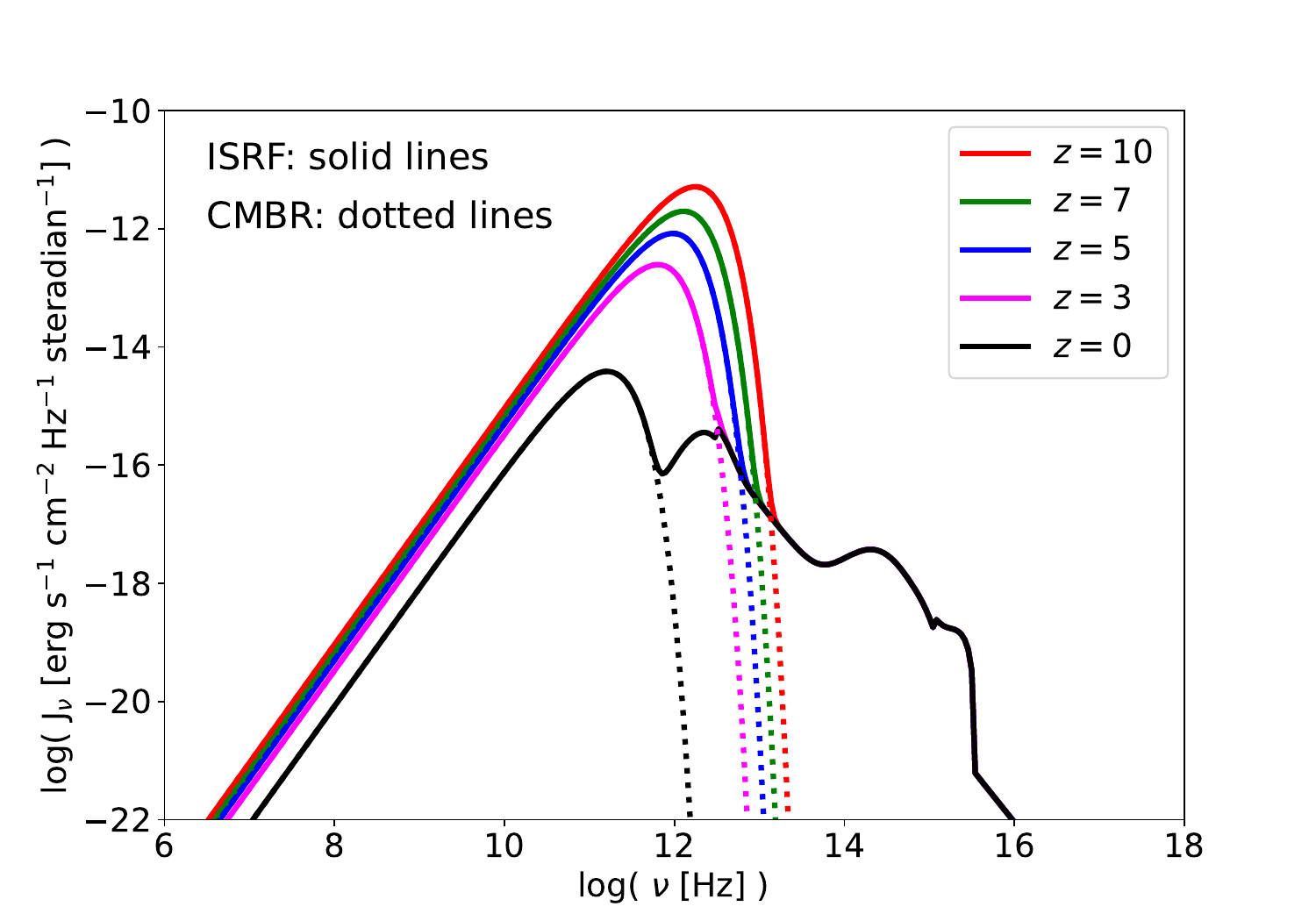} \vspace{-0.5cm}
\caption{The forms of the interstellar radiation field (ISRF) that are used for the calculations at redshifts $z=0, 3, 5, 7, 10$ (shown using different colours).   The solid lines show the full ISRFs (the same above frequency $\nu \approx 10^{13}$~Hz for all redshifts), while the dotted lines show the continuation of the contributions from the cosmic microwave background radiation (CMBR). It is assumed that the only difference for the ISRFs at different redshifts is that the CMBR has a greater temperature at a higher redshift.}
\label{ISRF}
\end{figure}

\begin{table*}
\begin{tabular}{lcccccccccccc}\hline
Redshift   & Metallicity & Objects & Brown  & Mass of Stars \&  & Mean  & Mean & Median & Stellar & L$_3$ Fit & Mass-to-light\\
& $Z$ & Formed & Dwarfs  & Brown Dwarfs & Mass & Log-Mass & Mass & Mergers & $\mu$ & Ratio over \\
&  Z$_\odot$ & & & M$_\odot$ & M$_\odot$ &M$_\odot$ & M$_\odot$ & & & Chabrier value \\ 
\hline
$z=0$ &   0.01 & 175  & $\leq 50$ & 81.8 & $0.47\pm0.07$ & $0.19\pm0.02$ & 0.20 &  23 & $0.19\pm0.02$  & 0.92   \\   
           &  0.1 & 209  & $\leq 49$ & 105.8 & $0.51\pm0.06$ & $0.20\pm0.02$ & 0.18 &  15 &  $0.19\pm0.02$ & 0.94  \\   
           &  1.0 & 272  & $\leq 76$ & 126.7 & $0.47\pm0.06$ & $0.18\pm0.01$ & 0.16 & 23 & $0.17\pm0.01$  & 1.18  \\  
           & 3.0 & 284  & $\leq 61$& 131.5 & $0.46\pm0.05$ & $0.20\pm0.02$ & 0.18 & 6 & $0.19\pm0.02$ & 0.85 \\   
\hline
$z=3$ & 0.01 & 236 & $\leq 78 $ &  75.8 & $0.32\pm0.04$  & $0.13\pm0.01$ & 0.12 & 25 & $0.13\pm0.01$ & 1.97 \\   
           &  0.1 &  168 & $\leq 42 $ & 100.9 & $0.60\pm0.08$ & $0.23\pm0.03$ & 0.24 & 13 & $0.23\pm0.02$ & 0.65 \\     
           &  1.0 &  191 & $\leq 39 $ & 117.1 & $0.61\pm0.09$ & $0.25\pm0.02$ & 0.23 &  9  & $0.24\pm0.02$ &  0.57  \\   
           &  3.0 &  142 & $\leq 21 $ & 121.3 & $0.85\pm0.13$ & $0.34\pm0.04$ & 0.30 &  7  & $0.32\pm0.04$ & 0.30 \\  
\hline
$z=5$ &  0.01 & 214 & $\leq 72$ & 80.6   & $0.38\pm0.04$ & $0.14\pm0.01$ & 0.13 & 22 & $0.14\pm0.01$  & 1.69   \\  
           &  0.1   & 139 & $\leq 38$ & 93.5   & $0.67\pm0.09$ & $0.23\pm0.03$ & 0.19 & 11 & $0.21\pm0.03$  & 0.71    \\   
           &  1.0   &   87 & $\leq 8$   & 98.2   & $1.13\pm0.19$ & $0.47\pm0.07$ & 0.47 & 4  & $0.46\pm0.07$  & 0.15  \\  
           &  3.0   &   71 & $\leq 1$   & 109.4 & $1.54\pm0.27$ & $0.73\pm0.11$ & 0.60 & 3   & $0.70\pm0.11$  &  0.062  \\  
\hline
$z=7$ & 0.01 & 166 & $\leq 58 $ & 69.7 & $0.42\pm0.07$ & $0.15\pm0.02$ & 0.13 &  24 & $0.14\pm0.02$ & 1.58 \\   
           &  0.1 & 104  & $\leq 26 $ & 87.2 & $0.84\pm0.13$ & $0.27\pm0.04$ & 0.23 &  13 & $0.25\pm0.04$  &  0.54 \\     
           &  1.0 & 61    & $\leq 1  $  & 93.0 & $1.53\pm0.26$ & $0.79\pm0.13$ & 0.75 &   0  & $0.79\pm0.12$  & 0.047    \\   
           &  3.0 & 29    & 0               & 89.0 & $3.07\pm0.57$ & $1.72\pm0.41$ & 2.20 &    2 & $1.84\pm0.46$  & 0.0083  \\  
\hline
$z=10$ & 0.01 & 117 & $\leq 36 $ & 71.7 & $0.61\pm0.11$ & $0.19\pm0.03$ & 0.15 & 9   & $0.18\pm0.03$ & 0.96 \\   
            &  0.1   &  77 & $\leq 11 $ & 69.8  & $0.91\pm0.13$ & $0.41\pm0.07$ & 0.44 & 11 & $0.42\pm0.07$ & 0.17  \\     
            &  1.0   & 42  & $0 $         & 83.2  & $1.98\pm0.29$ & $1.17\pm0.23$ & 1.60 & 2   & $1.28\pm0.27$   & 0.017 \\   
            &  3.0   & 21  &  $0$         & 84.8  & $4.04\pm0.75$ & $2.92\pm0.59$ & 3.02 & 2   & $3.13\pm0.79$   & 0.0026 \\  
\hline
\end{tabular}
\caption{\label{table1} The statistical properties of the stellar populations produced by each of the 20 radiation hydrodynamical calculations.  The calculations were run to 1.30~$t_{\rm ff}$ and use sink particles with $r_{\rm acc}=0.5$~au and no gravitational softening to model the stars and brown dwarfs.  Brown dwarfs are defined as those objects that have final masses less than 0.075 M$_\odot$.  The numbers of brown dwarfs are upper limits because some were still accreting at the end of the calculations.  At $z=0$, lower metallicities result in less gas being converted to stars and fewer objects being formed at the same time, but the mean and median masses of the stellar populations are statistically indistinguishable.  However, as the redshift increases, the highest metallicity calculations become increasingly bottom-light with the mean and median stellar masses increasing both with increasing redshift and increasing metallicity.  There is also a general trend that the average number of stellar mergers per star increases with decreasing metallicity (at all redshifts).}
\end{table*}

\subsection{Radiative transfer and the diffuse ISM model}
\label{hydro}

As for \cite{Bate2019,Bate2023}, the calculations used the \cite{BatKet2015} combined 
radiative transfer and diffuse ISM thermochemical method.  A brief summary of
this method is provided by \cite{Bate2019} and will not be repeated here since
the method is unchanged, except that the form of the interstellar radiation field (ISRF) is varied.  

For solar metallicity, the hydrogen and helium mass fractions are taken to be $X=0.70$ and 
$Y=0.28$, respectively, with the solar abundance taken to be ${\rm Z}_\odot=0.02$.
The equation of state of the gas takes into account the dissociation of molecular hydrogen and
ionisation of hydrogen and helium, but it ignores the contribution of metals.  Thus, the
equation of state of the gas is not changed for calculations with different metallicities. 
The simple chemical model of \cite{KetCas2008} is used to treat the evolution 
of the abundances of C$+$, neutral carbon, CO, and the depletion of CO on to dust grains.  
The atomic and molecular hydrogen abundances used the same molecular hydrogen 
formation and dissociation rates as \cite{Gloveretal2010}.  This simple chemical
model is only used to determine the atomic and molecular line cooling rates of the gas,
which are most important at low-densities.
When we change the metallicity between calculations, the abundances of gas-phase metals and
dust are all assumed to scale linearly with the metallicity.  This is, of course, a simplification
 (see, for example, the discussion of the distinction between metallicity and dust
abundance and how the latter may vary with the former in \cite{Whitworth_etal2024} 
in the context of the minimum stellar mass).
In the calculations presented in this paper the main effect of changing the metallicity is due 
to the change in the dust opacity, which shields the gas from the ISRF and
affects the radiation transport.   In reality, the dust abundance may not scale linearly 
with metallicity, and the properties of the dust grains themselves may vary with the
dust abundance.   In particular, going to low metallicities studies frequently find that the 
dust abundance decreases somewhat more rapidly than the metallicity, and at low
metallicities the dust abundances can vary substantially between different systems,
such as nearby galaxies and damped Lyman-$\alpha$ systems
\citep[e.g.][]{GalGalJon2018, Galliano_etal2021, DeVis_etal2019,Konstantopoulou_etal2024}. 
This should be kept in mind when interpreting the results of the low-metallicity calculations.
For example, if in fact the dust abundance at metallicity $Z=0.1~{\rm Z}_\odot$ is in fact 100 times
lower than the dust abundance at solar metallicity rather than 10 times lower (as has been 
measured in some nearby galaxies), then in fact the calculations presented here as being
appropriate for $Z=0.01~{\rm Z}_\odot$ may be more representative of $Z=0.1~{\rm Z}_\odot$ systems
than of $Z=0.01~{\rm Z}_\odot$ systems.

The star-forming molecular clouds are assumed to be bathed in an ISRF that contributes 
to the heating rate of dust grains and photoelectric heating of the gas.  
The ISRF is attenuated due to dust extinction inside the cloud, with the same opacities 
as in \cite{Bate2014, Bate2019, Bate2023}.
To describe the ISRF at redshift $z=0$ the analytic form of \cite*{ZucWalGal2001} is used, with a modification to
include the `standard' UV component from \cite{Draine1978} in the energy range $h\nu=5-13.6$~eV.
As in \cite{Bate2023}, the functional form of the component of the ISRF that is due to 
the cosmic microwave background radiation (CMBR) is modified to reflect different redshifts.  
The temperature of the CMBR scales as
\begin{equation}
T_{\rm CMBR}(z) = (1+z) T_{\rm CMBR}(0),
\label{eq:TCMBR}
\end{equation}
where $T_{\rm CMBR}(0)=2.73$~K, so that at $z=3, 5, 7, 10$, the $T_{\rm CMBR}(z) = 10.9, 16.4, 21.8, 30.0$~K, respectively.
In Figure \ref{ISRF}, we plot the forms of the ISRF that are used for the calculations at different redshifts, and the contributions of the CMBR to each of the ISRFs.  As in \cite{Bate2023}, we assume that the other contributions to the ISRF do not change with redshift (i.e., we assume that the star-forming clouds are in similar radiative environments, except for the contribution from the CMBR).  Of course, the ISRF may change with environment, for example, in a starburst environment there may be a stronger high-frequency component.  But, for the present paper we only consider changes to the CMBR.

\subsection{Sink particles}
\label{sec:sinks}

As in \cite{Bate2012, Bate2019, Bate2023}, the calculations followed the collapse of each protostar 
through the first hydrostatic core phase and into the second collapse phase (that begins at densities of
$\sim 10^{-7}$~g~cm$^{-3}$) due to the dissociation of molecular hydrogen \citep{Larson1969}.
Sink particles \citep{BatBonPri1995} were inserted when the density exceeded
$10^{-5}$~g~cm$^{-3}$, which is approximately two orders of magnitude before the
stellar core begins to form. The associated free-fall time at this density is only one week.

A sink particle is formed by turning the densest SPH gas particle in a region undergoing 
the second collapse (the first particle in that region to exceed $10^{-5}$~g~cm$^{-3}$), 
and all SPH gas particles contained within the accretion radius
$r_{\rm acc}=0.5$ au of that particle, into a point mass with their combined mass and momentum.  
The method by which this is done is as described in \cite{BatBonPri1995},
except that when radiative transfer is used (as for the calculations discussed in this paper) 
the checks on the boundedness of the gas within the SPH smoothing kernel that are
described in \cite{BatBonPri1995} are omitted.  Instead, we rely on the physical fact that 
when gas reaches $10^{-5}$~g~cm$^{-3}$ the formation of a `stellar core' \citep{Larson1969} 
is inevitable, because in calculations such as these such high densities can only 
be reached deep within a first hydrostatic core when the dissociation of molecular hydrogen is
almost complete.  It should be noted that, using this criterion, sink particles are inserted a substantial amount of 
time after a bound first hydrostatic core first forms (typically several thousand years), because 
after it is first formed a first hydrostatic core needs to accrete more mass and radiate away 
energy before its central region becomes hot enough to dissociate molecular hydrogen.
As described in \cite{BatBonPri1995},
gas that subsequently falls within the accretion radius is accreted by the sink particle 
if it is bound and its specific angular momentum is less than 
that required to form a circular orbit at radius $r_{\rm acc}$. 
The sink particles interact with the gas only via accretion and gravity.
The sink particles themselves do not contribute radiative feedback 
\citep[see][for detailed discussions of this limitation]{Bate2012,JonBat2018b}, 
although any hot dense gas in the discs surrounding them does radiate to the
surroundings.
The gravitational forces between sink particles are not softened at all, but
sink particles whose centres pass within 0.03 au of each other (i.e., $\approx 6$~R$_\odot$)
are merged.

\subsection{Initial conditions and resolution}
\label{initialconditions}

The same initial density and velocity structure is used for each calculation, which is also the same as that which was used by \cite{Bate2012, Bate2014, Bate2019, Bate2023}.  Only the redshift or metallicity is changed for each calculation, allowing close comparison of the stellar properties between the calculations.  
\cite{Bate2012} provides a detailed description of the initial conditions.  
Each calculation begins with a uniform-density, spherical, molecular cloud containing 
500 M$_\odot$ of gas, with a radius of 0.404 pc (83300 au).  Thus, the initial density is 
$1.2\times 10^{-19}$~g~cm$^{-3}$ (hydrogen number density, $n_{\rm H} \approx 6\times 10^4$~cm$^{-3}$) 
and the initial free-fall time is
$t_{\rm ff}=6.0\times 10^{12}$~s or $1.90\times 10^5$ years.    
Although the clouds have a uniform density, they have an initial 
supersonic `turbulent' velocity field that was generated in the same manner
as \citet*{OstStoGam2001} and \cite*{BatBonBro2003}.
This is a divergence-free random Gaussian velocity field with 
a power spectrum $P(k) \propto k^{-4}$, where $k$ is the wavenumber. 
It was generated on a $128^3$ uniform grid, with the velocities of the particles interpolated from the grid.
The same initial velocity field is used for all the calculations, 
and there is no imposed turbulent driving (i.e., the calculations assume `decaying turbulence').
The velocity field was normalised so that the kinetic energy 
of the turbulence was equal to the magnitude of the gravitational potential 
energy of the cloud.  
This initial level of kinetic energy is a somewhat arbitrary.
Significant initial kinetic energy is required to generate highly structured gas (similar to that observed
in molecular clouds) prior to the star formation beginning.  Conversely, if the initial velocity field is
too strong much of the gas will be unbound and will not be involved in star formation
\citep[see, for example,][who studied star formation in unbound clouds]{Clark_etal2005}.  
It should be noted that the initial thermal energy is comparatively small ($\approx 2-4$\% of the
initial kinetic/gravitational energies, depending on the calculation). 
When beginning with the kinetic energy equal to the magnitude of the gravitational potential energy,
the total kinetic energy drops quickly as structure is formed within the gas.  By $\approx 0.6~t_{\rm ff}$
the kinetic energy has dropped to around half of the magnitude of the gravitational potential energy.
Star formation begins at around $0.7-0.9~t_{\rm ff}$, depending on the calculation \citep[e.g.,][]{Bate2019,Bate2023}.

The initial gas and dust temperatures were set so that the dust was in thermal equilibrium with the ISRF.  Similarly, the gas was initially in thermal equilibrium such that heating due to the ISRF and cosmic rays was balanced by cooling from atomic and molecular line emission and collisional coupling with the dust.  The resulting clouds have dust and gas temperatures that are coolest at the centre and warmest on the outside.  

Each calculation used $3.5 \times 10^7$ SPH particles to model the cloud (the same resolution as in \citealt{Bate2019, Bate2023}), which is sufficient to resolve the local Jeans mass throughout the calculation and necessary to capture fragmentation correctly (\citealt{BatBur1997, Trueloveetal1997, Whitworth1998, Bossetal2000}; \citealt*{HubGooWhi2006}).  There is no consensus as to the resolution that is necessary and sufficient to model disc fragmentation (see Section 2.3 of \citealt{Bate2019} for discussion of this issue and further references).  This should be noted as a caveat for the results presented in this paper.

\begin{figure*}
\centering
    \includegraphics[width=17.5cm]{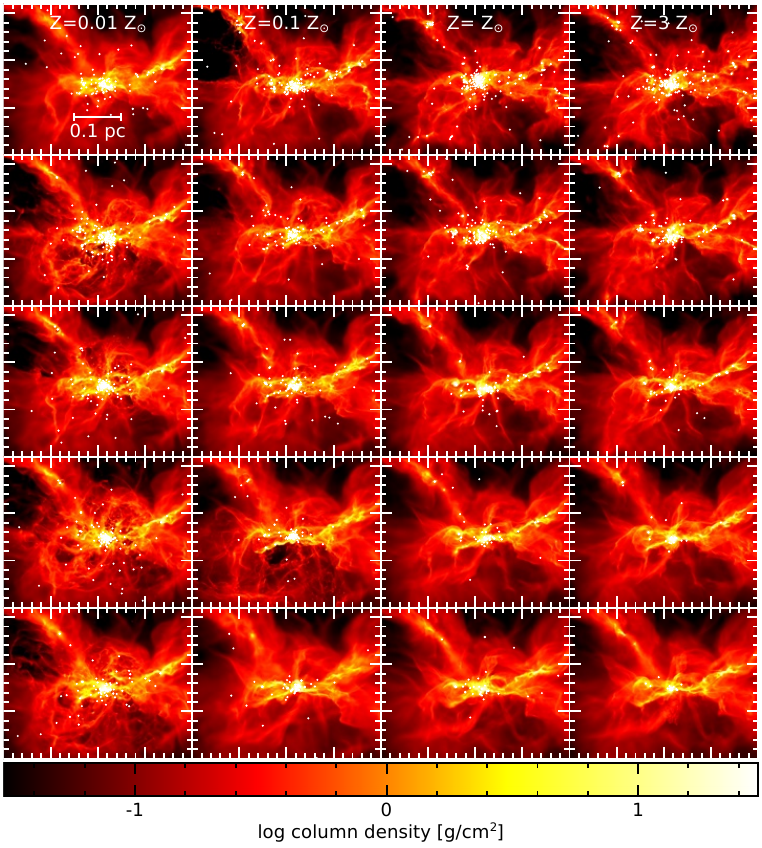} \vspace{-0.2cm}
\caption{Column density at the end of each of the twenty calculations ($t=1.30$~$t_{\rm ff}$ ).  From top to bottom, the rows are for calculations at redshifts $z=0, 3, 5, 7, 10$, and from left to right the calculations have metallicities of $Z=1/100$, 1/10, 1, 3 Z$_\odot$. The colour scale is logarithmic, ranging from $0.03-30$~g~cm$^{-2}$.  The stars and brown dwarfs are plotted using white circles. The overall cloud structure is similar in all the calculations, because the same initial turbulent velocity field was used for each calculation, but the structure is generally smoother with increasing redshift, particularly at high metallicity, due to the warmer CMBR temperature and reduced fragmentation.}
\label{fig:Density}
\end{figure*}

\begin{figure*}
\centering
    \includegraphics[width=17.5cm]{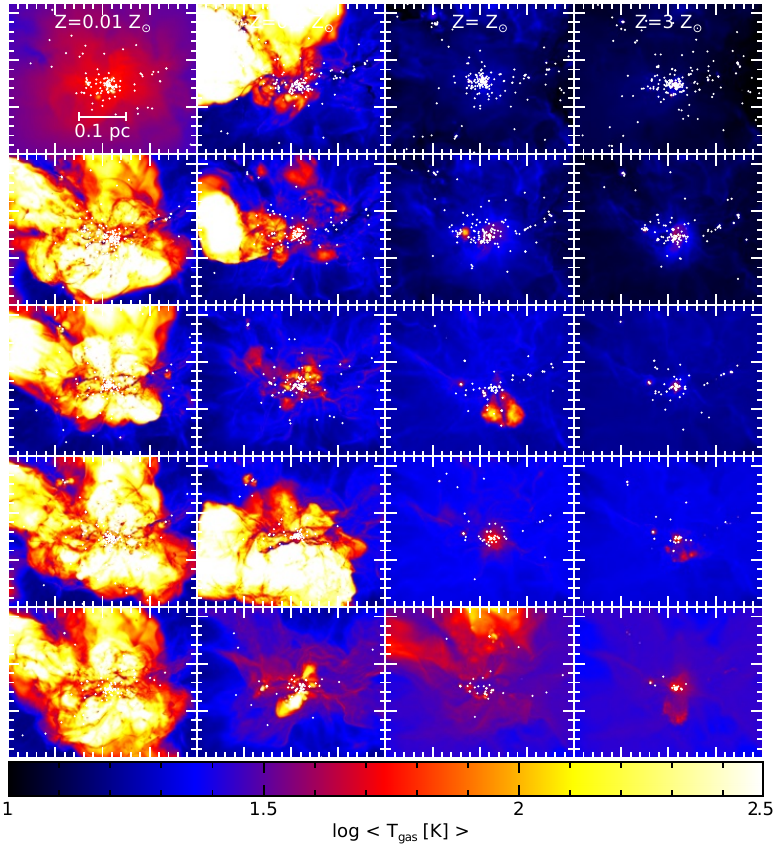} \vspace{0cm}
\caption{The mass-weighted gas temperature, $T_{\rm gas}$, at the end of each of the twenty calculations ($t=1.30$~$t_{\rm ff}$ ).  From top to bottom, the rows are for calculations at redshifts $z=0, 3, 5, 7, 10$, and from left to right the calculations have metallicities of $Z=1/100$, 1/10, 1, 3 Z$_\odot$. The colour scale is logarithmic, ranging from $T_\mathrm{gas} = 10-300$~K.  The stars and brown dwarfs are plotted using white circles. Generally, the gas temperatures are higher at lower metallicity (e.g., the second row), and somewhat higher with increasing redshift (e.g., the right column).  However, once star formation begins the gas temperature on large scales tends to vary greatly with time because of the radiation produced as mass falls into the gravitational potentials of the protostars.  In the low-metallicity (1/100 and 1/10~Z$_\odot$) calculations, in which the gas cools slowly, this can produce large warm gas `bubbles', particularly towards the end of the calculations.}
\label{fig:Tgas}
\end{figure*}

\begin{figure*}
\centering
    \includegraphics[width=17.5cm]{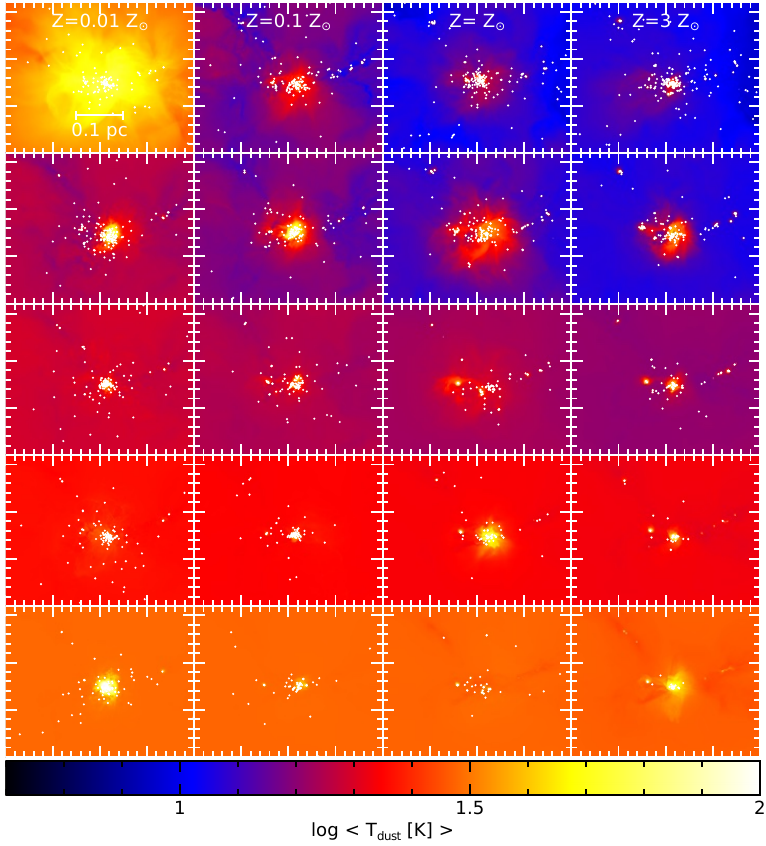} \vspace{0cm}
\caption{The mass-weighted dust temperature, $T_\mathrm{dust}$, at the end of each of the twenty calculations ($t=1.30$~$t_{\rm ff}$ ).  From top to bottom, the rows are for calculations at redshifts $z=0, 3, 5, 7, 10$, and from left to right the calculations have metallicities of $Z=1/100$, 1/10, 1, 3 Z$_\odot$. The colour scale is logarithmic, ranging from $T_\mathrm{dust} = 5-100$~K.  The stars and brown dwarfs are plotted using white circles. Generally, the dust temperatures are higher with increasing redshift due to the warmer CMBR.  At low redshifts ($z \lsim 5$), the dust tends to be cooler at higher metallicity because of the increasing extinction of the short-wavelength component of the interstellar radiation field (which dominates over the CMBR component at low redshift). }
\label{fig:Tdust}
\end{figure*}

\begin{figure*}
\centering
    \includegraphics[width=17.5cm]{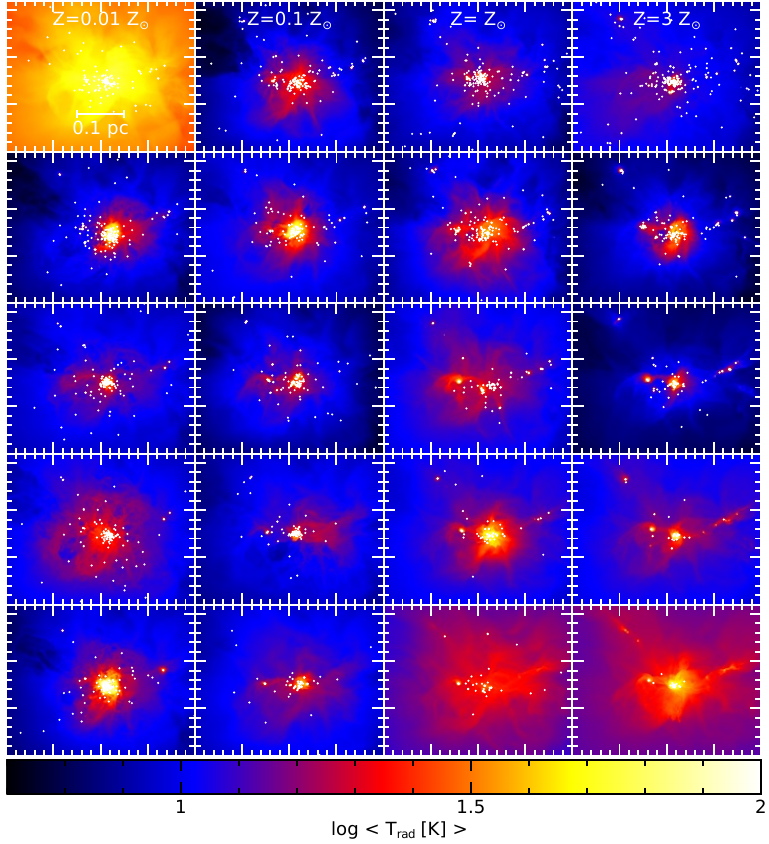} \vspace{0cm}
\caption{The mass-weighted temperature of the radiation field created by the star-formation activity, $T_{\rm rad}$, at the end of each of the twenty calculations ($t=1.30$~$t_{\rm ff}$ ).  From top to bottom, the rows are for calculations at redshifts $z=0, 3, 5, 7, 10$, and from left to right the calculations have metallicities of $Z=1/100$, 1/10, 1, 3 Z$_\odot$. The colour scale is logarithmic, ranging from $T_{\rm rad} = 5-100$~K.  These figures show the temperature of the radiation field produced by the gas falling onto the protostars.  This tends to be highly variable in time as the accretion rates fluctuate.  Generally the temperatures of the protostellar radiation field are highest near the centre of the forming clusters.  The protostellar radiation field $z=0$, $Z=0.01~\mathrm{Z}_\odot$ (top left panel) happens to gets hot towards the end of the calculation -- for much of its evolution it was more similar to the other panels.  The $z=10$, high-metallicity (solar and 3 times solar) calculations tend to have strong radiation fields because the protostars are quite massive and they are accreting at high rates (bottom right panels). }
\label{fig:Trad}
\end{figure*}

\section{Results}
\label{sec:results}

Twenty radiation hydrodynamical calculations were performed, covering four different gas metallicities ($Z=0.01, 0.1, 1, 3~\mathrm{Z}_\odot$) each at five different redshifts ($z=0, 3, 5, 7, 10$).  The initial conditions were identical except for the metallicity of the gas and the external radiation field that they were subjected to, with calculations at higher redshift having a warmer CMBR field.  By using the same initial conditions except for the metallicity and redshift, the effects of these two variables on the resulting stellar properties can be studied.

Seven of these calculations are continuations of calculations that have already been published.  \cite{Bate2019} presented the results from four calculations at $z=0$ with the same four metallicities.  \cite{Bate2023} presented the results from three calculations at $z=5$ with three metallicities ($Z=0.01, 0.1, 1~\mathrm{Z}_\odot$).  In both papers the calculations were evolved to 1.20 initial cloud free-fall times ($\approx 2.28 \times 10^5$ yrs).  For this paper, these calculations were extended to $t=1.30~t_\mathrm{ff}$ ($\approx 2.47\times 10^5$ yrs), and thirteen completely new calculations were also run to $t=1.30~t_\mathrm{ff}$ (four calculations at each of $z=3, 7, 10$, and one calculation at $z=5$ with $Z=3$~Z$_\odot$).  The calculations were evolved further than in the earlier papers to increase the numbers of objects formed, so improving the statistical characterisation of the stellar populations.

\begin{figure*}
\centering 
    \includegraphics[width=17.5cm]{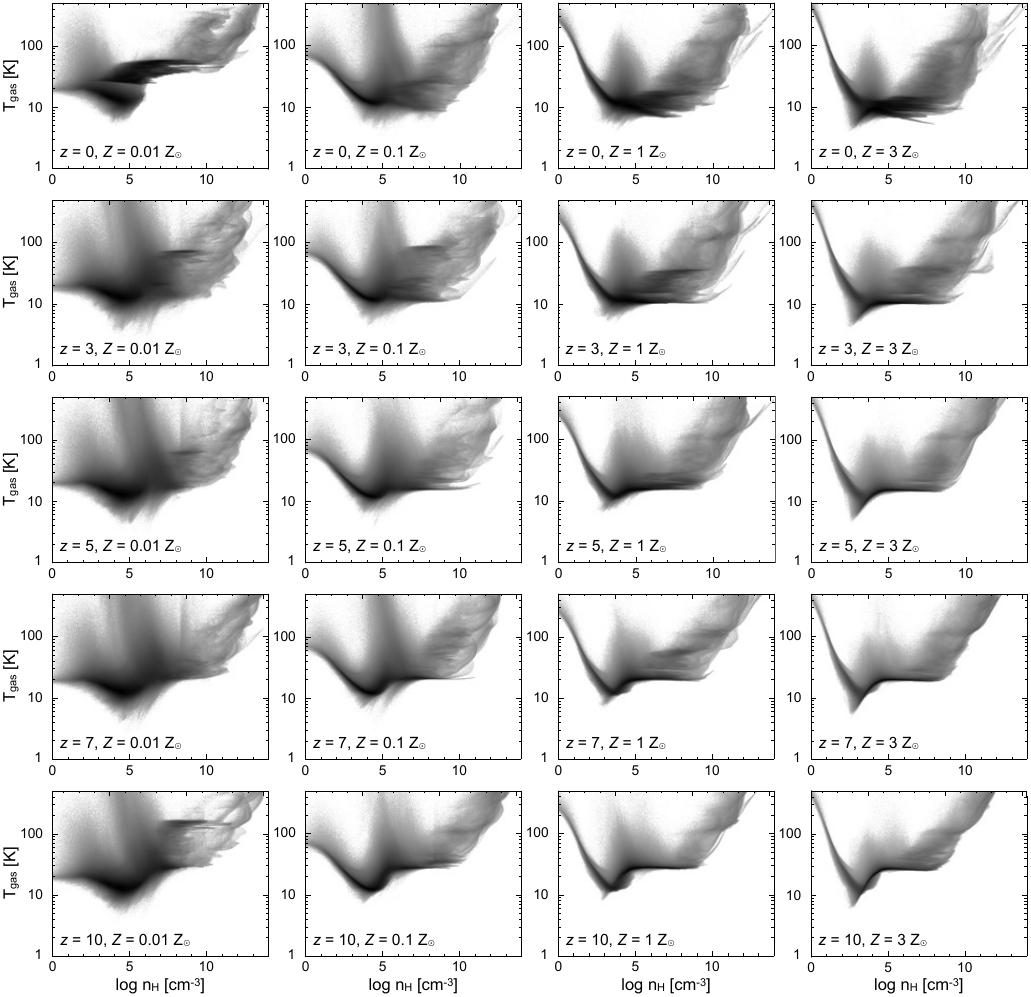} \vspace{0cm}
\caption{Phase diagrams of gas temperature vs molecular cloud density at the end of each of the twenty calculations ($t=1.30$~$t_{\rm ff}$). The grey scale is proportional to the logarithm of the number of fluid elements (i.e., darker regions contain more SPH particles).  The low-density gas ($n_{\rm H}\lsim 10^{8}~{\rm cm}^{-3}$) tends to be warmer at low metallicity than with high metallicity due to the poor cooling, but the high-density gas ($n_{\rm H}\gsim 10^{10}~{\rm cm}^{-3}$) tends to be cooler at low metallicity due to the reduced optical depths that allow it to cool more rapidly. At high densities, the gas temperatures become coupled to the dust temperatures.  }
\label{fig:pixgas}
\end{figure*}

\begin{figure*}
\centering 
    \includegraphics[width=17.5cm]{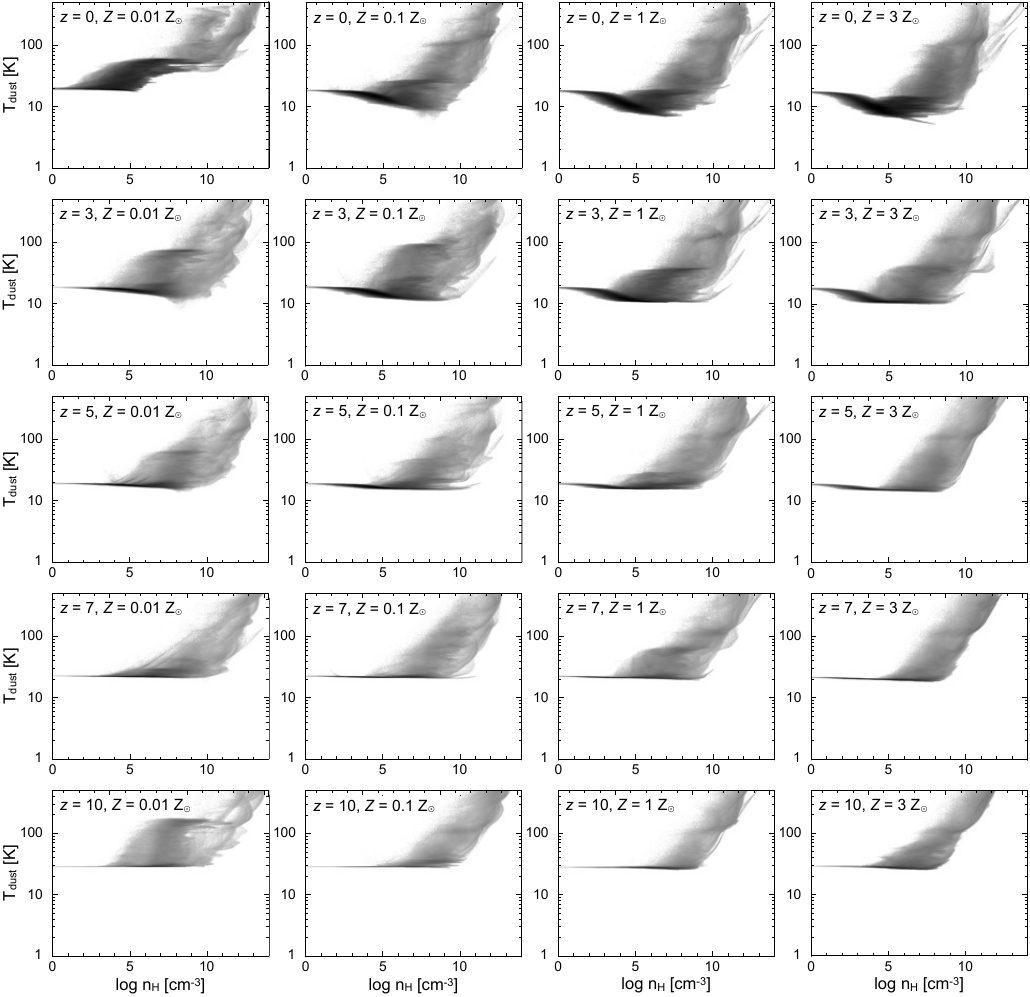} \vspace{0cm}
\caption{Phase diagrams of dust temperature vs molecular cloud density at the end of each of the twenty calculations ($t=1.30$~$t_{\rm ff}$). The grey scale is proportional to the logarithm of the number of fluid elements (i.e., darker regions contain more SPH particles).  The `temperature floor' that increases with increasing redshift due to the increasing temperature of the CMBR is clearly visible (from top to bottom).  At low redshifts ($z=0, 3$), the CMBR is relatively weak and much of the dust heating comes from the short-wavelength component of the ISRF.  Short-wavelength radiation does not penetrate well into dense gas with high-metallicity, so in the top two rows the minimum dust temperature clearly decreases with increasing metallicity.  For the present-day calculations \citep[see also][]{Bate2019} dust temperatures can be as low as $5-6$~K with metallicity $Z \gsim {\rm Z}_\odot$. But at higher redshifts the coolest dust temperatures are much higher.  Since dust cooling is crucial for gas cooling at high densities and metallicities, this leads to increasingly higher gas temperatures at high densities as the redshift increases for $Z \gsim 0.1 ~{\rm Z}_\odot$.  }
\label{fig:pixdust}
\end{figure*}

\begin{figure*}
\centering
    \includegraphics[width=17.5cm]{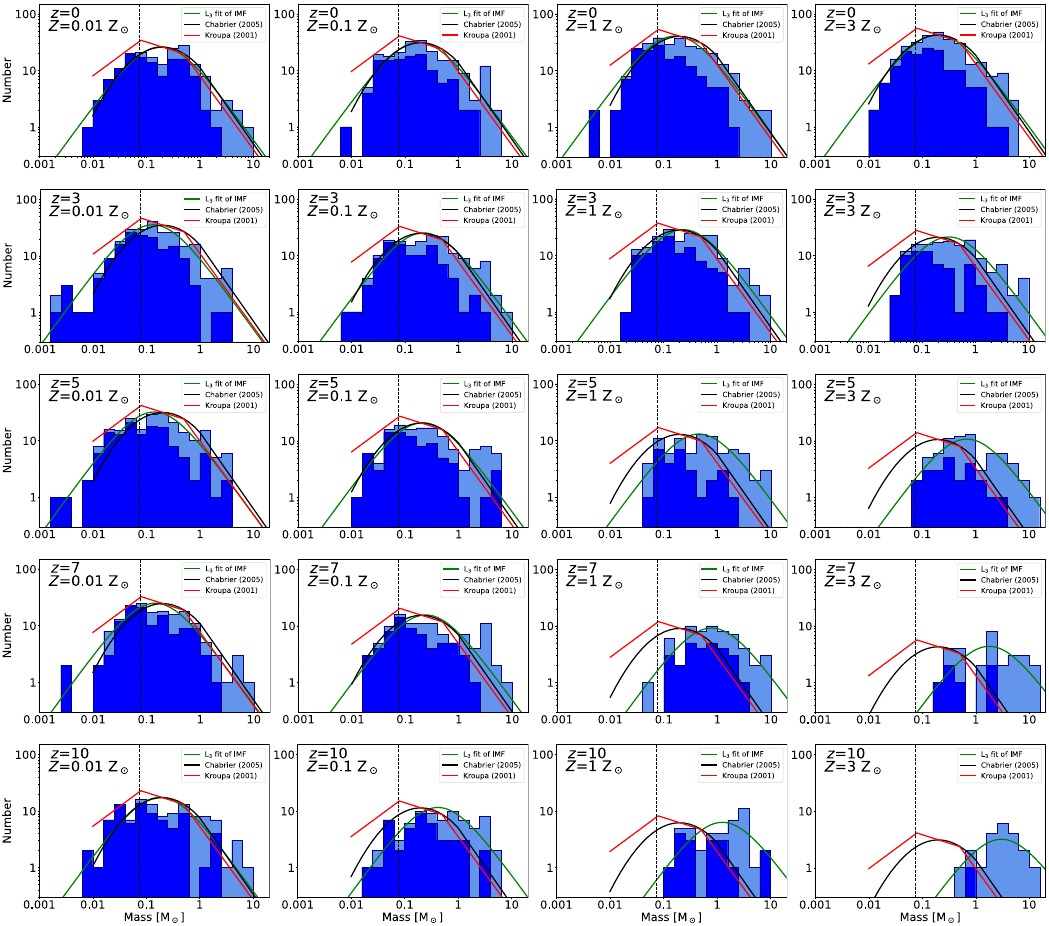} \vspace{0cm}
\caption{The differential mass functions at the end of each of the 20 calculations:  from top to bottom the rows are for redshifts $z=0,3,5,7,10$, and from left to right the columns are for metallicities $Z=0.01, 0.1, 1, 3~{\rm Z}_\odot$, respectively. The dark blue histograms show those objects which have reached their final masses, defined as having accretion rates less than $10^{-7}$~M$_\odot \mathrm{yr}^{-1}$, while the light blue histograms include those objects that are still accreting significantly when the calculation was stopped. Also over plotted are the Kroupa (2001) IMF (red lines), the Chabrier (2005) stellar IMF (black lines), and the L$_3$ functional fit to the numerical mass functions (green lines).  The vertical dashed line denotes the star/brown dwarf boundary (0.075~M$_\odot$). The mass function becomes increasingly bottom-light with increasing redshift at high metallicity.}
\label{fig:IMFdiff}
\end{figure*}

\begin{figure*}
\centering \vspace{-0.2cm}
    \includegraphics[width=17.5cm]{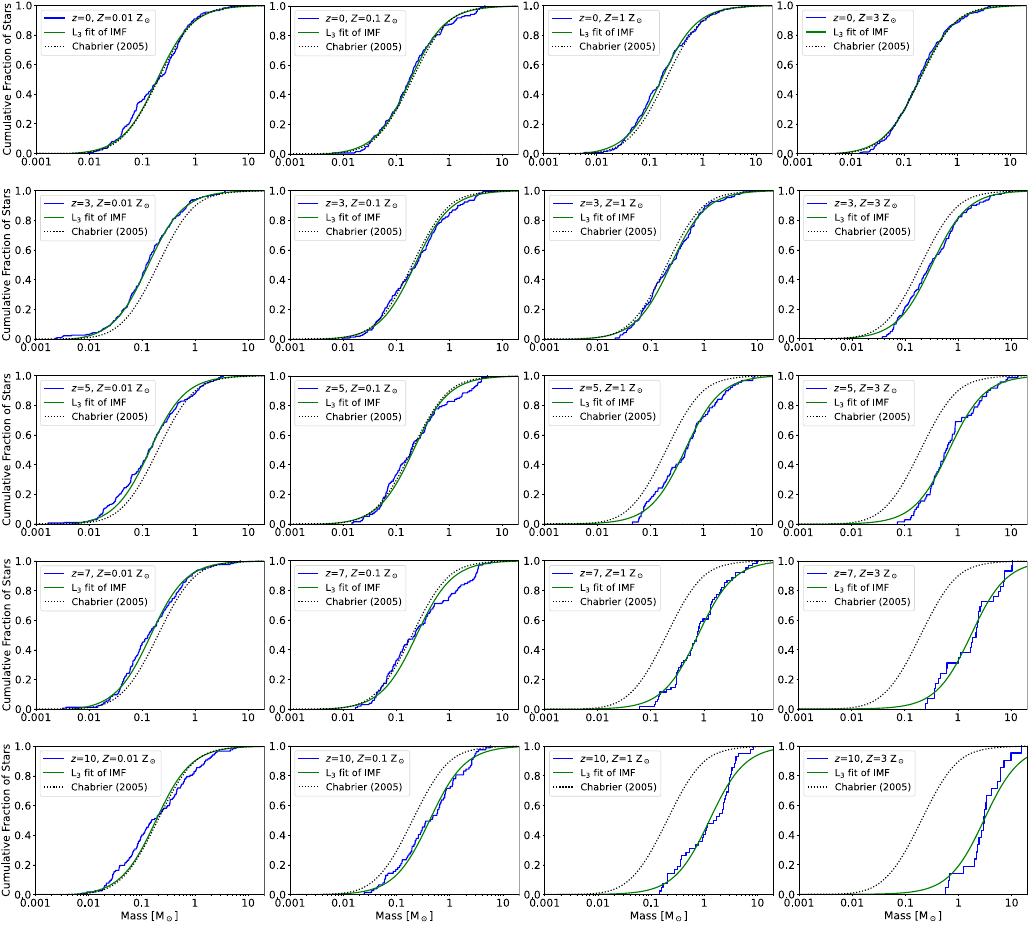} \vspace{-0.2cm}
\caption{The cumulative mass functions (stepped, blue solid lines) at the end of each of the 20 calculations:  from top to bottom the rows are for redshifts $z=0,3,5,7,10$, and from left to right the columns are for metallicities $Z=0.01, 0.1, 1, 3~{\rm Z}_\odot$, respectively.  Also over plotted are the cumulative Chabrier (2005) stellar IMF (black dotted lines), and the L$_3$ functional fit to the numerical mass functions (smooth, green lines).  The characteristic (median) stellar mass increases with increasing redshift at high metallicity. }
\label{fig:IMFcum}
\end{figure*}

\subsection{Cloud evolution and star formation}
\label{sec:clouds}

As the calculations evolve, the initial `turbulent' velocity field applied to the clouds generates density and velocity structure and dense regions collapse to form stars and brown dwarfs (modelled using sink particles with accretion radii of 0.5~au).  Table \ref{table1} lists the statistical properties of the stellar populations at the end of each of the twenty calculations.  For each redshift and metallicity, it gives the total number of objects (stars or brown dwarfs) formed, the number that are brown dwarfs (masses $<0.075~\mathrm{M}_\odot$, given as an upper limit since some are still accreting when the calculations were stopped), the total mass of the stars and brown dwarfs (recall that the initial cloud mass is 500~M$_\odot$ for all calculations), the mean mass of the objects, the mean mass computed using the logarithm of the mass, the median mass, the number of stellar mergers that occurred during each calculation, the value of the parameter, $\mu$, of the L$_3$ function used to fit the stellar mass distribution (see Section \ref{sec:fitting}), and the zero-age mass-to-light ratio of the stellar population (using the L$_3$ fit) relative to the Chabrier value (see Section \ref{sec:MLratio}).

Along with the a summary of the statistical properties of the stellar population produced by each of the calculations, we provide images of the density and temperature structure at the end of each of the calculations.  Figure \ref{fig:Density} displays images of the gas column density and stellar positions, Figure \ref{fig:Tgas} provides renderings of the mass-weighted gas temperature ($T_\mathrm{gas}$), Figure \ref{fig:Tdust} provides renderings of the mass-weighed dust temperature ($T_\mathrm{dust}$), and Figure \ref{fig:Trad} provides renderings of the mass-weighed temperature of the radiation field ($T_\mathrm{rad}$) produced by the star formation activity (this radiation field excludes the interstellar radiation field, which is treated separately).

Many of the general trends with metallicity and redshift that are apparent in these figures have already been discussed by \cite{Bate2019, Bate2023}, respectively.
Briefly, at a given time and a given redshift, lower metallicity clouds convert less gas into stars.  Similarly for a given time and metallicity, clouds at higher redshift convert less gas into stars, but the magnitude of this effect reduces with decreasing metallicity so that at $Z=0.01~\mathrm{Z}_\odot$ this effect is very weak.  These effects are a direct result of the typical gas temperatures in the clouds, with warmer gas having greater thermal pressure support against gravity, delaying the onset of star formation in warmer clouds and thereby resulting in less stellar mass at a given time.  Gas with lower metallicity does not cool as well as high metallicity gas, so it tends to be warmer.  With increasing redshift, the CMBR is warmer, producing a `temperature floor' that increases with increasing redshift.  At the lowest metallicity, because the gas is poor at cooling, most of the gas is hotter than the `temperature floor' provided by the CMBR regardless of the redshift, thus explaining why there is less variation with redshift with lower metallicity gas.  

\subsection{The effects of redshift and metallicity on the gas and dust temperatures}
\label{sec:temperatures}

The general trends of how the gas temperatures vary with metallicity and redshift can be seen in Figure \ref{fig:Tgas} with hotter gas with decreasing metallicity (right panels to left panels) and/or increasing redshift (particularly in the right two columns).  Care needs to be taken in interpreting these renderings because they give single snapshots at the end of each calculation, when many stars have already formed and the radiation released by rapidly accreting protostars can fluctuate greatly with time.  Nevertheless, the broad trends in the gas temperature are apparent.  

Despite the very different gas temperatures, the density structure is relatively similar between each of the calculations at the same time (Figure \ref{fig:Density}), due to the use of the same initial velocity field for each calculation.  At the lowest metallicity, there is little difference in the structure or the number of objects formed.  Similarly at low redshifts ($z=0, 3$) there is little clear difference.  But at high redshift and high metallicity, the density structure in the gas tends to become smoother and there are noticeably fewer objects formed.  This is a result of the gas temperatures tending to be warmer with increasing redshift, and less structured at high metallicity (the low metallicity calculations tend to produce hot bubbles of gas, as seen in Fig.~\ref{fig:Tgas}, which produce lots of fine structures in the column density plots).

The dust temperatures in the molecular clouds are initially determined by the heating from the ISRF, including the CMBR.  This gives higher dust temperatures at higher redshifts (due to the warmer CMBR).  At low redshift, the radiative power provided by the ISRF is dominated by the short-wavelength component.  The short-wavelength radiation is effective at heating the dust throughout the cloud at low metallicity due to the low extinction, but this radiation does not penetrate into regions with high metallicity and density, resulting in colder dust temperatures with increasing metallicity at low redshift.  At high redshift, the radiative power of the ISRF becomes dominated by the CMBR which has long wavelengths and penetrates well into the clouds even at high metallicity, so there is much less difference in the dust temperatures with different metallicities.  

Once star formation begins, the dust temperatures in the calculations continue to be dominated by the heating from the ISRF on large scales.  See, for example, the generally increasing dust temperatures in the outskirts of the panels going down each of the columns in Fig.~\ref{fig:Tdust}.  However, in the vicinity of the star formation the radiation released by accretion flows can dominate.  This can be seen by comparing the dust temperatures in Fig.~\ref{fig:Tdust} with the temperature of the radiation field that is produced by the star formation in Fig.~\ref{fig:Trad}.  The combination of the radiation from protostar formation and that of the ISRF sets the dust temperatures.    As emphasised above, the radiation field due to the star formation is highly variable in time as the strengths of the accretion flows fluctuate.  This results in `flickering' of the temperature fields in the vicinity of the protostars.  This was first emphasised by \cite{Bate2009b} in radiation hydrodynamical simulations of lower-mass clouds.  Only snapshots at the end of each of the calculations are provided in the figures in this paper.  In particular, note that at the end of the calculation with $z=0$ and $Z=0.01~\mathrm{Z}_\odot$ (the top left panels), the radiation field generated by the star formation happens to dominate the dust temperatures throughout region that is rendered (this is not generally the case earlier in the calculation).

To illustrate further the effects of the increasing CMBR temperature with redshift and the effects of different metallicities on the gas and the dust temperatures, we plot temperature-density phase diagrams at the end of each of the calculations \citep[similar to those plotted by][]{Bate2019, Bate2023}.  In Fig.~\ref{fig:pixgas} we provide phase diagrams of the gas temperature vs cloud density.  In Fig.~\ref{fig:pixdust} we provide phase diagrams of the dust temperature vs cloud density.  

At high densities, the gas and dust temperatures are well coupled, by collisions between gas molecules and dust particles.  This coupling occurs at lower densities with higher metallicities, so for example, at $Z=0.01~\mathrm{Z}_\odot$, tight gas-dust temperature coupling doesn't occur until molecular hydrogen number densities of $n_\mathrm{H2} \gsim 10^{8}$~cm$^{-3}$, whereas at $Z=3~\mathrm{Z}_\odot$ the dust and gas temperatures are well coupled at  $n_\mathrm{H2} \gsim 10^{6}$~cm$^{-3}$.  Moreover, in all cases, the denser gas tends to be hotter.  This is a result of the star formation:  very dense molecular gas that is collapsing onto protostars on the dynamical timescale can't cool quickly enough to remain isothermal so it heats up as it collapses.  In addition, due to radiative transfer, gas and dust that are near to the regions containing accreting protostars are heated by radiation from hotter dust.

At lower densities, the gas temperatures have much greater dispersions at a given cloud density as the metallicity decreases.  This is because low-metallicity gas is very poor at cooling, so gas that is strongly compressed heats up quickly and doesn't cool down again easily.  By contrast, the minimum gas temperature doesn't vary much with different metallicities, so the net effect is that the average temperature of this low-density gas increases with decreasing metallicity.

At low densities, there are two main effects seen in the dust temperatures (Fig.~\ref{fig:pixdust}).  First, at low-redshifts the dust temperatures tend to decrease with increasing density.  This is because, in the absence of (nearby) star formation, the dust is in thermal equilibrium with the ISRF.  For the local Galactic ISRF (which is what is assumed for $z=0$), much of the radiative power comes from short-wavelength radiation which is attenuated by the dust.  Thus, particularly at high metallicity, the densest regions of the cloud prior to star formation beginning can get very cold, as low as $T_\mathrm{dust} \approx 6$~K.  The long-wavelength CMBR penetrates well into the cloud, but it is cold (2.73~K).  Second, as the redshift is increased, the CMBR increases in temperature and so the dust temperatures at low densities become more dependent on the CMBR and less dependent on the short-wavelength component ISRF.  Despite its increasing temperature the CMBR is still able to penetrate well into the cloud, even at $z=10$, so a dust temperature `floor' becomes clearly visible for $z \gsim 5$, and this temperature floor increases as $T_\mathrm{dust} \propto (1+z)$ as expected.

\subsection{The stellar populations}
\label{sec:stellarpop}

As discussed above, when the calculations are stopped at $t = 1.30~t_\mathrm{ff}$, the amount of gas that has been converted into stars and brown dwarfs generally decreases with increasing redshift and/or decreasing metallicity due to the warmer gas temperatures.  The magnitude of this difference between calculations at different redshifts is diminished at the lowest metallicities (in the fifth column of Table \ref{table1}, the total mass in stars/brown dwarfs only varies between 70 and 82 M$_\odot$ regardless of redshift) since at low metallicities the low-density gas tends to be hot regardless of the redshift because it is poor at cooling and there is little extinction of the ISRF.  Similarly, the magnitude of the difference between calculations with different metallicities is diminished at the highest redshifts (for the $z=10$ calculations, the total mass in stars/brown dwarfs only varies between 70 and 85 M$_\odot$ regardless of metallicity; Table \ref{table1}) because the gas is warm at high redshifts regardless of the metallicity due to the warmer CMBR.  In contrast, at $z=0$ there is a large increase in the gas mass that is converted into stars with increasing metallicity because more metal-rich gas tends to be colder.  Similarly, for metal-rich gas there is a large increase in the gas mass that is converted into stars with decreasing redshift because of the cooler gas.
 
For the numbers of stars and brown dwarfs that have been produced by the end of the calculations, the numbers of protostars also tend to decrease with increasing redshift (except for the lowest metallicity at $z=0$).  However, there is a change in the trend with metallicity at around $z=3$.  For present-day star formation at $z=0$, the number of protostars increases with increasing metallicity in the same manner that the amount of gas that is converted into stars increases.  This results in a mean stellar mass (given in the sixth column in Table \ref{table1}) that does not vary significantly with metallicity for star formation at $z=0$ \citep[as has been found in past calculations;][]{Myers_etal2011, Bate2014, Bate2019}.
By contrast, for $z \gsim 3$, the numbers of protostars that are formed decreases quite significantly with increasing metallicity, whereas the amount of gas being converted into stars still tends to increase with increasing metallicity.  The result is that the mean protostellar mass tends to increase with increasing redshift and with increasing metallicity \citep[see also][]{Bate2023}.  Moreover, the magnitude of this increase is greater with increasing redshift. 

The same trends are true for the mean stellar mass when computed in the logarithm of mass (see the seventh column of Table \ref{table1}), and the median mass (the eighth column of Table \ref{table1}).  Except at $z=0$, the mean log-mass and the median stellar mass increase with increasing metallicity, and for metallicity $Z \gsim \mathrm{Z}_\odot$ these characteristic masses increase consistently with increasing redshift.  At $Z = 0.1~\mathrm{Z}_\odot$ the median stellar mass does not appear to increase with redshift until $z \gsim 7$.  

For the lowest metallicity, $Z = 0.01~\mathrm{Z}_\odot$, there does not appear to be any significant trend with redshift. Averaging over all five calculations the median mass of $\approx 0.15$~M$_\odot$ is slightly lower than the median mass of the \cite{Chabrier2005} stellar IMF with is 0.20~M$_\odot$, perhaps pointing to a slightly bottom-heavy mass function at the lowest metallicity (for redshifts $z \lsim 10$).  By contrast, the median stellar mass at $Z = 0.1~\mathrm{Z}_\odot$ is in good agreement with the Chabrier value for redshifts $z \lsim 5$.

Fig.~\ref{fig:IMFdiff} plots the differential mass functions at the end of all twenty calculations.  The dark blue histograms give the numbers of stars or brown dwarfs in each mass bin whose accretion rates are less than $10^{-7}$~M$_\odot$~yr$^{-1}$ (i.e., barring a new phase of accretion, they have essentially reached their final masses).  The light blue histograms give the numbers of stars that have accretion rates higher than this value at the end of the calculations.  The masses of these latter objects are therefore lower limits, in that if the calculations were continued these objects would continue to grow.  The evolution of the mass distributions is discussed further below.

Overlaid on the histograms in Fig.~\ref{fig:IMFdiff} are the parameterisations of the Galactic stellar IMF by \cite{Kroupa2001} (red solid line) and \cite{Chabrier2005} (black solid line), along with fits to the numerical mass distributions of the L$_3$ function (green solid lines) proposed by \cite{Maschberger2013} (see Section \ref{sec:fitting} for more discussion).  The stellar mass distributions obtained from the radiation hydrodynamical calculations at low redshift and/or low metallicity are very similar to the parameterisations of the Galactic IMF.  However, for increasing redshift the numerical mass functions clearly become increasingly `bottom-light' (i.e., there is a deficit of brown dwarfs and low-mass stars compared to the observed Galactic IMF), with the magnitude of the effect increasing with increasing metallicity.

Fig.~\ref{fig:IMFcum} gives the corresponding cumulative mass functions at the end of all twenty calculations (blue solid lines).  Also plotted for comparison are the cumulative stellar IMF of \cite{Chabrier2005} (black dotted lines), and the cumulative L$_3$ functions (green solid lines; using the same fits as those depicted in Fig.~\ref{fig:IMFdiff}).  Again, it can be seen that the numerical mass distributions are very similar to the \cite{Chabrier2005} mass function for low redshift and/or low metallicity, but for high redshift and/or metallicity the median stellar mass shifts to higher masses, with the shift beginning at a lower redshift for higher metallicity.  So, for example, with a metallicity $Z=3~\mathrm{Z}_\odot$ even at $z=3$ the median mass has increased and for $z \geq 7$ the median mass is more than a order of magnitude greater than for the \cite{Chabrier2005} IMF.  For metallicity $Z=\mathrm{Z}_\odot$ the median stellar mass doesn't begin to increase until $z \approx 4$, while at metallicity $Z=0.1~\mathrm{Z}_\odot$, the median mass isn't found to increase until  $z \approx 10$.  Finally, as noted above, most of the lowest metallicity calculations ($Z=0.01~\mathrm{Z}_\odot$) are slighly bottom-heavy, whereas all of the calculations with $Z=0.1~\mathrm{Z}_\odot$ (except the one at $z=10$) are in good agreement with the \cite{Chabrier2005} IMF.

Note that in all cases except for the highest redshift and metallicity (the bottom right panel of Fig.~\ref{fig:IMFcum}), the slope of the cumulative mass functions from the radiation hydrodynamical simulations (i.e., the blue lines) are very similar to that of the \cite{Chabrier2005} mass function.  Thus, the breath of the differential mass functions (i.e., the overall mass ranges) are very similar.  For the case with the highest redshift ($z=10$) and metallicity ($Z=3~\mathrm{Z}_\odot$) the blue line is significantly steeper, or in other words the distribution of masses is noticeably narrower.  This is due to the fact that the fragmentation is so greatly inhibited by the high gas temperatures in this calculation that the protostars don't usually form close enough to dynamically interact.  Without such dynamical interactions `kicking' protostars out of the dense gas and stopping their growth at low masses, there is little to stop these protostars accreting.  Hence, in the bottom right panel of Fig.~\ref{fig:IMFdiff}, only a single object has stopped accreting when the calculation is stopped.

\begin{table*}
\begin{tabular}{lcccccccccc}\hline
Redshift   & Metallicity & \multicolumn{4}{c}{$t=1.10~t_{\rm ff}$} & ~~~ & \multicolumn{4}{c}{$t=1.30~t_{\rm ff}$} \\
\cmidrule(lr){3-6}\cmidrule(lr){8-11}
   & $Z$         & Number of & Mean  & Mean       & Median  & & Number of & Mean & Mean & Median \\
   &                & Objects & Mass  & Log-Mass & Mass      & & Objects & Mass  & Log-Mass & Mass  \\
&  Z$_\odot$ &             & M$_\odot$ & M$_\odot$ & M$_\odot$ & &             & M$_\odot$ & M$_\odot$& M$_\odot$ \\ 
\hline
$z=0$  &  0.01  &  61  &  $0.28\pm0.04$  &  $0.15\pm0.02$  &  0.16  & &  175  &  $0.47\pm0.07$  &  $0.19\pm0.02$  &  0.20  \\
            &  0.1  &  105  &  $0.35\pm0.05$  &  $0.16\pm0.02$  &  0.15  & &  209  &  $0.51\pm0.06$ &  $0.20\pm0.02$  &  0.18 \\
            &  1  &  180  &  $0.28\pm0.03$  &  $0.14\pm0.01$  &  0.14 &  &  272  &  $0.47\pm0.06$  &  $0.18\pm0.01$  &  0.16 \\
            &  3  &  192  &  $0.28\pm0.03$  &  $0.15\pm0.01$  &  0.14  & &  284  &  $0.46\pm0.05$  &  $0.20\pm0.02$  &  0.18 \\ \hline
$z=3$  &  0.01  &  49  &  $0.27\pm0.04$  &  $0.16\pm0.03$  &  0.18  & &  236  &  $0.32\pm0.04$  &  $0.13\pm0.01$  &  0.12 \\
            &  0.1  &  88  &  $0.31\pm0.04$  &  $0.17\pm0.02$  &  0.16  & &  168  &  $0.60\pm0.08$  &  $0.23\pm0.03$  &  0.24 \\
            &  1  &  89  &  $0.45\pm0.07$  &  $0.24\pm0.03$  &  0.24  & &  191  &  $0.61\pm0.09$  &  $0.25\pm0.02$  &  0.23 \\
            &  3  &  72  &  $0.57\pm0.10$  &  $0.30\pm0.04$  &  0.29  & &  142  &  $0.85\pm0.13$  &  $0.34\pm0.04$  &  0.30 \\ \hline
$z=5$  &  0.01  &  46  &  $0.33\pm0.05$  &  $0.17\pm0.04$  &  0.23  & &  214  &  $0.38\pm0.05$  &  $0.14\pm0.01$  &  0.13 \\
            &  0.1  &  50  &  $0.44\pm0.08$  &  $0.21\pm0.04$  &  0.19  & &  139  &  $0.67\pm0.09$  & $ 0.23\pm0.03 $ &  0.19 \\
            &  1  &  39  &  $0.78\pm0.16$  &  $0.42\pm0.08$  &  0.40  & &  87  &  $1.13\pm0.19$  &  $0.47\pm0.07$  &  0.47 \\ 
            &  3  &  30  &  $0.98\pm0.20$  &  $0.63\pm0.11$  &  0.61  & &  71  &  $1.54\pm0.27$  &  $0.73\pm0.11$  &  0.60 \\ \hline
$z=7$  &  0.01  &  37  &  $0.30\pm0.06$  &  $0.15\pm0.03$  &  0.14  & &  166  &  $0.42\pm0.07$  &  $0.15\pm0.02$  &  0.13 \\
            &  0.1  &  43  &  $0.34\pm0.07$  &  $0.15\pm0.03$  &  0.11  & &  104  &  $0.84\pm0.13$  &  $0.27\pm0.04$  &  0.23 \\
            &  1  &  24  &  $0.87\pm0.18$  &  $0.58\pm0.12$  &  0.65  &&   61  &  $1.52\pm0.26$  &  $0.79\pm0.13$  &  0.75 \\
            &  3  &  14  &  $1.11\pm0.31$  &  $0.75\pm0.19$  &  0.68  & &  29  &  $3.07\pm0.57$  &  $1.72\pm0.41$  &  2.20 \\ \hline
$z=10$  &  0.01  &  16  &  $0.54\pm0.14$  &  $0.25\pm0.11$  &  0.23  & &  117  &  $0.61\pm0.11$  &  $0.19\pm0.03$  &  0.15 \\
            &  0.1  &  16  &  $0.51\pm0.10$  &  $0.33\pm0.10$  &  0.40  & &  77  &  $0.91\pm0.13$  & $ 0.41\pm0.07$  &  0.44 \\
            &  1  &  10  &  $1.30\pm0.23$  & $ 0.94\pm0.34$  &  1.27  & &  42  &  $1.98\pm0.28$  &  $1.17\pm0.23$  &  1.60 \\
            &  3  &  7  &  $1.62\pm0.50$  &  $1.23\pm0.39$  &  1.42  & &  21  &  $4.04\pm0.75$  &  $2.92\pm0.59$  &  3.02 \\
\hline
\end{tabular}
\caption{\label{table2} The variation of the statistical properties of the stellar populations produced by the calculations between times $t=1.10~t_{\rm ff}$ and $t=1.30~t_{\rm ff}$.  The period between these two times is approximately 38,000~yrs.  For each calculation the mean mass increases between these two times, but for most of the calculations the mean of the logarithms of the masses does not change significantly.  Similarly, for most of the calculations the median masses do not change substantially (sometimes they go up slightly and sometimes they go down slightly).  Thus, for these calculations the characteristic (mean log-mass or median) stellar mass has reached a quasi-steady state --- although some objects continue to accrete mass, many have stopped accreting substantially, and new objects (with low initial masses) continue to be formed.  The exceptions are the calculations at both high redshift and high metallicity:  those at $z=7,10$ and $Z=1-3~{\rm Z}_\odot$.  In these calculations, with few objects and a large fraction that are still accreting when the calculations are stopped, the characteristic stellar masses  are best thought of as lower limits. }
\end{table*}

\subsubsection{Evolution of the stellar mass distributions}
\label{sec:IMFevolution}

The fact that some objects are still accreting when the calculations are stopped raises questions as to how much the mass distributions would continue to evolve if the calculations were continued further.  For the low-mass end of the stellar mass distribution (the focus of this paper), we are mostly concerned with determining the characteristic stellar mass.  The characteristic stellar mass can be defined in many ways.  For mass distributions that are similar to a log-normal at low masses the median stellar mass, the mean of the logarithm of the stellar masses, and the peak in the d$\log{N}/$d$\log{m}$ distribution, and the $\mu$ parameter in our chosen L$_3$ functional fit (Section \ref{sec:fitting}), all have similar values (see the 7th and 8th columns of Table \ref{table1} and Fig.~\ref{fig:IMFdiff}).   Because of this, in this paper we may refer to any of these interchangeably.

Since we are most interested in the characteristic stellar mass, the question becomes how much this may evolve with time.  Clearly if some objects are still accreting when the calculations are stopped, and if the calculations were continued, it is likely that more high-mass stars would be produced.  However, this is not such a problem for determining the characteristic masses of the mass functions for two reasons.  First, in most of the calculations, many objects have stopped accreting significantly when the calculations are stopped (for example, if more than half were to have stopped accreting, the median mass wouldn't evolve at all).  Second, in most of the calculations, while some objects are accreting to higher masses, new (low-mass) objects are continually being formed.  As discussed in previous similar studies (e.g., in \citealt{Bate2012, Bate2014, Bate2019}), this results in the median stellar mass (or equivalently one of the other measures of the characteristic stellar mass) essentially being invariant regardless of the point during a calculation when it is examined.  For example, \cite{Bate2012} showed that in the calculation discussed in that paper the median mass varied by no more than a factor of two (non-monotonically) between $t=0.90 - 1.20$~t$_{\rm ff}$.  For the particular calculations discussed in this paper, Table 1 of \cite{Bate2023} gives the mean, mean log-mass, and median masses at time $t=1.20~t_\mathrm{ff}$ for the $z=0$ calculations and the $z=5$ calculations with $Z \leq \mathrm{Z}_\odot$.  For all of these calculations, the mean log-mass values at $t=1.20~t_\mathrm{ff}$ are within the statistical uncertainties for the values at $t=1.30~t_\mathrm{ff}$, and the median masses are also very similar.  To emphasise this further, in Table \ref{table2} for each of the twenty calculations we provide the number of objects, the mean mass, the mean of the logarithm of the masses, and the median mass at both $t=1.10~t_\mathrm{ff}$ and $t=1.30~t_\mathrm{ff}$ (the latter values are repeated from Table \ref{table1} to make comparison easier). 
Comparing the values between the two times it can be seen that the mean stellar mass always increases, despite the fact that many more objects may have been formed between these two times.  This shows that indeed the most massive stars are continuing to accrete to higher masses.  Nevertheless, because some stars have their accretion terminated during this period, and new low mass stars are continuing to form the characteristic stellar mass (either the median stellar mass, or mean of the logarithm of the masses; Table \ref{table2}) remains almost invariant in most of the calculations (in some cases the exact values go up slightly, and in others they go down slightly, but in most cases the changes are small).  In a real star-forming region, star formation may continue until it is stopped by some event such as photoionisation and dispersal of the molecular cloud.  At this point the stellar mass distribution will be `frozen', and it will become the initial mass function for that region.  If the star formation proceeds such that the characteristic stellar mass is essentially constant in time, as in most of the calculations discussed here, then it does not matter when the star formation is terminated -- the characteristic stellar mass will be the same.  This approximate invariance of the median stellar mass is also in agreement with the analytic theory of the IMF developed by \cite{ClaWhi2021} based on a contest between accretion and fragmentation.  It should be noted from Table \ref{table2}, however, that continued accretion is a significant problem for determining the characteristic stellar mass in the calculations at the highest redshifts and metallicities: at $z=7,10$  with $Z = 3~\mathrm{Z}_\odot$, and to a lesser extent with $Z = \mathrm{Z}_\odot$.  In the two highest metallicity calculations most of the protostars are still accreting significantly when the calculations are stopped, very few have reached their final masses, and there is little ongoing star formation.  Therefore, for these two calculations in particular, the characteristic stellar masses (i.e., the median mass, the mean log-mass, or the value of $\mu$) should be viewed as lower limits.

Another question is how much variation there might be in the characteristic stellar mass if a different random seed was used to generate the initial velocity field that was used for the calculations.  For these twenty calculations all used the same initial velocity field so as to limit the differences between the stellar populations, as much as possible, to those arising just from the variation of the two parameters (redshift and metallicity) under investigation.  However, since the calculations involve chaotic dynamical interactions between protostars, some level of stochastic variation is to be expected.  Running all the calculations with multiple random seeds is too expensive (each calculation took between 1.5 to 5 million core hours).  However, \cite{JonBat2018a} did report the results of one such investigation.  They performed three radiation hydrodynamical calculations of star cluster formation, with three different cloud densities.  Their medium density calculation was identical to the radiation hydrodynamical calculation of \cite{Bate2012} except that they used a different random seed to generate the initial velocity field.  They found that for these two calculations the two mass distributions were statistically indistinguishable (using a Kolmogorov-Smirnov test); their discussion can be found at the end of Section 3.2 of \cite{JonBat2018a}, and the two cumulative mass functions are compared in their Figure~3.  Both calculations were run to $t=1.20~t_{\rm ff}$.  The \cite{Bate2012} calculation produced 183 objects with a median mass of 0.21~M$_\odot$, while the medium-density calculation of \cite{JonBat2018a} produced 233 with a median masses of 0.18~M$_\odot$.  This level of variation in the median mass between the two calculations is similar to the formal uncertainties that are found in the mean log-mass in Tables \ref{table1} and \ref{table2} for the calculations in this paper that produce in excess of 100 objects.  Thus, the expected level of `stochastic variation' between different calculations is likely to be negligible.  Further evidence for this is that the twenty calculations reported here demonstrate very clear and consistent trends in the variation of the characteristic stellar masses with both redshift and metallicity.  The existence of the clear trends is evidence that the variations are not dominated by stochastic variation.

\subsection{The effects of redshift and metallicity on the stellar mass distribution}
\label{sec:temperatures}

Examining the thermal behaviour of the gas is important for understanding how the resulting stellar mass distribution varies with redshift and metallicity.  At low redshift ($z=0-3$), there are two opposing effects of different metallicities on the ability of the gas to cool that combine to produce a relatively invariant IMF.  The first effect is that low-density, low-metallicity gas is poor at cooling. As discussed in the previous sections, the average temperature of the low-density gas tends to be higher in the low-metallicity calculations.  This increases the pressure support, delays the collapse of the cloud as a whole compared to higher metallicity clouds, and increases the Jeans mass within the cloud.  On their own, the warmer gas temperatures would be expected to result in higher-mass stars.  However, lower metallicity gas also has a lower opacity due to the reduced dust abundance.  Therefore, the second effect is that once dense gas begins to collapse dynamically, it can collapse to higher densities before becoming optically thick, trapping thermal energy, and heating up (the so-called `opacity limit for fragmentation'; \citealt{LowLyn1976, Rees1976}).  Since the Jeans mass and length are inversely proportional to the square root of density, this means that fragmentation can occur at higher densities, on smaller scales, and produce fragments with lower initial masses at lower metallicities.  Conversely, with higher metallicities, the gas tends to be colder both because it is better able to cool, and because dense gas within a molecular cloud is shielded from the short-wavelength component of the ISRF due to the stronger dust extinction.  This leads to lower gas pressures, and smaller Jeans lengths and masses which, on their own, would be expected to give more fragmentation and produce stars with characteristically lower masses.  However, the higher metallicity also means that collapsing gas reaches the opacity limit more quickly, at lower densities, which reduces small-scale fragmentation.
At low redshift, at least for metallicities in which dust dominates the cooling at high densities ($Z \gsim 0.01~\mathrm{Z}_\odot$), these two competing effects (poorer cooling of low-density gas at low metallicity, but enhanced cooling of high-density gas at low metallicity) essentially cancel each other out and the resulting IMFs have been found to be essentially independent of the metallicity \citep{Myers_etal2011, Bate2014, Bate2019}.

When moving to higher redshifts, for low metallicity gas ($Z \gsim 0.01~\mathrm{Z}_\odot$) there is little change in the IMF because even at redshift $z=0$ the typical gas temperature is hotter than the temperature of the CMBR, at least up to $z=10$ (see Fig.~\ref{fig:pixgas}).  So essentially nothing changes as the redshift increases.

However, with higher metallicity gas at $z=0$, the densest gas (deeply embedded within the molecular cloud) can be very cold, as cold as $T_\mathrm{gas} \approx 5-6$~K at super-solar metallicities.  For an identical cloud, but at redshift $z=3$, the CMBR is $T_\mathrm{CMBR} = 2.73 \times (1+z) \approx 11$~K, and the CMB radiation has long enough wavelengths that it can penetrate the entire cloud.  Thus, the densest gas cannot cool to as lower temperatures at intermediate or high redshift as it could at $z=0$.  This raises the thermal Jeans mass, inhibits small-scale fragmentation, reduces the number of objects that are formed for a given amount of gas and, thereby, raises the typical stellar mass.  As the redshift is increased, this effect on the stellar mass distribution acts first at the highest metallicity because, in the absence of the warm CMBR, this gas is the coldest.  As the redshift is increased further, the star formation in lower metallicity gas is affected as the `temperature floor' starts to significantly raise the temperature of the dense gas.  As is clearly seen from the simulations, the result is that the stellar mass distribution is essentially invariant at low metallicity and/or low redshift, but it becomes increasingly `bottom-light' at high redshift and/or high metallicity.

\subsection{Empirical fitting of the stellar mass functions}
\label{sec:fitting}

In the absence of an analytic model to describe the form of the IMF and its variation with redshift and metallicity it is useful to find an analytical function that can be used to fit the numerical results.  Such a function could be used, for example, to set the stellar populations in simulations of galaxy formation to study the effects of a variable IMF.

Many analytic forms have been proposed to describe the IMF in the past.  One function that has a number of advantages is the L$_3$ function proposed by \cite{Maschberger2013}.  This function has analytic equations for both the differential and cumulative forms of the IMF and other quantities such as the mean and median stellar mass.  It also has relatively few parameters -- the basic shape of the function is described by 3 parameters (essentially a high-mass slope, a low-mass slope, and a characteristic mass), and there are (optional) high-mass and low-mass cut-offs.  This function has been used to fit the results of numerical mass functions, for example, \cite*{GusHopGra2019}.

For convenience, we give a brief summary of the analytic equations here.  More details can be found in \cite{Maschberger2013}.   The form of the IMF is based on an auxilliary function
\begin{equation}
G(m) = \left( 1+ \left( \frac{m}{\mu} \right)^{1-\alpha} \right)^{1-\beta},
\label{eq:L3aux}
\end{equation}
where $m$ is the stellar mass, and $\alpha, \beta, \mu$ are parameters that describe the form of the mass function.  The probability density function (PDF) is given by
\begin{equation}
p_\mathrm{L3}(m) = \frac{(1-\alpha)(1-\beta)}{\mu \left[ G(m_\mathrm{u}) - G(m_\mathrm{l}) \right]} \left( \frac{m}{\mu} \right) \left( 1+ \left( \frac{m}{\mu} \right)^{1-\alpha} \right)^{-\beta},
\label{eq:L3diff}
\end{equation}
and the cumulative distribution function (CDF) is given by
\begin{equation}
P_\mathrm{L3}(m) = \frac{ G(m) - G(m_\mathrm{l}) }{ G(m_\mathrm{u}) - G(m_\mathrm{l}) },
\label{eq:L3cum}
\end{equation}
where $m_\mathrm{l}$ and $m_\mathrm{u}$ are the lower and upper mass cut-offs, respectively.  The median mass can be expressed as
\begin{equation}
\widetilde{m} = \mu \left(  \left[  \frac{1}{2} \left(  G(m_\mathrm{u}) + G(m_\mathrm{l})  \right)  \right]^{\frac{1}{1-\beta}} - 1  \right)^{\frac{1}{1-\alpha}}.
\label{eq:median}
\end{equation}
The peak mass (the maximum when plotting the PDF in a d$N$/dlog$(m)$ graph) is given by
 \begin{equation}
m_\mathrm{p} = \mu (\beta - 1)^{\frac{1}{1-\alpha}}.
\label{eq:peak}
\end{equation}
The mean mass (i.e., the expectation value) can be expressed using the incomplete Beta function
\begin{equation}
B(x; p, q) = \int^x_0 t^{p-1} (1-t)^{q-1} \mathrm{d}t.
\label{eq:incompleteBeta}
\end{equation}
Using this function, the mean mass can be written as
\begin{equation}
\begin{aligned}
\overline{m} & = \int^{m_\mathrm{u}}_{m_\mathrm{l}} m \; p_\mathrm{L3}(m) \; \mathrm{d}m, \\
& =  \mu (1-\beta) \frac{ B(t(m_\mathrm{u}); a, b) - B(t(m_\mathrm{l}); a, b) }{ G(m_\mathrm{u}) - G(m_\mathrm{l})  }, 
\end{aligned}
\label{eq:mean}
\end{equation}
where $a=(2-\alpha)/(1-\alpha)$ and $b=\beta - \alpha$.  This expression can be derived using equation \ref{eq:L3diff} by making the substitutions $y=m/\mu$, and $t(m) = y^{1-\alpha}/(1+y^{1-\alpha})$.  

The \cite{Chabrier2005} single-star IMF, is defined using a power-law slope of 2.35 for $M\geq 1$~M$_\odot$ and a log-normal distribution for lower masses (with a mean of 0.20~M$_\odot$ and a variance of $\sigma=0.55$).  A close fit to this can be obtained using the L$_3$ function with parameters:  $\mu=0.20$, $\alpha=2.3$, $\beta=2.0$.  \cite{Maschberger2013} also gives values that provide close fits to the \cite{Kroupa2001} and \cite{Chabrier2003} IMFs.

In fitting the numerical mass functions obtained from the twenty radiation hydrodynamical simulations discussed in this paper, we make several simplifications.  Fundamentally we make these simplifications because none of the simulations produce very large numbers of stars and brown dwarfs.  The greatest number of objects produced by any of the calculations is less than 300.  This means that although the simulations constrain the median mass (or the mass at which the peak of the mass distribution occurs in a plot of $\log{N}$ vs $\log{m}$) quite well, they do not well constrain the high-mass slope, the low-mass slope, or either of the high-mass or low-mass cut-offs.  Therefore, first, we fix the value of $\alpha=2.3$ which is the value used by several Galactic mass functions \citep[e.g.][]{Kroupa2001} and similar to the original Salpeter value.  Second, we fix the high-mass cut-off to be 150~M$_\odot$ and the low-mass cut-off to be 0.005~M$_\odot$.  These limits do not affect the basic shape of the mass function, only its normalisation.  Finally, we fix the value of $\beta=2.0$.  This last choice is somewhat arbitrary.  In fitting the numerical mass distributions with the $L_3$ function we first tried leaving both $\mu$ and $\beta$ as free parameters, but we found that, given the comparatively small numbers of objects, these parameters were somewhat degenerate.  Similarly good fits could be obtained by increasing the value of $\beta$ and decreasing the value of $\mu$.  To reduce this degeneracy, the mass distributions would need to contain many more objects which is numerically intractable currently.  We decided to fix the value at $\beta=2.0$ for three main reasons: a) this value provides reasonable fits for all of the twenty calculations; b) this value provides a good fit to the \cite{Chabrier2005} mass function (see the black and green lines in the top row of Fig.~\ref{fig:IMFdiff}); c) for this value of $\beta$, the values of $\mu$, the median mass, $\overline{m}$, and the peak mass, $m_\mathrm{p}$ are all very similar (a somewhat annoying feature of the L$_3$ function is that the parameter $\mu$ can differ quite significantly from the median and peak values if $\beta$ is very different from 2; see equation \ref{eq:peak}).

To fit the L$_3$ function to the distributions of masses obtained from each of the numerical simulations, we use maximum likelihood estimation \citep{Eliason1993}.  Essentially, we search a grid of $\mu$ values to find the minimum of the quantity $\sum_i \left[ -\log{p_{L3}(m_i)} \right]$, where the sum is over all the masses of the objects, $m_i$ at the end of a given simulation (this is equivalent to calculating the probability of each mass from the PDF and multiplying the probabilities together, but avoids having to deal with small numbers).  The basic idea here is that if there is a high probability of a given mass, this will contribute a low value to the sum, while conversely if there is a low probability of obtaining a given mass this contributes a high value to the sum.  Thus, minimising the sum of values provides the best fit.  The uncertainty in the best fit value is obtained from a grid search of $\mu$ as described by (\citealt{Naylor2009}, Section 7).

The resulting L$_3$ function fits to the numerical mass functions are plotted using solid green lines in Fig.~\ref{fig:IMFdiff} on top of the histograms and in Fig.~\ref{fig:IMFcum} on top of the cumulative mass functions.  The value of $\mu$ used for each fit is given in Table \ref{table1}, along with the 68 percent uncertainty.  The L$_3$ function provides reasonable fits in all cases.  For the calculations at low redshift and/or low-metallicity it is clear that the best fit L$_3$ function (green lines) are very similar to the \cite{Chabrier2005} mass function (black solid lines in Fig.~\ref{fig:IMFdiff} or the black dotted lines in Fig.~\ref{fig:IMFcum}).  However, for high redshift and/or high metallicity the mass functions (and the $\mu$ values) move towards higher masses as the redshift or metallicity increase.

To compute the form of the mass function at any redshift ($z \in [0,10]$) and metallicity ($Z \in [0.01,3]~\mathrm{Z}_\odot$) requires interpolation between the results obtained from the twenty calculations.  In particular, since we have fit the numerical mass functions using the L$_3$ function with the single free parameter, $\mu$, we can parameterise the IMF with redshift and metallicity if we can find an empirical function that does a good job of fitting the values of $\mu$ from the tenth column of Table \ref{table1}.

We require a function that provides a value of $\mu$ that increases with increasing redshift, but for which the increase begins at a higher redshift with decreasing metallicity.  Without an analytical theoretical expectation for how the IMF varies with redshift and metallicity, any function will do, but it is pragmatic to choose something relatively simple.  Furthermore, since we have argued that much of the variation of the IMF found in the calculations is driven by the temperature of the CMBR which scales as $T_\mathrm{CMBR} \propto (1+z)$, it seems reasonable to include a term of this form in the empirical function. We chose
\begin{equation}
\mu(z,Z) = \mu_0 + a(1+z)^b Z,
\label{eq:fit}
\end{equation}
where $\mu_0$ is the value of $\mu$ at low redshift and/or metallicity, and $a$ and $b$ are two constants.  

\begin{figure}
\centering
    \includegraphics[width=8.5cm]{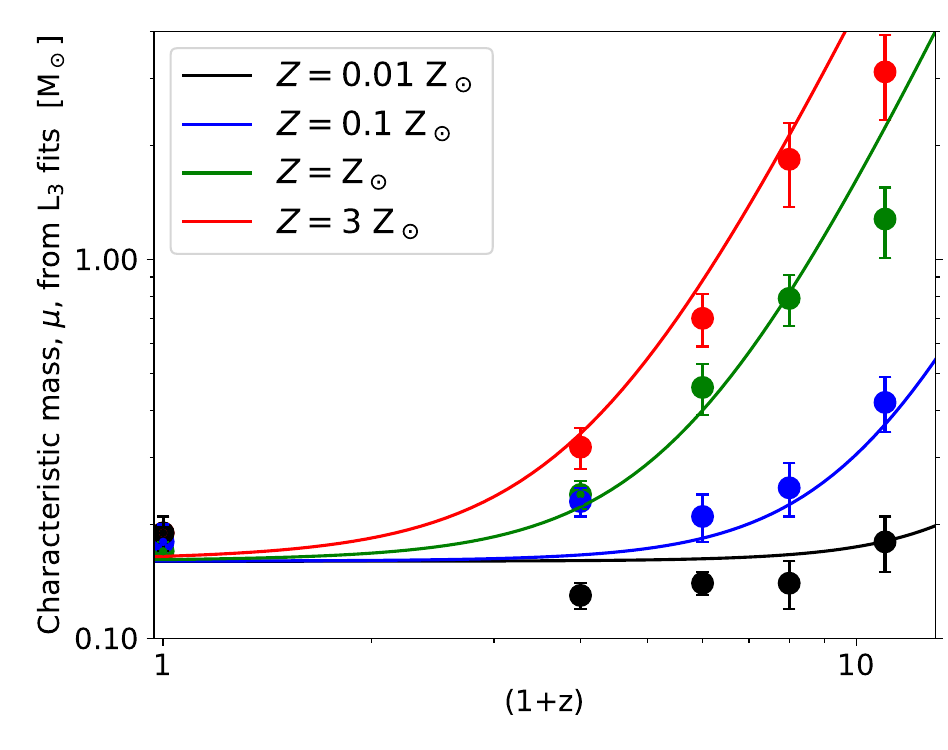} \vspace{0cm}
\caption{The characteristic stellar masses, given by the $\mu$ parameter of the L$_3$ function fit to the numerical stellar mass distributions, from the twenty radiation hydrodynamical calculations are plotted as a function of redshift ($1+z$) using coloured circles, along with their uncertainties (see the tenth column of Table \ref{table1}).  The $\mu$ value is essentially equal to the median stellar mass, and also the peak of the IMF when plotting d$N$/dlog$(m)$. Different colours are used to denote the four metallicities: 1/100 (black), 1/10 (blue), 1 (green), and 3 (red) times solar metallicity.  Also plotted are four solid lines (one for each metallicity) given by equation \ref{eq:fit} with parameters: $\mu_0=0.16$, $a=10^{-4}$ and $b=4$.  These lines provide a reasonable fits to the numerical results and allow the generation of stellar IMFs at arbitrary redshift and metallicity values.}
\label{fig:L3muFit}
\end{figure}

\begin{figure*}
\centering
    \includegraphics[width=8.5cm]{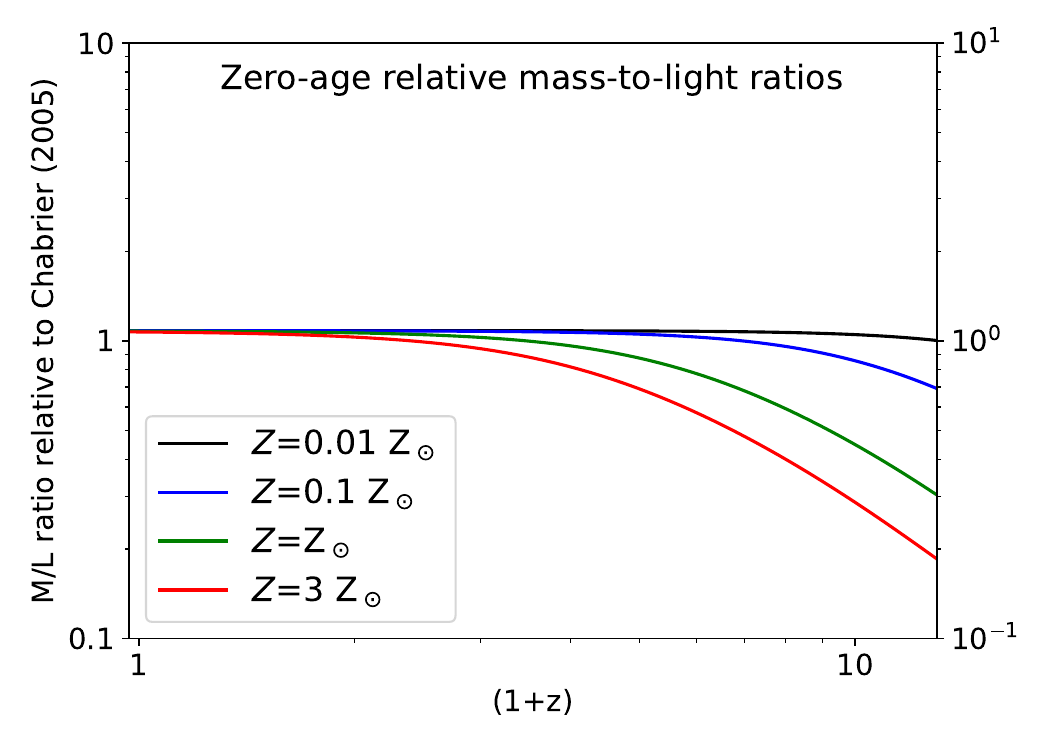} \vspace{0cm}
    \includegraphics[width=8.5cm]{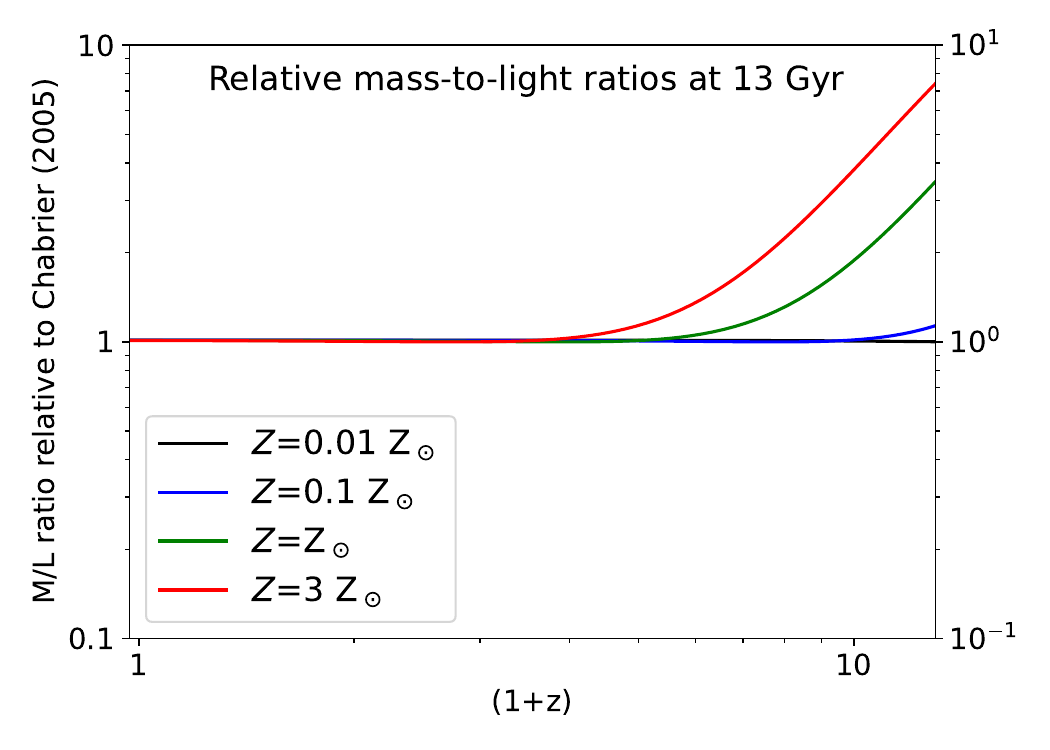} \vspace{0cm}
\caption{The predicted variation of the mass-to-light ratios of stellar populations with bottom-light IMFs as predicted in this paper, relative to the mass-to-light ratio of a typical Galactic IMF \citep{Chabrier2005}, as functions of redshift and metallicity (different coloured lines are used to denote four metallicities: 1/100 (black), 1/10 (blue), 1 (green), and 3 (red) times solar metallicity).  In the left panel, we give the variation of the mass-to-light ratio with redshift and metallicity for zero-age populations, with the mass-to-light ratios decreasing more with increasing metallicity.  In the right panel, we give the variation of the mass-to-light ratio for stellar populations that are 13 Gyr old (i.e., those that only contain main-sequence stars with masses less than 0.9~M$_\odot$), with the mass-to-light ratios increasing more with increasing metallicity.  The assumptions going into these models are described in Section \ref{sec:MLratio}. }
\label{fig:MLR}
\end{figure*}

A reasonable fit is obtained with $a=10^{-4}$, and $b=4$, and this function is plotted as a function of $(1+z)$ in Fig.~\ref{fig:L3muFit} for the four metallicities for which numerical mass functions have been obtained.  The circles with uncertainties are the values of $\mu$ given in Table \ref{table1} that were obtained from the L$_3$ fits of the numerical mass distributions.  It can be seen that the above empirical function provides a reasonable fit to the mass functions derived from the radiation hydrodynamical simulations.  The value of $\mu_0$ has been estimated from the $z=0$ and $Z=0.01~\mathrm{Z}_\odot$ results.  The results of fitting the L$_3$ function to the $z=0$ calculations give values of $\mu$ that vary from $\mu=0.17$ to 0.19 with uncertainties of $\approx 0.02$.  The $Z=0.01~\mathrm{Z}_\odot$ results give slightly lower values of $\mu$ for the $z=3,5,7$ calculations (ranging from $\mu=0.13$ to 0.14, with similar uncertainties).  With the limited precision, it is difficult to know whether any of these values differ significantly from the others, so we take a value of $\mu_0=0.16$, although this should be taken to have a similar level of uncertainty of $\pm 0.02$.  We note again that a good L$_3$ fit to the \cite{Chabrier2005} IMF is obtained with $\mu=0.20$, so our empirical formula will give a very slightly bottom-heavy IMF for low metallicity or at $z=0$ compared to the \cite{Chabrier2005} IMF.

The empirical fit gives a reasonable fit for the behaviour of $\mu$ as the redshift is increased for metallicities $Z \geq 0.1~\mathrm{Z}_\odot$ (Fig.~\ref{fig:L3muFit}), except that it tends to over predict the value of $\mu$ for the $z=10$ calculations with solar and super-solar metallicities.  However, these values of $\mu$ have the largest uncertainties and, as discussed at the end of Section \ref{sec:stellarpop}, in these two calculations most of the stars were still accreting when the calculations were stopped (see the lower right panels of Fig.~\ref{fig:IMFdiff}) so it is likely that of all the calculations these two are the most likely to significantly underestimate the eventual characteristic stellar masses.  Thus, the fact that the empirical fit somewhat exceeds these two values seems appropriate.

There is considerable uncertainty in the values of $a$ and $b$.  In particular, reasonable fits can be obtained using values of $b \in [3,4]$.  If $b=3$ is used, with an appropriate value of $a$, this tends to make the increase in $\mu$ with redshift shallower, over-estimating the values of $\mu$ at $z=3$ for the solar and super-solar metallicity calculations, but fitting the $z=10$ results more closely.  However, we believe that $\mu$ values obtained from the $z=3$ calculations are likely to be more precise than the $z=10$ values because of the larger numbers of stars produced and the larger fraction stars that have finished accreting in the $z=3$ calculations.  Running even larger simulations that produce many more stars would be the only way to significantly improve the precision of the results (which is not computational feasible at the current time).

\subsection{Mass-to-light ratios}
\label{sec:MLratio}

The mean mass of a stellar mass distribution described by the L$_3$ function is given by equation \ref{eq:mean}, which is the integral of the stellar mass times the probability of that mass.  The luminosity-mass relation for main-sequence stars can be expressed as
\begin{equation}
\frac{L_*}{L_\odot} \propto \left( \frac{M_*}{M_\odot} \right)^{\Gamma},
\end{equation}
where, for example, for masses from $M_* \approx 0.43-2$~M$_\odot$ the exponent is $\Gamma \approx 4$, and for masses $M_* \approx 0.43-2$~M$_\odot$ the exponent $\Gamma \approx 3.5$.  

Expressing all masses and luminosities in units of M$_\odot$ and L$_\odot$, respectively, we can compute an approximate mean luminosity for a population of main-sequence stars that is described by the L$_3$ PDF as 
\begin{equation}
\overline{L}  =  \int^{m_\mathrm{u}}_{m_\mathrm{l}} m^{\Gamma} \; p_\mathrm{L3}(m) \; \mathrm{d}m.
\label{eq:light}
\end{equation}
This is not as simple to solve as the equation for the median mass, but it can still be solved analytically.  We can write
\begin{equation}
\begin{aligned}
\overline{L} & =   \int^{m_\mathrm{u}}_{m_\mathrm{l}} m^{\Gamma} \; p_\mathrm{L3}(m) \; \mathrm{d}m, \\
& = \mu^{\Gamma} \frac{(1-\alpha) (1-\beta)}{ G(m_\mathrm{u}) - G(m_\mathrm{l})  } \int^{y_\mathrm{u}}_{y_\mathrm{l}}  y^{\Gamma-\alpha} (1+ y^{1-\alpha} )^{-\beta} \mathrm{d}y, 
\end{aligned}
\label{eq:light1}
\end{equation}
where again we have made the substitution $y=m/\mu$, with the integration limits $y_\mathrm{l}=m_\mathrm{l}/\mu$ and $y_\mathrm{u}=m_\mathrm{u}/\mu$.  The solution of the integral can be written using the hypergeometric function, ${}_2\mathrm{F}_1(a, b; c; x)$, so that we finally obtain
\begin{equation}
\begin{aligned}
\overline{L}  = & \mu^{\Gamma} \frac{(1-\alpha) (1-\beta)}{ (\Gamma+1-\alpha) \left[ G(m_\mathrm{u}) - G(m_\mathrm{l}) \right]  }  \times \\
&  \left[  (y_\mathrm{u}^{\Gamma+1-\alpha}) {~}_2\mathrm{F}_1 \left( \frac{\alpha-(\Gamma+1)}{\alpha-1}, \beta; \frac{2\alpha-(\Gamma+2)}{\alpha-1}; -y_\mathrm{u}^{1-\alpha} \right) \right. \\
& \left.  - {~} (y_\mathrm{l}^{\Gamma+1-\alpha}) {~}_2\mathrm{F}_1\left( \frac{\alpha-(\Gamma+1)}{\alpha-1}, \beta; \frac{2\alpha-(\Gamma+1)}{\alpha-1}; -y_\mathrm{l}^{1-\alpha} \right)   \right].
\end{aligned}
\label{eq:light2}
\end{equation}
The hypergeometric function can be evaluated, for example, using the Python function scipy.special.hyp2f1.

The mass-to-light ratio of the stellar mass population can then be approximated analytically as the ratio of equation \ref{eq:mean} (expressed in solar masses) to equation \ref{eq:light2}, i.e., $\overline{M}/\overline{L}$, expressed in solar units M$_\odot$/L$_\odot$.  

In this paper, rather than present actual mass-to-light ratios in solar units, it is more instructive to compare the mass-to-light ratio of the varying IMFs with the mass-to-light ratio of a typical Galactic stellar population.  We choose to do this by fitting the \cite{Chabrier2005} mass function with an L$_3$ function ($\alpha=2.3$, $\beta=2.0$, and $\mu=0.20$), computing its mass-to-light ratio as above, and then using this value, $(M/L)_\mathrm{Chabrier}$, to normalise the other mass-to-light ratios.

\subsubsection{Zero-age mass-to-light ratios}

For zero-age stellar populations, a bottom-light mass function gives an increase in the mean stellar mass, but it increases the mean stellar luminosity at a higher rate because stellar luminosity increases rapidly with increasing stellar mass.  Therefore, relative to a typical Galactic IMF, the mass-to-light ratio is lower with a bottom-light IMF.  In other words, for the same total mass in stars, there will be more light.  Conversely, for a given observed luminosity, there will be less mass in stars.

The left panel of Fig.~\ref{fig:MLR} shows approximately how the zero-age mass-to-light ratio depends on redshift and metallicity, relative to the \cite{Chabrier2005} IMF.  For simplicity, we have assumed $\Gamma=3.5$ for all stellar masses, for which $(M/L)_\mathrm{Chabrier}=3.3\times 10^{-4}$~M$_\odot$/L$_\odot$.  There is little change in the graph if $\Gamma=3$ or $\Gamma=4$ is used instead.  

The left panel illustrates that the predicted variation in the IMF has a moderate effect on the zero-age mass-to-light ratio of a stellar population.  If solar metallicity gas forms stars at $z=10$, their zero-age mass-to-light ratio is about 2.5 times lower than for a typical Galactic IMF.  At $z=7$, the mass-to-light ratio is about 1.5 times lower.  There is little effect for lower-metallicity gas --- at $z=10$ the mass-to-light ratio for $Z=0.1~\mathrm{Z}_\odot$ is only reduced by $\approx 25$\%.

\subsubsection{Mass-to-light ratios of old stellar populations}

As a stellar population ages, the more massive stars die leaving stellar remnants.  Since they had provided most of the light, the mass-to-light ratio increases.  Relative to a typical Galactic IMF, the massive stars are of more importance for a bottom-light IMF, so for an old bottom-light IMF the mass-to-light ratio increases even more.  For example, in the extreme case of an IMF that is so bottom-light that there are no stars with masses $M_*< \mathrm{M}_\odot$, after 10 Gyr there would be no light (at least from main sequence stars) so the mass-to-light ratio would tend towards infinity.

The right panel of Fig.~\ref{fig:MLR} shows approximately how the mass-to-light ratio of a 13 Gyr old stellar population depends on redshift and metallicity, relative to a similarly old \cite{Chabrier2005} IMF.   To produce this plot we have assumed that the total mass of the population doesn't change, but there is no light from stars that had masses $M_*> 0.9~\mathrm{M}_\odot$.  In other words, when evaluating the integral in equation \ref{eq:mean} we use our usual value of $m_\mathrm{u}=150~\mathrm{M}_\odot$, whereas when evaluating equation \ref{eq:light} we use an upper limit $m_\mathrm{u}=0.9~\mathrm{M}_\odot$. 

Again we find that that the predicted variation in the low-mass IMF with redshift and metallicity has a moderate effect on the mass-to-light ratios of a very old stellar population.  There is essentially no effect for metallicities $Z\leq 0.1~\mathrm{Z}_\odot$, while for a solar metallicity stellar population formed at $z=10$ the mass-to-light ratio at 13 Gyr is approximately a factor of two greater than for a \cite{Chabrier2005} IMF.  If a super-solar metallicity ($Z = 3~\mathrm{Z}_\odot$) population managed to form at $z=10$, by 13 Gyr it would have a mass-to-light ratio four times greater than for a \cite{Chabrier2005} IMF.

\section{Discussion}
\label{sec:discussion}

The idea that fewer stars may form from the same amount of at gas at higher redshift due to the warmer CMBR, and that these stars may tend to have greater characteristic masses has been discussed for a long time, based on the assumption that the Jeans mass would be expected to rise as the CMBR temperature increases with increasing redshift \citep[e.g.,][]{Larson1998}.  However, as shown first by \cite{Bate2023}, it is not this simple: the magnitude of this effect depends on the metallicity of the gas.  Moreover, for higher metallicities, the departure from a `normal' IMF occurs at lower redshifts as the metallicity is increased (Fig.~\ref{fig:L3muFit}). For example, with low-metallicity ($Z=0.01~\mathrm{Z}_\odot$) gas, the characteristic stellar mass does not depend on the redshift despite the warmer CMBR (at least for $z \lsim 10$) because the gas is poor at cooling regardless of the redshift and even at $z=0$ it tends to be hotter than the CMBR would be at $z=10$.  So the change in the CMBR from $z=0$ to $z=10$ has little effect.  But for solar metallicity gas, the temperatures of the densest star-forming gas at $z=0$ drop as low as $T_\mathrm{g} \approx 6$~K and a significant fraction of the densest gas has temperature lower than 10~K.  Therefore, at redshifts where the temperature of the CMBR starts to provide a `temperature floor' greater than this (e.g., even at $z=3$, the CMBR temperature is $T_\mathrm{CMBR} \approx 11$~K), the fragmentation of the gas is inhibited and this begins to increase the characteristic stellar mass.

\subsection{Comparison with previous theoretical work}

\subsubsection{Comparison with the predictions of Bate (2023)}

Based on the results of only $z=0$ and $z=5$ calculations, evolved to $t=1.20~t_\mathrm{ff}$, \citeauthor{Bate2023} (\citeyear{Bate2023}, Section 4.2) made some comments on how the IMF may change for low-redshifts ($z \approx 1-4$) and also for higher redshifts.  We can now compare these predictions with the new results.  For the explored range of metallicities ($Z=0.01-3~\mathrm{Z}_\odot$), \cite{Bate2023} expected little variation of the IMF for $z \approx 1-3$, because the CMBR temperatures at these redshifts are lower than the bulk of the gas temperatures at $z=0$, except for a slight increase in the characteristic stellar mass at super-solar metallicities approaching $z=3$.  This is essentially what we find (Fig.~\ref{fig:L3muFit}), although the increase is somewhat larger, with $\mu$ approximately doubling, than the increase of $\approx 30$\% that was predicted for the super-solar case.  \citeauthor{Bate2023} predicted that a significant increase in $\mu$ should be found for solar-metallicity gas at $z=4$, but not for much lower metallicities; this is indeed what we find here.

However, in the limit of large redshifts, \cite{Bate2023} anticipated that the characteristic stellar mass would scale $\propto (1+z)^{3/2}$ (because the Jeans mass scales as $\propto T^{3/2}$), and that eventually the variation with metallicity should diminish (as the CMBR temperature dominates other over other heating and cooling processes).  Neither of these predictions are compatible with the new results.  Instead, in the limit of high redshift we find a much more rapid increase in the characteristic stellar mass with redshift, and the dependence on metallicity remains (equation \ref{eq:fit}: for large redshifts, $\mu \propto (1+z)^b Z$, with $b \approx 4$).

The persistent metallicity dependence is presumably due to the ability of collapsing gas to keep cooling rapidly to higher densities when it has a lower metallicity (i.e., it remains optically thin to higher densities).  \cite{Bate2019} identified this as the reason for both the metallicity independence of the IMF at $z=0$ (at lower metallicity, enhanced small-scale fragmentation at high density negates the effect of the typically higher temperatures on large-scales increasing the thermal Jeans mass), and the cause of the anti-correlation of the close binary frequency with metallicity.  The latter has been observed for solar-type stars \citep{Badenes_etal2018, ElBRix2019, MoeKraBad2019} and found in simulations of present-day star formation at different metallicities \citep{Bate2019}.  

In the present context, although the increasing temperature of the CMBR with increasing redshift produces a `temperature floor' that is independent of metallicity, the thermodynamic behaviour of this gas under compression still depends on its opacity.  In particular, gas that is compressed on a timescale faster than its cooling time (i.e., because it is optically thick) will behave approximately adiabatically, while optically thin gas may behave approximately isothermally.  For a gravitationally-unstable clump of gas (i.e., the compression is caused by collapse), if the effective polytropic index describing its thermal behaviour ($T_\mathrm{g} \propto \rho^{\gamma}$) is $\gamma < 4/3$ it will continue to collapse (e.g., nearly isothermal), whereas if $\gamma > 4/3$ it may heat up quickly enough to stop collapsing.  Thus, even with a hot CMBR, if gravitationally unstable clumps of gas are able to form, the degree to which they fragment as they collapse will depend on their metallicity, with lower metallicities producing more fragments and stars with lower characteristic stellar masses.  Note that this discussion assumes that the gas is still in the dust-dominated cooling regime (i.e., not dust-free star-formation).

\subsubsection{Comparison with the accretion/ejection model for the IMF}
\label{sec:accdy}

The steeper than anticipated dependence of the characteristic stellar mass on redshift at high redshift (i.e., $b \approx 4$, rather than $b \approx 3/2$) might be explained by models of the origin of the shape of the IMF.  For example, \cite{BatBon2005} proposed that the shape of the IMF is due to a competition between accretion and dynamical interactions between protostars in which dynamical interactions can stop protostars from accreting by ejecting them from regions with dense molecular gas.  In this model, most stars accrete for only a short period of time before they are ejected from the dense gas, while a few stars are able to accrete to much higher masses over (comparatively) long timescales. The characteristic stellar mass is given by the product of the typical accretion rate and the typical ejection timescale.  \cite{BatBon2005} argue that the former is expected to scale $\propto c_\mathrm{s}^3/G$ (where $c_\mathrm{s}$ is the typical sound speed in the molecular gas), while the latter is expected to scale with the dynamical timescale of the proto-cluster, which in turn scales with the mass density, $\rho$, as $\propto \rho^{-1/2}$.  Together these two scalings imply that the characteristic stellar mass scales with the  Jeans mass.  This relatively simple model is able to reproduce the form of the stellar mass distributions obtained from numerical simulations \citep*[e.g.][]{Bate2005, Bate2009a, MatFedSet2023}.  However, it does not explicitly consider other thermodynamic effects such as radiative feedback from protostars, or in the present context, the effects of metallicity or a temperature floor.  Indeed, \cite{Bate2009b} retained this basic model, but argued that the thermal feedback from accreting protostars modifies the Jeans mass, increasing it locally.  If the protostars form closer together (i.e., the mass density is greater), the thermal feedback will have a greater effect, reducing the amount of fragmentation and increasing the timescale between dynamical encounters between protostars, thus leading to protostars accreting for longer before their accretion is stopped.  \cite{Bate2009b} argued that this effect of thermal feedback could offset the simple $\propto \rho^{-1/2}$ scaling, reducing the dependence of the characteristic stellar mass on the cloud's mean density \citep[see also][]{JonBat2018a}.

In applying such an accretion/ejection model in the context of different redshifts and metallicities, there are several new effects to consider.  First, the effect of the `temperature floor' caused by the CMBR would imply that the Jeans mass scales $\propto (1+z)^{3/2}$.  This would be expected to increase the typical accretion rates by the same factor (due to the $c_\mathrm{s}^3/G$ scaling).  But as we have seen, this scaling alone is not strong enough to explain the rapid increase in the characteristic stellar mass with increasing redshift.  However, at the same time as increasing the accretion rates, the warmer gas temperatures would be expected to decrease the amount of fragmentation, producing a lower number density of protostars.  Indeed, this effect of fewer protostars being formed with increasing redshift at a given metallicity is clear in the third column of Table \ref{table1} for $Z=0.1-3~\mathrm{Z}_\odot$.  This in turn leads to longer timescales between protostars interacting dynamically, allowing them to accrete for longer and therefore also acting to increase the characteristic stellar mass.  Because both of these effects act to increase stellar masses with increasing redshift (i.e., both increasing the accretion rates and increasing the typical accretion timescales), it is not surprising that the characteristic stellar mass increases rapidly with $\propto (1+z)^b$, with $b > 3/2$, though exactly why $b \approx 4$ is unclear.

The second effect is that of lower metallicities leading to greater small-scale fragmentation.  In the context of the accretion/ejection model, this would decrease the timescale between dynamical encounters, giving the typical protostar less time to accrete gas before it is ejected from the dense gas, and in turn leading to lower characteristic stellar masses with decreasing metallicity, as is seen.  Qualitatively, this describes the observed decrease of the characteristic stellar mass with decreasing metallicity at high redshift.  

For present-day star formation, there is no strong metallicity dependence of the characteristic stellar mass \citep{Bate2019}.  This can also be understood by considering the effects of different metallicities on the accretion rates and dynamical encounter timescales.  For example, at lower metallicities the low-density molecular gas tends to be hotter because it is inefficient at cooling.  This should lead to higher typical accretion rates (due to the $c_\mathrm{s}^3/G$ scaling).  However, this effect is offset by the enhanced small-scale fragmentation at high densities, due to the reduced opacity.  This will reduce the timescale between dynamical encounters, giving the typical protostar less time to accrete mass.  These two effects essentially cancel each other out at $z=0$, leading to the metallicity-independent IMF for metallicities $Z=0.01-3~\mathrm{Z}_\odot$.

\begin{figure*}
\centering
    \includegraphics[width=8.5cm]{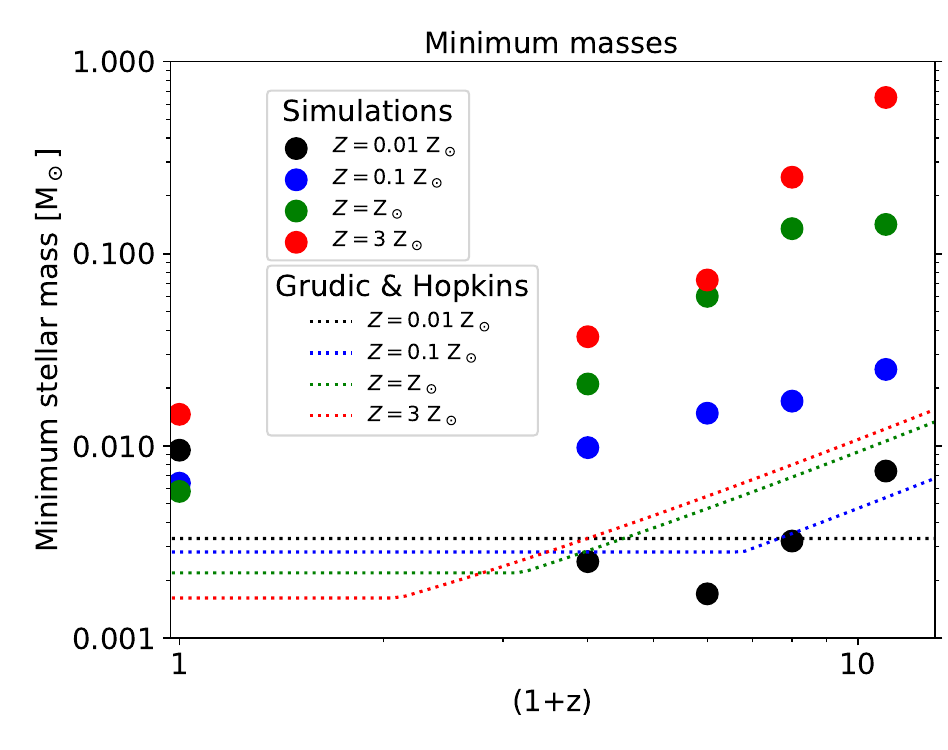} \vspace{0cm}
    \includegraphics[width=8.5cm]{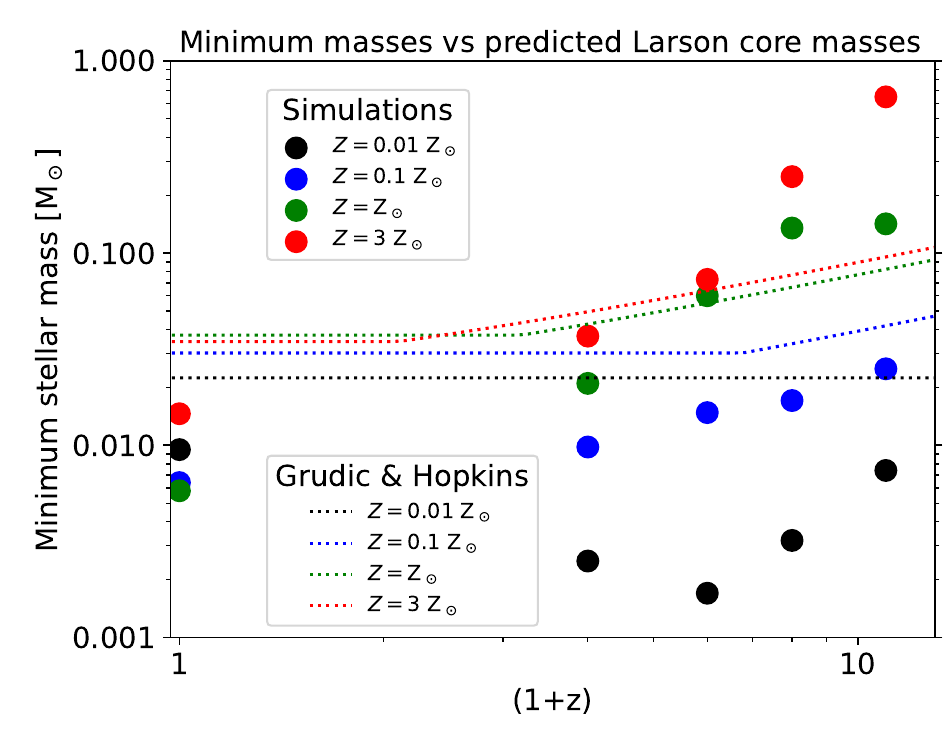} \vspace{0cm}
    
    \includegraphics[width=8.5cm]{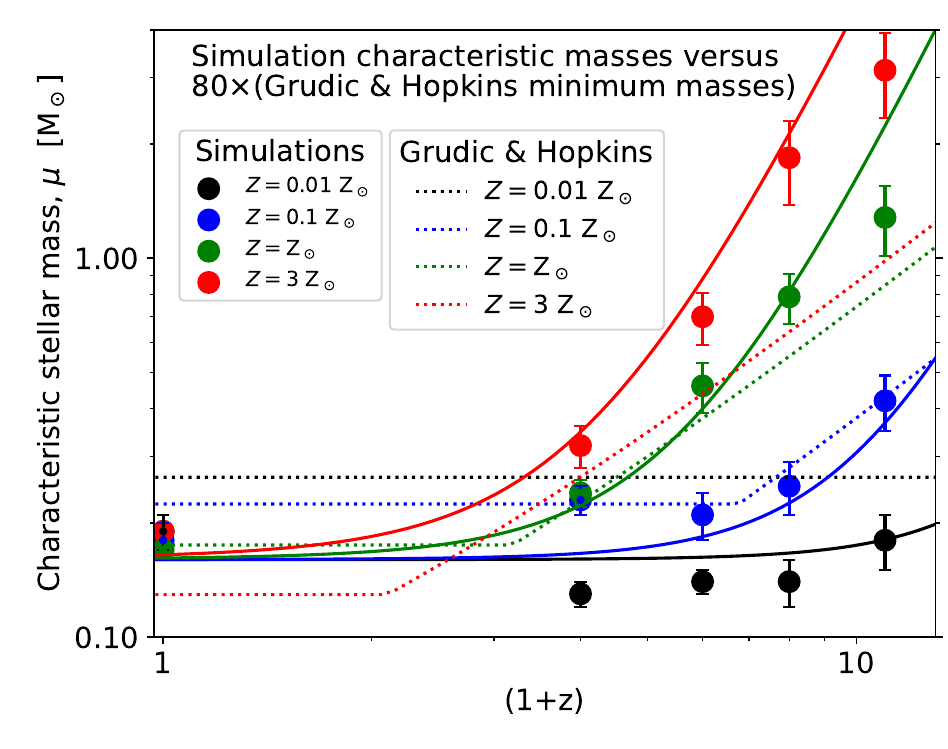} \vspace{0cm}
    \includegraphics[width=8.5cm]{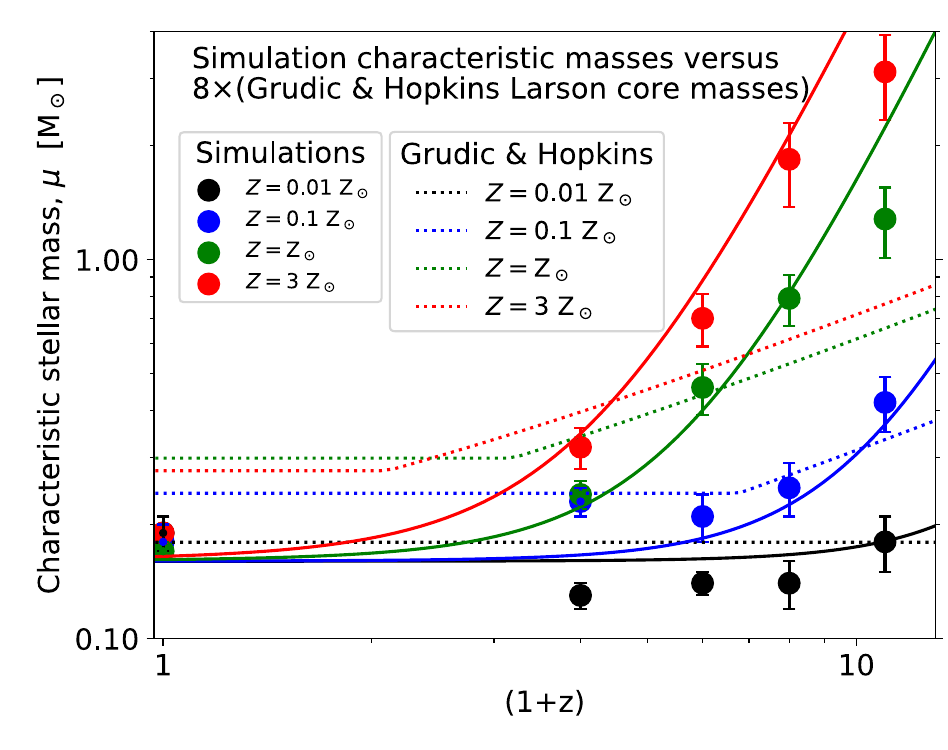} \vspace{0cm}
\caption{Comparison of the variation of the minimum stellar mass (top row) and the characteristic stellar mass (bottom row) with redshift and metallicity that is obtained from the simulations discussed in this paper with the minimum masses (left column) and first hydrostatic core (FHSC) masses (right column) predicted by the analytic models of Grudi\'c \& Hopkins (2023).  In the top-left panel, we directly compare the minimum stellar masses obtained from the simulations (coloured circles) with the predicted minimum stellar masses (dotted lines), with different metallicities denoted using different colours.  In the top-right panel, we directly compare the minimum stellar masses obtained from the simulations with the predicted FHSC masses (dotted lines).  In the lower two panels we arbitrarily scale up the analytic predictions by factors of 80 (for the minimum masses; bottom-left panel) and 8 (for the FHSC masses; bottom-right panel).  In the bottom panels we also plot the solid lines given by equation \ref{eq:fit} with parameters: $\mu_0=0.16$, $a=10^{-4}$ and $b=4$ that provide reasonable fits to the numerical results. Although the Grudi\'c \& Hopkins (2023) models do predict minimum stellar masses and FHSC masses that are constant at low redshift but change to increasing values at higher redshift, in qualitative agreement with the simulations, there is generally poor agreement in terms of actual values. }
\label{fig:Grudic}
\end{figure*}

\subsubsection{Comparison with models for the minimum stellar mass, and associated IMF models}

Theory predicts that there is a minimum stellar mass, the so-called `opacity limit for fragmentation', below which a clump of dusty molecular gas cannot collapse dynamically under its own gravity without the heating due to compression stopping that collapse \citep{Hoyle1953, LowLyn1976, Rees1976, BoyWhi2005}.  During the collapse of a more massive cloud, the mass of the first fragment initially has this mass \citep{Larson1969}, but being embedded in an infalling envelope it continues to grow rapidly producing what \cite{Larson1969} called the `first hydrostatic core' (FHSC).  This large (radius $\sim 5$~au) pressure-supported object eventually undergoes a second dynamical collapse when its central temperature exceeds that required to dissociate molecular hydrogen, and a stellar core with an initial mass of a few Jupiter masses is formed within the remnant of the first core \citep[e.g.,][]{Larson1969, Bate1998}.  The mass of the first hydrostatic core just before this second collapse depends on many factors, including the infall rate of gas on to it, its rotation rate and opacity, but is typically of order 0.03~M$_\odot$.
Some authors have argued that the characteristic stellar mass is associated with the minimum mass stellar mass or with the mass of the FHSC.  For example, \cite*{LeeHen2018b} and \cite*{HenLeeCha2019} argue that the characteristic stellar mass is about 5--10 times the FHSC mass due to further accretion from the envelope that is eventually halted by the formation of new fragments.  On the other hand, other authors argue that the characteristic stellar mass is not closely linked to the minimum stellar mass \citep[e.g.,][]{Bate2005}.  

Two papers have recently explored, analytically, the dependence of the minimum mass for star formation on redshift and metallicity \citep{Whitworth_etal2024, GruHop2023}.   They discuss the effects of some of the same processes when considering the minimum stellar mass that we have discussed when considering the characteristic stellar mass.  In particular, they also find that the CMBR provides a temperature floor that is only important above a critical redshift, and that this critical redshift has a lower value when the dust abundance is higher.  They also find that the dependence of the minimum stellar mass on redshift is stronger at higher dust abundances.  Thus, the qualitative behaviour that is found in the simulations presented in this paper is in broad agreement with the predictions of these analytic models.  We note that the simulations assume that the dust opacity scales in direct proportion to the metallicity.  This may well not be the case (see the discussion in Section \ref{hydro}).  Furthermore, in the simulations only the CMBR component of the ISRF is changed.  The short-wavelength component of the ISRF is taken to be that found in the solar vicinity and is not changed (see Section \ref{hydro} and Fig.~\ref{ISRF}).  \cite{Whitworth_etal2024} find that an enhanced local radiation field tends to increase the minimum stellar mass and so also moves the critical redshift at which the minimum mass increases with redshift to higher values. In the following discussion we delve a little deeper into how the analytic predictions compare with the results from the simulations.  

We first discuss the analytic results of \cite{Whitworth_etal2024}.  At $z=0$ with a local ISRF, they obtain minimum stellar masses that range from 0.007 M$_\odot$ at $Z=0.01~\mathrm{Z}_\odot$ to 0.003 at $Z=3~\mathrm{Z}_\odot$ (i.e., they decrease slightly with increasing metallicity).  They compare their results to the minimum stellar masses obtained from the numerical simulations presented by \cite{Bate2019} which they state are all $\sim 0.01$~M$_\odot$: they range from $\approx 0.005$ to $\approx 0.01$~M$_\odot$ (see also Fig.~\ref{fig:IMFdiff} and the upper panels of Fig.~\ref{fig:Grudic}) without any consistent trend with metallicity (the lowest mass is from the solar-metallicity calculation, while the highest mass is from the super-solar metallicity calculation). It should be noted that numerical results from small clusters of stars are always likely to produce higher masses than the true minimum simply because of small number statistics and the fact that most objects accrete to final masses that maybe one to three orders of magnitude greater than the minimum mass. Thus, the analytic results and the numerical results are in good agreement at $z=0$.

\cite{Whitworth_etal2024} also compares their predictions at $z=5$ to the simulations of \cite{Bate2023}.  At this redshift, they find that their predicted minimum masses range from 0.007 and 0.006~M$_\odot$ from $Z=0.01-1~\mathrm{Z}_\odot$.  In comparing these minimum masses to \cite{Bate2023} they make a mistake: they state that the two highest metallicity simulations have minimum masses $\approx 0.01$~M$_\odot$ and the lowest metallicity calculation has a minimum mass of $\approx 0.1$~M$_\odot$.  However, they got this backwards --- it is the highest metallicity case (solar metallicity) that has the greater minimum mass.  This result is not in agreement with their analytic model, and neither is the trend.  They find that the minimum mass should decrease with increasing metallicity, whereas in the simulation at $z=5$ and solar metallicity the minimum mass increases substantially compared to the lower metallicity simulations.  \cite{Whitworth_etal2024} do predict that the minimum mass increases  with increasing redshift above $z=3$, but only gradually from about 0.003 form $z<3$ to 0.011 at $z=8$ (with their fiducial background level of non-CMB radiation).  As can be seen in Fig.~\ref{fig:IMFdiff}, we find a much greater variation in the minimum stellar mass at high redshift and high metallicity.  This does not necessarily mean that the \cite{Whitworth_etal2024} model is incorrect, but simply that the minimum stellar mass may also depend on other effects.  In particular, the minimum mass for collapse is just that: a minimum.  Such objects will accrete to higher masses than this minimum if they are embedded in dense gas, unless something prevents them from doing so \citep[c.f.,][]{BatBon2005}.  From this point of view, it should be expected that all of the predicted minimum masses from \cite{Whitworth_etal2024} are smaller than or comparable to all of minimum stellar masses obtained from the simulations of \cite{Bate2019, Bate2023} and the calculations discussed in this paper and this is, indeed, the case. 

In the second paper, \cite{GruHop2023} present analytic models for the scalings of both the minimum stellar mass and the FHSC mass with redshift and metallicity.  Helpfully, they also provide a Python code to generate the values.  In Fig.~\ref{fig:Grudic} we use this code to compare their values with the results of the simulations.  In the upper panels, we compare their predicted minimum masses (left panel) and FHSC masses (right panel) with the masses of the lowest mass object that has finished accreting (accretion rate $<10^{-7}$~M~yr$^{-1}$) in each of the simulations.  As mentioned above, the general trends of a minimum mass that increases beyond a critical redshift and a greater effect at higher metallicity are in agreement with the results from the simulations.  However, the analytic values are a poor fit to the lowest stellar masses obtained in the simulations.  At the lowest metallicity ($Z=0.01~\mathrm{Z}_\odot$), the predicted minimum masses (top left panel) are in reasonable agreement with the simulation results, but for all other metallicities the simulations give substantially larger masses, particularly at high redshift.  Furthermore, at low redshift ($z \lsim 2$) \cite{GruHop2023} predict that the minimum mass should decrease with increasing metallicity (also the case for the \cite{Whitworth_etal2024} models), but the simulations don't produce any definitive trend.  Similarly, the predicted FHSC masses (top right panel) are a poor fit to the lowest masses produced by the simulations.  At low metallicities the predicted values are an order of magnitude too high, and at high metallicity and high redshift the predicted values are too low.  All that can really be said is that the ordering (increasing FSHC mass with increasing metallicity) is in agreement with the simulation minimum masses, and both increase with redshift.

In the lower two panels of Fig.~\ref{fig:Grudic}, we test the idea that the characteristic stellar mass of the stellar initial mass function (i.e., the median or peak mass, $\mu$) might be described by a scaling up of the minimum mass or the Larson FHSC mass \citep[as proposed by][]{LeeHen2018b, HenLeeCha2019}.  In the bottom left panel, we arbitrarily scale up the \cite{GruHop2023} minimum masses by a factor of 80.  The factor of 80 gives a reasonable fit to the characteristic stellar masses that have been obtained from the simulations at intermediate metallicities ($Z=0.1-1~\mathrm{Z}_\odot$; the blue \& green points and dotted lines).  However, as mentioned above, the scaled dotted lines give lower masses with increasing metallicity at low redshift, whereas the simulations give a characteristic stellar mass that does not significantly depend on metallicity.  Furthermore, the dotted line for $Z=3~\mathrm{Z}_\odot$ is too low, particularly at high redshift, and the dotted line for $Z=0.01~\mathrm{Z}_\odot$ is too high.  Moreover, the factor of 80 is entirely arbitrary.  In the bottom right panel, we arbitrarily scale up the \cite{GruHop2023} Larson FHSC masses by a factor of 8.  The factor of 8 gives a reasonable fit to the characteristic stellar masses for the low metallicity calculations ($Z=0.01-0.1~\mathrm{Z}_\odot$; the black \& blue points and dotted lines).  Also, the scaled dotted lines generally give higher masses with increasing metallicity in general agreement with the simulations.  However, there is still too much variation predicted in the characteristic stellar masses at low redshift ($z \lsim 1$) compared to the simulations, and the rate of increase of the characteristic stellar mass with increasing redshift that is predicted by scaling up the FHSC masses is much too low compared to the results from the solar and super-solar metallicity simulations.  Moreover, once again the factor of 8 is arbitrary (although it does fall within the range of $5-10$ proposed by \citealt{HenLeeCha2019}).

In summary, there are a number of similarities between the results from the simulations presented here and the analytic models of the scalings of the minimum stellar masses with metallicity and redshift, according to both \cite{Whitworth_etal2024} or \cite{GruHop2023}, and the scalings of the Larson core mass, according to to \cite{GruHop2023}.  In both the analytic models and in the simulations there is a critical redshift below which both the minimum stellar mass and the characteristic stellar mass are approximately constant.  Above this critical redshift, the value of which tends to decrease with increasing metallicity, both the minimum stellar mass and the characteristic stellar mass increase with increasing redshift.  However, there are also a number of ways in which the minimum stellar masses from the simulations differ from those predicted by the analytic models.  Similarly, the median (characteristic) stellar masses obtained from the simulations are not well described by scaling either the analytic minimum masses or first core masses by simple factors. The minimum stellar masses at low redshift are predicted to depend on metallicity in both of the analytic models (decreasing with increasing metallicity), but there is no evidence for this from the simulations (if anything the trend is opposite, although an exact cut-off is difficult to determine with the limited number of stars that are produced).  Furthermore, the rate of increase with redshift is slower than is found in the simulations.  To try to reproduce the characteristic stellar masses that are obtained from the simulations, the best option is to scale the \cite{GruHop2023} Larson FHSC values up by a factor of $\approx 8$.  This provides a reasonable level of agreement at low metallicities, but not for solar or super-solar metallicities.  For the high metallicities, the magnitude of the increase of masses with increasing redshift is much too small (the \cite{GruHop2023} Larson FHSC values increase by a factor of two between $z=0$ and $z=10$, whereas the characteristic stellar masses from the simulations increase by an order of magnitude).  These differences between the analytic models and the results from the simulations strongly suggest that there is not a simple mapping from the minimum stellar mass, or the Larson core mass, to the characteristic stellar mass.  Instead, it is likely necessary to consider additional processes (such as the roles of accretion and dynamical interactions, as discussed in Section \ref{sec:accdy}).  Furthermore, as mentioned above, the minimum stellar mass is a minimum.  The vast majority of objects will accrete to substantially higher masses. For an object to end up with a mass near the absolute minimum mass in a star-forming region may require very unusual circumstances to stop it from accreting to a higher mass (e.g., dynamical ejection from a multiple system immediately after it is formed).  So in a typical star-forming region, and in numerical simulations such as those presented here, the {\em lowest mass object} is likely to have a mass significantly greater than the nominal minimum mass.

\subsection{Comparison with observations of galaxies and stellar clusters}

For the variation of the low-mass end of the IMF that we predict in this paper to have a significant effect on the stellar populations of galaxies or stellar clusters a sufficient fraction of the stars need to form at a high enough redshift with a large enough metallicity.  Since previous generations of stars are required to increase the metallicity of star-forming gas, and this takes time, it is not clear whether or not this will lead to observable effects.

\subsubsection{JWST high-redshift galaxies}

Over the past two years, observations of high-redshift galaxies made using the James Webb Space Telescope (JWST) have identified relatively massive galaxies at redshifts $z \gsim 10$, with unexpectedly bright, seemingly massive, galaxies being discovered at $z>7$ \citep[e.g.,][]{Naidu_etal2022, Labbe_etal2023, Xiao_etal2023}.  If the IMFs of these galaxies were top-heavy or bottom-light, their inferred masses would be reduced \citep[e.g.,][]{Steinhardt_etal2023, Woodrum_etal2023, Wang_etal2024}.

The current spectroscopically-confirmed high-redshift record holders are two galaxies at redshifts $z=14.3$ and $z=13.9$ with estimated masses $\sim 10^8$~M$_\odot$ \citep{Carniani_etal2024}.  At this redshift, the Universe is only $\approx 300$~Myr old, and star formation in these galaxies is inferred to have been ongoing for $\approx 100$~Myr.

Extrapolating the results of this paper to $z=14$ using equation \ref{eq:fit}, it is found that the characteristic (median or peak) stellar mass should be $\mu \approx 0.7$~M$_\odot$ for stars with metallicities $Z\approx 0.1~\mathrm{Z}_\odot$ (compared to the present-day Galactic IMF, for which $\mu = 0.2$~M$_\odot$), and if solar-metallicity gas is available in these high-redshift galaxies the characteristic stellar mass of the stars that it is producing may be as high as $\mu \approx 5$~M$_\odot$.  In terms of the zero-age mass-to-light ratio, assuming the high-mass end of the IMF remains Salpeter, these $z=14$ bottom-light IMFs would reduce the mass-to-light ratio relative to the \cite{Chabrier2005} value by a factor of two for $Z\approx 0.1~\mathrm{Z}_\odot$ or by a factor of 5 for solar-metallicity star formation.  In other words, if a typical Galactic IMF is used to infer the total stellar masses of these young galaxies whereas in fact the IMF is bottom-light, the total stellar mass will be being over-estimated.  This factor could be significant if a large enough fraction of the stellar light is coming from the stars that formed from moderate to high metallicity molecular gas, though a better estimate would require knowledge of the fractions of the stars that formed with different metallicities (e.g., from galaxy formation simulations).

For star formation at $z<10$, the effects of the change in the low-mass end of the IMF on the overall zero-age mass-to-light ratio are likely to be small (Fig.~\ref{fig:MLR}).  Even for stars formed from solar-metallicity gas at $z=10$, the mass-to-light ratio is only a factor of two lower than with a typical Galactic IMF, and most of the stars in such galaxies are likely to have formed at lower metallicities.  Furthermore, if there is a change of the power-law slope of the high-mass end of the IMF at high redshift \citep[e.g.,][]{Chon_etal2022} this could easily overwhelm the effect from the change in the low-mass end of the IMF.

\subsubsection{Centres of early-type galaxies}

A variety of observations over many years point to possible changes of the IMF in the centres of massive early-type galaxies \cite[see][for a review]{Smith2020}.  In particular, the central regions of early-type galaxies apparently have high mass-to-light ratios (\citealt*{GuzLucBow1993, JorFRaKja1996}; \citealt{Burstein_etal1997, Thomas_etal2011, Cappellari_etal2012, Cappellari_etal2013}; \citealt*{DutMenSim2012};,\citealt{Wegner_etal2012, Dutton_etal2013}; \citealt*{TorRomNap2013}; \citealt{McDermid_etal2014}; \citealt*{DavMcD2017, Li_etal2017, Shetty_etal2020}) that increase with increasing central velocity dispersion (i.e., mass), a result that has been strengthened by using gravitational lensing to measure the total projected mass within the Einstein radius to obtain the mass-to-light ratios (\citealt*{FerSahWil2005, FerSahBur2008}; \citealt{Auger_etal2010, Ferreras_etal2010, Treu_etal2010, Barnabe_etal2011, Posacki_etal2015, Spiniello_etal2015, Newman_etal2017}).
Furthermore, the central regions of early-type galaxies are observed to have high metallicities (\citealt*{Faber1973,WorFabGon1992}; \citealt{Vazdekis_etal1996,Vazdekis_etal1997, Trager_etal2000,Gallazzi_etal2006}; \citealt*{GraFabSch2009a,GraFabSch2009b}; \citealt{Kuntschner_etal2010,McDermid_etal2015}).  This is in contrast to the outskirts of such galaxies, which are generally composed of metal-poor stars \citep[e.g.,][]{Greene_etal2015}.

Historically, these observations, along with high values of [$\alpha$/Fe] observed in massive early-type galaxies, have been used argue for a top-heavy IMF \citep[e.g.][]{Larson1998}.  However, since the mass-to-light ratios of old stellar populations depend sensitively on the behaviour of the IMF in the vicinity of a solar-mass (because stars above $\approx 0.8$~M$_\odot$ have evolved into dark stellar remnants, while the light is generated by stars below this mass), \cite{Larson1998} has argued that the high mass-to-light ratios could also be explained if the IMF had a Salpeter-type slope for high-mass stars, but a varying (characteristic) mass below which the IMF flattens or turns over (e.g., increasing the fraction of very low-mass stars (i.e.\ bottom-heavy), or increasing the excess of stellar remnants (e.g., either top-heavy or bottom-light).  

Spectral analysis of stellar populations can be used to constrain the fraction of low-mass stars in early-type galaxies (\citealt*{Spinrad1962, Cohen1978, FabFre1980, CarVisPic1986, HarCou1988, DelHar1992}; \citealt{Cenarro_etal2003, Falcon_etal2003}; \citealt*{vanCon2010, vanCon2011, Convan2012a, Convan2012b, SmiLucCar2012}; \citealt{Spiniello_etal2012, Spiniello_etal2014};  \citealt{Ferreras_etal2013,  Ferreras_etal2015}; \citealt{LaBarbera_etal2013, LaBarbera_etal2017, MartinNavarro_etal2015, vanDokkum_etal2017}). 
Such studies frequently indicate substantial populations of low-mass stars \citep[e.g.,][]{vanCon2010} that point to bottom-heavy stellar populations, although not in all early-type galaxies (\citealt*{SmiLuc2013, SmiLucCon2015}; \citealt{Leier_etal2016}).  

In attempting to reconcile such (sometimes apparently contradictory) observations, several groups have proposed varying IMFs.  For example, to produce both the high metallicity and enhanced [$\alpha$/Fe] of early-type galaxies and a bottom-heavy IMF, the IMF maybe time-dependent IMF such that it changes from top- to bottom-heavy at early times \citep{Vazdekis_etal1997, NarDav2013, Weidner_etal2013, Ferreras_etal2015}.  Alternately, \citet{Gunawardhana_etal2011} and \citet{Weidner_etal2013} proposed that the IMF becomes more top-heavy with increasing star-formation rate.  Similarly, \citet{Baugh_etal2005} proposed that the IMF becomes more top-heavy and \cite{Whitworth_etal2024} have proposed that the IMF becomes bottom-light in star burst environments in order to reproduce galaxy luminosity functions \citep[see also][]{Gunawardhana_etal2011}.   Top-heavy IMFs have also been proposed to explain a variety of other observations \citep[e.g.,][]{NarDav2012, Zhang_etal2018}.  \cite{BarCraSch2018, BarSchCra2019a, BarSchCra2019b} investigated the effects of bottom-heavy and top-heavy IMFs on galaxy formation simulations.  Recently, \cite{vanDokCon2024} proposed a `concordance' IMF that can be simultaneously bottom-heavy, with a steep slope at low stellar masses, and top-heavy, with a shallow slope at high masses, that applies to the most massive galaxies, with stellar velocity dispersions $\sigma \sim 300$~km~s$^{-1}$, and their progenitors.

While bottom-light IMFs similar to those obtained from the star formation calculations discussed in this paper can potentially help to explain the observed high light-to-mass ratios in the centres of massive early-type galaxies, they are apparently in contradiction with the bottom-heavy stellar populations that are inferred from spectral analysis.  On the other hand, as discussed by \cite{Bate2023}, these results do provide a potential mechanism for metal-rich gas to produce time-dependent IMFs that changes from top-heavy to a standard IMF as redshift decreases \citep[similar to that postulated by][]{Vazdekis_etal1997, Weidner_etal2013, Ferreras_etal2015}.  

It may also be possible to get both bottom-light and bottom-heavy populations by varying the redshift, metallicity, {\it and} the molecular cloud density.  Several studies have proposed that the characteristic stellar mass may decrease at high molecular cloud densities (e.g., \citealt*{Hopkins2013, ChaHenCha2014}, \citealt{JonBat2018a, TanKruFed2022, TanKru2024}).  For example, for present-day low-mass star formation \cite{JonBat2018a} find from radiation-hydrodynamical calculations of star cluster formation that the characteristic stellar mass scales with the mean molecular cloud density as $\rho^{-1/5}$.   This is less sensitive to density than a naive Jeans mass argument (in which $M_{\rm J} \propto T^{3/2} \rho^{-1/2}$), helping to explain why the IMF does not seem to vary greatly in our galaxy, but nevertheless such a density dependence leaves open the possibility of significant variations in the characteristic stellar mass in extreme (high-density) environments.  Thus, there is more work to be done in the future to map out how the low-mass IMF may depend on a wider variety of parameters.

\subsubsection{Stellar mass functions of stellar clusters}

Another possibility to observe variations of the IMF is to study the stellar populations of stellar clusters, such as old globular clusters, lensed clusters at high-redshift, or nearby clusters in extreme environments.  Stellar clusters are less complex than galaxies: they are not thought to contain significant dark matter, they don't contain super-massive black holes and active galactic nuclei, and they are (largely) formed in a single star formation event (i.e., their stars have a single age).  Ideally this makes observations of them easier to interpret, and in nearby clusters individual stars can be imaged directly.

One complication is the effects of dynamical evolution.  In particular, for old globular clusters the lowest mass stars are lost preferentially.  But if this can be taken into account, they can in principle be used to study the form of the low-mass ($M\lsim 0.8$~M$_\odot$) IMF at high redshift.  For example, \citet*{StrCalSet2011} studied more than a hundred globular clusters in M31, most of which are very old (i.e., they formed at high redshift).  Dividing their sample into several metallicity bins (the most metal-rich sample having metallicities [Fe/H]$~>-0.4$), they found that the more metal-rich globular clusters are increasingly bottom-light, even accounting for the effects of dynamical evolution.  \citet{ShaGie2015} questioned this result, arguing that the masses of the M31 globular clusters may have been systematically underestimated with increasing metallicity.  But four of the M31 metal-rich globular clusters were also studied using spectral analysis by \citet{Convan2012b} to determine the distributions of low-mass stars and, in agreement with  \citet{StrCalSet2011}, they found that these globular clusters had a significant deficit of low-mass stars relative to a Galactic Kroupa IMF.  This observational result is consistent with the expectation from this paper of an increasingly bottom-light IMF for globular clusters with increasing metallicity assuming they formed at $z \gsim 7$.

In some cases, JWST now allows individual stellar clusters to be studied at redshifts, $z\gsim 4$, using gravitational lensing \citep[e.g.][]{Mowla_etal2022, Mowla_etal2024, Claeyssens_etal2023, Vanzella_etal2022, Vanzella_etal2023a}.  The highest redshift case to date is imaging of the Cosmic Gems Arc \citep{Salmon_etal2018} which contains 5 young massive star clusters \citep{Adamo_etal2024} at $z=10.2$ \citep{Bradley_etal2024}.  Although these clusters are thought to have a low metallicity, perhaps around 1\% of the solar value, future studies of such systems may provide more IMF constraints.

Finally, the study of young stellar clusters in our Galaxy or in nearby galaxies in extreme environments may also help to constrain the low-mass end of the IMF.  Regions of interest may include: near the Galactic centre, where the molecular cloud densities, turbulence, and magnetic fields may differ compared to local star forming clouds, or in the outer Galaxy or in satellite galaxies where the metallicity may be lower.  Although the results presented in this paper \citep[along with those of][]{Myers_etal2011, Bate2014, Bate2019} do not predict a strong dependence of the low-mass end of the IMF with metallicity ($Z \gsim 0.01$~Z$_\odot$) for present-day star formation, the calculations of \cite{ChoOmuSch2021, Chon_etal2022} did predict a greater fraction of low-mass stars and brown dwarfs with decreasing metallicity ($Z = 0.1 - 10^{-4}$~Z$_\odot$), although this effect is largely erased when they include radiative feedback in their most recent calculations \citep{Chon_etal2024}.  Thus, observational constraints on the low-mass end of the IMF at different metallicities, even for present-day star formation, would be very useful to help constrain models.  Past studies of star clusters in the outer Galaxy have struggled to reach the sensitivities required to reach below the turnover in the IMF \citep[e.g.,][]{Yasui_etal2023}.  But recent results with the JWST demonstrate that it is now possible to investigate the low-mass end of the IMF in the outer galaxy into the brown dwarf regime \citep{Yasui_etal2024}.  This study of a star-forming region with metallicity $Z \approx 0.2$~Z$_\odot$ indicates that it may have a bottom-heavy IMF, although this conclusion depends on the assumed age and distance of the region.

\section{Conclusions}
\label{conclusions}

We have presented results from twenty radiation hydrodynamical simulations of low-mass star cluster formation that are subjected to the cosmic microwave background radiation appropriate for redshifts $z=0, 3, 5, 7, 10$.  At each of the five redshifts, four calculations were performed with metallicities of 1/100, 1/10, 1, and 3 times the solar value (Z$_\odot$).  The calculations resolve the opacity limit for fragmentation, protoplanetary discs (radii $\gsim 1$~au), and multiple stellar systems.  The calculations include a thermochemical model of the diffuse ISM and treat gas and dust temperatures separately, which is particularly important for setting the temperatures at low densities.  These calculations allow us to study the dependence of the low-mass end of the IMF on redshift and metallicity.

We draw the following conclusions:
\begin{enumerate}

\item We find that the stellar mass distributions obtained from the calculations become increasingly bottom light as the redshift and/or metallicity are increased.  The mass functions obtained for present-day star formation ($z=0$) do not depend significantly on metallicity and are similar to the observed Galactic IMF, in agreement with previous studies \citep{Myers_etal2011, Bate2014, Bate2019}.  Also for the lowest-metallicity (1/100 solar) at all redshifts up to $z=10$ the stellar mass distributions are comparable to the observed Galactic IMF, or possibly slightly bottom-heavy.  However, for higher metallicities there is a critical redshift above which the mass distributions become more bottom-light with increasing redshift (i.e., there is a deficit of brown dwarfs and low-mass stars compared to a typical Galactic IMF) and the critical redshift above which the IMF becomes increasingly bottom-light decreases with increasing metallicity (see Figs.~\ref{fig:IMFdiff} and \ref{fig:IMFcum}).  This behaviour was first identified by \cite{Bate2023}, but here we map out the IMF variation of the low-mass end of the IMF in much greater detail.

\item This dependence of the low-mass end of the IMF on redshift and metallicity is a result of metal-rich gas being unable to cool to as lower temperatures at higher redshift compared to low redshift due to the warmer cosmic microwave background radiation.  The warmer CMB radiation produces a `temperature floor' that increases with increasing redshift, inhibiting fragmentation of the star-forming gas so that those protostars that do form tend to accrete more gas (i.e.\ they have a greater characteristic stellar mass). The reason that there is little dependence of the IMF on redshift ($z \lsim 10$) at low metallicity ($Z=0.01$~Z$_\odot$) is that the bulk of the gas is warm regardless of the redshift because it is poor at cooling, so the increasing temperature of the CMB radiation with increasing redshift has little on the star-forming molecular clouds.

\item We fit the numerical stellar mass functions using the analytic L$_3$ function that was proposed by \cite{Maschberger2013} as a convenient way to parameterise IMFs.  We use this parameterisation in such a way that it has a single free parameter, $\mu$, that is equal to the median or peak mass in d$N$/dlog($m)$.  We then propose an empirical parameterisation that describes how the median mass varies with redshift and metallicity (equation \ref{eq:fit}).  This parameterisation may be used to study the effects of variable low-mass IMFs in other areas of astrophysics, for example, in simulations of galaxy formation.  

\item A bottom-light mass function reduces the zero-age mass-to-light ratio compared to a typical Galactic stellar initial mass function (i.e., a given amount of luminosity is generated from a lower total mass in stars).  Therefore, if the high-redshift galaxies recently observed by JWST have bottom-light IMFs this would reduce their estimated masses compared to those derived by assuming a typical Galactic IMF.  The observed trend of old globular clusters in M31 being more bottom-light with increasing metallicity \citep{StrCalSet2011, Convan2012b} is consistent with the dependencies of the IMF on redshift and metallicity that we find from the numerical simulations.  

\item At the lowest metallicity ($Z=0.01$~Z$_\odot$) the stellar mass functions obtained from the calculations may be slightly bottom-heavy compared to a typical Galactic IMF.  This effect does not appear to be strong enough to explain the bottom-heavy IMFs that are apparently observed in the centres of many massive elliptical galaxies.  However, when combined with a characteristic stellar mass that decreases with increasing molecular cloud density (as has been proposed in some studies), this may help to explain the results obtained from observations of massive elliptical galaxies.  

\end{enumerate}

Further numerical studies of star formation that continue to map out the behaviour of the IMF with different initial conditions and in different environments, and also constrain the high-mass end of the IMF are required in the future.

\section*{Acknowledgements}

MRB thanks Tim Naylor for discussions on how best to fit the L$_3$ functions to the stellar mass distributions, and for providing codes to compute the best fit values and their uncertainties.  MRB also thanks Mike Grudi\'c for discussions about the minimum stellar mass and providing a link to the Python code associated with his paper.
MRB also thanks the referee, Ant Whitworth, for raising a number of points that resulted in improvements to the paper. 

This work was partially supported by the European Research Council under the European Commission's Seventh Framework Programme (FP7/2007-2013 Grant Agreement No. 339248), and by the Science and Technology Facilities Council (STFC) Small Award ST/Y001907/1.  The calculations discussed in this paper were performed on the University of Exeter Supercomputer, Isca, and on the DiRAC Complexity, DIAL-2 and DIAL-3 systems, operated by the University of Leicester IT Services, which forms part of the STFC DiRAC HPC Facility (www.dirac.ac.uk). The latter equipment is funded by BIS National E-Infrastructure capital grant ST/K000373/1 and STFC DiRAC Operations grant ST/K0003259/1. DiRAC is part of the National E-Infrastructure.  Some of the figures were produced using SPLASH \citep{Price2007}, an SPH visualization tool publicly available at http://users.monash.edu.au/$\sim$dprice/splash.

For the purpose of open access, the author has applied a Creative Commons Attribution (CC BY) licence to any Author Accepted Manuscript version arising.

\section*{Data availability}

Data from the SPH calculations performed for this paper is available from the 
University of Exeter's Open Research Exeter (ORE) repository \citep{Bate2024_data}. 
This data set includes the initial conditions and input files for the 13 SPH calculations that were performed specifically for this
paper, the final SPH dump files for each of the twenty calculations (at $t=1.30~t_{\rm ff}$), data files giving sink particle 
properties throughout each of the calculations, and data and Python scripts used to generate the figures.
For the seven calculations that were followed up to $t=1.20~t_{\rm ff}$ for earlier papers \citep{Bate2019,Bate2023}, 
the data from the early parts of the calculations are also available from the ORE repository  \citep{Bate2019_data,Bate2022b_data}.
Any other data underlying this article will be shared on reasonable request to the corresponding author.

\bibliography{/Users/mbate/Tex/mbate}

\begin{thebibliography}{}

\bibitem[\protect\citeauthoryear{{Adamo}, {Bradley}, {Vanzella}, {Claeyssens},
  {Welch}, {Diego}, {Mahler}, {Oguri}, {Sharon}, {Abdurro'uf}, {Hsiao}, {Xu},
  {Messa}, {Lassen}, {Zackrisson}, {Brammer}, {Coe}, {Kokorev}, {Ricotti},
  {Zitrin}, {Fujimoto}, {Inoue}, {Resseguier}, {Rigby}, {Jim{\'e}nez-Teja},
  {Windhorst}, {Hashimoto} \& {Tamura}}{{Adamo} et~al.}{2024}]{Adamo_etal2024}
{Adamo} A.,  {Bradley} L.~D.,  {Vanzella} E.,  {Claeyssens} A.,  {Welch} B.,
  {Diego} J.~M.,  {Mahler} G.,  {Oguri} M.,  {Sharon} K.,  {Abdurro'uf} {Hsiao}
  T. Y.-Y.,  {Xu} X.,  {Messa} M.,  {Lassen} A.~E.,  {Zackrisson} E.,
  {Brammer} G.,  {Coe} D.,  {Kokorev} V.,  {Ricotti} M.,  {Zitrin} A.,
  {Fujimoto} S.,  {Inoue} A.~K.,  {Resseguier} T.,  {Rigby} J.~R.,
  {Jim{\'e}nez-Teja} Y.,  {Windhorst} R.~A.,  {Hashimoto} T.,    {Tamura} Y.,
  2024, \nat, 632, 513

\bibitem[\protect\citeauthoryear{{Auger}, {Treu}, {Bolton}, {Gavazzi},
  {Koopmans}, {Marshall}, {Moustakas} \& {Burles}}{{Auger}
  et~al.}{2010}]{Auger_etal2010}
{Auger} M.~W.,  {Treu} T.,  {Bolton} A.~S.,  {Gavazzi} R.,  {Koopmans}
  L.~V.~E.,  {Marshall} P.~J.,  {Moustakas} L.~A.,    {Burles} S.,  2010, \apj,
  724, 511

\bibitem[\protect\citeauthoryear{{Badenes}, {Mazzola}, {Thompson}, {Covey},
  {Freeman}, {Walker}, {Moe}, {Troup}, {Nidever}, {Allende Prieto} \& {et
  al.}}{{Badenes} et~al.}{2018}]{Badenes_etal2018}
{Badenes} C.,  {Mazzola} C.,  {Thompson} T.~A.,  {Covey} K.,  {Freeman} P.~E.,
  {Walker} M.~G.,  {Moe} M.,  {Troup} N.,  {Nidever} D.,  {Allende Prieto} C.,
    {et al.} 2018, \apj, 854, 147

\bibitem[\protect\citeauthoryear{{Barber}, {Crain} \& {Schaye}}{{Barber}
  et~al.}{2018}]{BarCraSch2018}
{Barber} C.,  {Crain} R.~A.,    {Schaye} J.,  2018, \mnras, 479, 5448

\bibitem[\protect\citeauthoryear{{Barber}, {Schaye} \& {Crain}}{{Barber}
  et~al.}{2019a}]{BarSchCra2019a}
{Barber} C.,  {Schaye} J.,    {Crain} R.~A.,  2019a, \mnras, 482, 2515

\bibitem[\protect\citeauthoryear{{Barber}, {Schaye} \& {Crain}}{{Barber}
  et~al.}{2019b}]{BarSchCra2019b}
{Barber} C.,  {Schaye} J.,    {Crain} R.~A.,  2019b, \mnras, 483, 985

\bibitem[\protect\citeauthoryear{{Barnab{\`e}}, {Czoske}, {Koopmans}, {Treu} \&
  {Bolton}}{{Barnab{\`e}} et~al.}{2011}]{Barnabe_etal2011}
{Barnab{\`e}} M.,  {Czoske} O.,  {Koopmans} L. V.~E.,  {Treu} T.,    {Bolton}
  A.~S.,  2011, \mnras, 415, 2215

\bibitem[\protect\citeauthoryear{{Bate}}{{Bate}}{1998}]{Bate1998}
{Bate} M.~R.,  1998, \apjl, 508, L95

\bibitem[\protect\citeauthoryear{{Bate}}{{Bate}}{2005}]{Bate2005}
{Bate} M.~R.,  2005, \mnras, 363, 363

\bibitem[\protect\citeauthoryear{{Bate}}{{Bate}}{2009a}]{Bate2009a}
{Bate} M.~R.,  2009a, \mnras, 392, 590

\bibitem[\protect\citeauthoryear{{Bate}}{{Bate}}{2009b}]{Bate2009b}
{Bate} M.~R.,  2009b, \mnras, 392, 1363

\bibitem[\protect\citeauthoryear{{Bate}}{{Bate}}{2012}]{Bate2012}
{Bate} M.~R.,  2012, \mnras, 419, 3115

\bibitem[\protect\citeauthoryear{{Bate}}{{Bate}}{2014}]{Bate2014}
{Bate} M.~R.,  2014, \mnras, 442, 285

\bibitem[\protect\citeauthoryear{{Bate}}{{Bate}}{2018}]{Bate2018}
{Bate} M.~R.,  2018, \mnras, 475, 5618

\bibitem[\protect\citeauthoryear{{Bate}}{{Bate}}{2019a}]{Bate2019_data}
{Bate} M.~R.,  2019a, The statistical properties of stars and their dependence
  on metallicity (Dataset), doi:10.24378/exe.1123, Available at:
  https://doi.org/10.24378/exe.1123.
University of Exeter, UK

\bibitem[\protect\citeauthoryear{{Bate}}{{Bate}}{2019b}]{Bate2019}
{Bate} M.~R.,  2019b, \mnras, 484, 2341

\bibitem[\protect\citeauthoryear{{Bate}}{{Bate}}{2022}]{Bate2022b_data}
{Bate} M.~R.,  2022, The statistical properties of stars and their dependence
  on redshift (Dataset), doi:10.24378/exe.4324, Available at:
  https://doi.org/10.24378/exe.4324.
University of Exeter, UK

\bibitem[\protect\citeauthoryear{{Bate}}{{Bate}}{2023}]{Bate2023}
{Bate} M.~R.,  2023, \mnras, 519, 688

\bibitem[\protect\citeauthoryear{{Bate}}{{Bate}}{2024}]{Bate2024_data}
{Bate} M.~R.,  2024, Variation of the low-mass end of the stellar initial mass
  function with redshift and metallicity (Dataset), doi:10.24378/exe.5426,
  Available at: https://doi.org/10.24378/exe.5426.
University of Exeter, UK

\bibitem[\protect\citeauthoryear{{Bate} \& {Bonnell}}{{Bate} \&
  {Bonnell}}{2005}]{BatBon2005}
{Bate} M.~R.,  {Bonnell} I.~A.,  2005, MNRAS, 356, 1201

\bibitem[\protect\citeauthoryear{{Bate}, {Bonnell} \& {Bromm}}{{Bate}
  et~al.}{2003}]{BatBonBro2003}
{Bate} M.~R.,  {Bonnell} I.~A.,    {Bromm} V.,  2003, MNRAS, 339, 577

\bibitem[\protect\citeauthoryear{{Bate}, {Bonnell} \& {Price}}{{Bate}
  et~al.}{1995}]{BatBonPri1995}
{Bate} M.~R.,  {Bonnell} I.~A.,    {Price} N.~M.,  1995, MNRAS, 277, 362

\bibitem[\protect\citeauthoryear{{Bate} \& {Burkert}}{{Bate} \&
  {Burkert}}{1997}]{BatBur1997}
{Bate} M.~R.,  {Burkert} A.,  1997, \mnras, 288, 1060

\bibitem[\protect\citeauthoryear{{Bate} \& {Keto}}{{Bate} \&
  {Keto}}{2015}]{BatKet2015}
{Bate} M.~R.,  {Keto} E.~R.,  2015, \mnras, 449, 2643

\bibitem[\protect\citeauthoryear{{Baugh}, {Lacey}, {Frenk}, {Granato}, {Silva},
  {Bressan}, {Benson} \& {Cole}}{{Baugh} et~al.}{2005}]{Baugh_etal2005}
{Baugh} C.~M.,  {Lacey} C.~G.,  {Frenk} C.~S.,  {Granato} G.~L.,  {Silva} L.,
  {Bressan} A.,  {Benson} A.~J.,    {Cole} S.,  2005, \mnras, 356, 1191

\bibitem[\protect\citeauthoryear{{Benz}}{{Benz}}{1990}]{Benz1990}
{Benz} W.,  1990, in {Buchler} J.~R.,  ed., Numerical Modelling of Nonlinear
  Stellar Pulsations Problems and Prospects. Kluwer, Dordrecht, p.~269

\bibitem[\protect\citeauthoryear{{Benz}, {Cameron}, {Press} \& {Bowers}}{{Benz}
  et~al.}{1990}]{Benzetal1990}
{Benz} W.,  {Cameron} A.~G.~W.,  {Press} W.~H.,    {Bowers} R.~L.,  1990, \apj,
  348, 647

\bibitem[\protect\citeauthoryear{{Boss}, {Fisher}, {Klein} \& {McKee}}{{Boss}
  et~al.}{2000}]{Bossetal2000}
{Boss} A.~P.,  {Fisher} R.~T.,  {Klein} R.~I.,    {McKee} C.~F.,  2000, \apj,
  528, 325

\bibitem[\protect\citeauthoryear{{Boyd} \& {Whitworth}}{{Boyd} \&
  {Whitworth}}{2005}]{BoyWhi2005}
{Boyd} D.~F.~A.,  {Whitworth} A.~P.,  2005, \aap, 430, 1059

\bibitem[\protect\citeauthoryear{{Bradley}, {Adamo}, {Vanzella}, {Sharon},
  {Brammer}, {Coe}, {Diego}, {Kokorev}, {Mahler}, {Oguri}, {Abdurro'uf},
  {Bhatawdekar}, {Christensen}, {Fujimoto}, {Hashimoto}, {Y. -Y Hsiao},
  {Inoue}, {Jim{\'e}nez-Teja}, {Messa}, {Norman}, {Ricotti}, {Tamura},
  {Windhorst}, {Xu} \& {Zitrin}}{{Bradley} et~al.}{2024}]{Bradley_etal2024}
{Bradley} L.~D.,  {Adamo} A.,  {Vanzella} E.,  {Sharon} K.,  {Brammer} G.,
  {Coe} D.,  {Diego} J.~M.,  {Kokorev} V.,  {Mahler} G.,  {Oguri} M.,
  {Abdurro'uf} {Bhatawdekar} R.,  {Christensen} L.,  {Fujimoto} S.,
  {Hashimoto} T.,  {Y. -Y Hsiao} T.,  {Inoue} A.~K.,  {Jim{\'e}nez-Teja} Y.,
  {Messa} M.,  {Norman} C.,  {Ricotti} M.,  {Tamura} Y.,  {Windhorst} R.~A.,
  {Xu} X.,    {Zitrin} A.,  2024, arXiv e-prints, p. arXiv:2404.10770

\bibitem[\protect\citeauthoryear{{Burstein}, {Bender}, {Faber} \&
  {Nolthenius}}{{Burstein} et~al.}{1997}]{Burstein_etal1997}
{Burstein} D.,  {Bender} R.,  {Faber} S.,    {Nolthenius} R.,  1997, \aj, 114,
  1365

\bibitem[\protect\citeauthoryear{{Cappellari}, {McDermid}, {Alatalo}, {Blitz},
  {Bois}, {Bournaud}, {Bureau}, {Crocker}, {Davies} \& {et al.}}{{Cappellari}
  et~al.}{2012}]{Cappellari_etal2012}
{Cappellari} M.,  {McDermid} R.~M.,  {Alatalo} K.,  {Blitz} L.,  {Bois} M.,
  {Bournaud} F.,  {Bureau} M.,  {Crocker} A.~F.,  {Davies} R.~L.,    {et al.}
  2012, \nat, 484, 485

\bibitem[\protect\citeauthoryear{{Cappellari}, {McDermid}, {Alatalo}, {Blitz},
  {Bois}, {Bournaud}, {Bureau}, {Crocker}, {Davies} \& {et al.}}{{Cappellari}
  et~al.}{2013}]{Cappellari_etal2013}
{Cappellari} M.,  {McDermid} R.~M.,  {Alatalo} K.,  {Blitz} L.,  {Bois} M.,
  {Bournaud} F.,  {Bureau} M.,  {Crocker} A.~F.,  {Davies} R.~L.,    {et al.}
  2013, \mnras, 432, 1862

\bibitem[\protect\citeauthoryear{{Carniani}, {Hainline}, {D'Eugenio},
  {Eisenstein}, {Jakobsen}, {Witstok}, {Johnson}, {Chevallard}, {Maiolino},
  {Helton} \& {et al.}}{{Carniani} et~al.}{2024}]{Carniani_etal2024}
{Carniani} S.,  {Hainline} K.,  {D'Eugenio} F.,  {Eisenstein} D.~J.,
  {Jakobsen} P.,  {Witstok} J.,  {Johnson} B.~D.,  {Chevallard} J.,  {Maiolino}
  R.,  {Helton} J.~M.,    {et al.} 2024, arXiv e-prints, p. arXiv:2405.18485

\bibitem[\protect\citeauthoryear{{Carter}, {Visvanathan} \& {Pickles}}{{Carter}
  et~al.}{1986}]{CarVisPic1986}
{Carter} D.,  {Visvanathan} N.,    {Pickles} A.~J.,  1986, \apj, 311, 637

\bibitem[\protect\citeauthoryear{{Cenarro}, {Gorgas}, {Vazdekis}, {Cardiel} \&
  {Peletier}}{{Cenarro} et~al.}{2003}]{Cenarro_etal2003}
{Cenarro} A.~J.,  {Gorgas} J.,  {Vazdekis} A.,  {Cardiel} N.,    {Peletier}
  R.~F.,  2003, \mnras, 339, L12

\bibitem[\protect\citeauthoryear{{Chabrier}}{{Chabrier}}{2003}]{Chabrier2003}
{Chabrier} G.,  2003, \pasp, 115, 763

\bibitem[\protect\citeauthoryear{{Chabrier}}{{Chabrier}}{2005}]{Chabrier2005}
{Chabrier} G.,  2005, in {E.~Corbelli, F.~Palla, \& H.~Zinnecker} ed., The
  Initial Mass Function 50 Years Later Vol.~327 of Astrophysics and Space
  Science Library, {The Initial Mass Function: from Salpeter 1955 to 2005}.
Springer, Dordrecht, pp 41--50

\bibitem[\protect\citeauthoryear{{Chabrier}, {Hennebelle} \&
  {Charlot}}{{Chabrier} et~al.}{2014}]{ChaHenCha2014}
{Chabrier} G.,  {Hennebelle} P.,    {Charlot} S.,  2014, \apj, 796, 75

\bibitem[\protect\citeauthoryear{{Chon}, {Hosokawa}, {Omukai} \&
  {Schneider}}{{Chon} et~al.}{2024}]{Chon_etal2024}
{Chon} S.,  {Hosokawa} T.,  {Omukai} K.,    {Schneider} R.,  2024, \mnras, 530,
  2453

\bibitem[\protect\citeauthoryear{{Chon}, {Omukai} \& {Schneider}}{{Chon}
  et~al.}{2021}]{ChoOmuSch2021}
{Chon} S.,  {Omukai} K.,    {Schneider} R.,  2021, \mnras, 508, 4175

\bibitem[\protect\citeauthoryear{{Chon}, {Ono}, {Omukai} \& {Schneider}}{{Chon}
  et~al.}{2022}]{Chon_etal2022}
{Chon} S.,  {Ono} H.,  {Omukai} K.,    {Schneider} R.,  2022, \mnras, 514, 4639

\bibitem[\protect\citeauthoryear{{Claeyssens}, {Adamo}, {Richard}, {Mahler},
  {Messa} \& {Dessauges-Zavadsky}}{{Claeyssens}
  et~al.}{2023}]{Claeyssens_etal2023}
{Claeyssens} A.,  {Adamo} A.,  {Richard} J.,  {Mahler} G.,  {Messa} M.,
  {Dessauges-Zavadsky} M.,  2023, \mnras, 520, 2180

\bibitem[\protect\citeauthoryear{{Clark}, {Bonnell}, {Zinnecker} \&
  {Bate}}{{Clark} et~al.}{2005}]{Clark_etal2005}
{Clark} P.~C.,  {Bonnell} I.~A.,  {Zinnecker} H.,    {Bate} M.~R.,  2005,
  \mnras, 359, 809

\bibitem[\protect\citeauthoryear{{Clark} \& {Whitworth}}{{Clark} \&
  {Whitworth}}{2021}]{ClaWhi2021}
{Clark} P.~C.,  {Whitworth} A.~P.,  2021, \mnras, 500, 1697

\bibitem[\protect\citeauthoryear{{Cohen}}{{Cohen}}{1978}]{Cohen1978}
{Cohen} J.~G.,  1978, \apj, 221, 788

\bibitem[\protect\citeauthoryear{{Conroy} \& {van Dokkum}}{{Conroy} \& {van
  Dokkum}}{2012a}]{Convan2012a}
{Conroy} C.,  {van Dokkum} P.,  2012a, \apj, 747, 69

\bibitem[\protect\citeauthoryear{{Conroy} \& {van Dokkum}}{{Conroy} \& {van
  Dokkum}}{2012b}]{Convan2012b}
{Conroy} C.,  {van Dokkum} P.~G.,  2012b, \apj, 760, 71

\bibitem[\protect\citeauthoryear{{Cunningham}, {Krumholz}, {McKee} \&
  {Klein}}{{Cunningham} et~al.}{2018}]{Cunningham_etal2018}
{Cunningham} A.~J.,  {Krumholz} M.~R.,  {McKee} C.~F.,    {Klein} R.~I.,  2018,
  \mnras, 476, 771

\bibitem[\protect\citeauthoryear{{Davis} \& {McDermid}}{{Davis} \&
  {McDermid}}{2017}]{DavMcD2017}
{Davis} T.~A.,  {McDermid} R.~M.,  2017, \mnras, 464, 453

\bibitem[\protect\citeauthoryear{{De Vis}, {Jones}, {Viaene}, {Casasola},
  {Clark}, {Baes}, {Bianchi}, {Cassara}, {Davies}, {De Looze}, {Galametz},
  {Galliano}, {Lianou}, {Madden}, {Manilla-Robles}, {Mosenkov}, {Nersesian},
  {Roychowdhury}, {Xilouris} \& {Ysard}}{{De Vis}
  et~al.}{2019}]{DeVis_etal2019}
{De Vis} P.,  {Jones} A.,  {Viaene} S.,  {Casasola} V.,  {Clark} C.~J.~R.,
  {Baes} M.,  {Bianchi} S.,  {Cassara} L.~P.,  {Davies} J.~I.,  {De Looze} I.,
  {Galametz} M.,  {Galliano} F.,  {Lianou} S.,  {Madden} S.,  {Manilla-Robles}
  A.,  {Mosenkov} A.~V.,  {Nersesian} A.,  {Roychowdhury} S.,  {Xilouris}
  E.~M.,    {Ysard} N.,  2019, \aap, 623, A5

\bibitem[\protect\citeauthoryear{{Delisle} \& {Hardy}}{{Delisle} \&
  {Hardy}}{1992}]{DelHar1992}
{Delisle} S.,  {Hardy} E.,  1992, \aj, 103, 711

\bibitem[\protect\citeauthoryear{{Draine}}{{Draine}}{1978}]{Draine1978}
{Draine} B.~T.,  1978, \apjs, 36, 595

\bibitem[\protect\citeauthoryear{{Dutton}, {Macci{\`o}}, {Mendel} \&
  {Simard}}{{Dutton} et~al.}{2013}]{Dutton_etal2013}
{Dutton} A.~A.,  {Macci{\`o}} A.~V.,  {Mendel} J.~T.,    {Simard} L.,  2013,
  \mnras, 432, 2496

\bibitem[\protect\citeauthoryear{{Dutton}, {Mendel} \& {Simard}}{{Dutton}
  et~al.}{2012}]{DutMenSim2012}
{Dutton} A.~A.,  {Mendel} J.~T.,    {Simard} L.,  2012, \mnras, 422, L33

\bibitem[\protect\citeauthoryear{{El-Badry} \& {Rix}}{{El-Badry} \&
  {Rix}}{2019}]{ElBRix2019}
{El-Badry} K.,  {Rix} H.-W.,  2019, \mnras, 482, L139

\bibitem[\protect\citeauthoryear{{Eliason}}{{Eliason}}{1993}]{Eliason1993}
{Eliason} S.~R.,  1993, {Maximum likelihood estimation}.
SAGE Publications, Inc

\bibitem[\protect\citeauthoryear{{Elsender} \& {Bate}}{{Elsender} \&
  {Bate}}{2021}]{ElsBat2021}
{Elsender} D.,  {Bate} M.~R.,  2021, \mnras, 508, 5279

\bibitem[\protect\citeauthoryear{{Elsender}, {Bate}, {Lakeland}, {Jensen} \&
  {Lubow}}{{Elsender} et~al.}{2023}]{Elsender_etal_2023}
{Elsender} D.,  {Bate} M.~R.,  {Lakeland} B.~S.,  {Jensen} E. L.~N.,    {Lubow}
  S.~H.,  2023, \mnras, 523, 4353

\bibitem[\protect\citeauthoryear{{Faber}}{{Faber}}{1973}]{Faber1973}
{Faber} S.~M.,  1973, \apj, 179, 731

\bibitem[\protect\citeauthoryear{{Faber} \& {French}}{{Faber} \&
  {French}}{1980}]{FabFre1980}
{Faber} S.~M.,  {French} H.~B.,  1980, \apj, 235, 405

\bibitem[\protect\citeauthoryear{{Falc{\'o}n-Barroso}, {Peletier}, {Vazdekis}
  \& {Balcells}}{{Falc{\'o}n-Barroso} et~al.}{2003}]{Falcon_etal2003}
{Falc{\'o}n-Barroso} J.,  {Peletier} R.~F.,  {Vazdekis} A.,    {Balcells} M.,
  2003, \apjl, 588, L17

\bibitem[\protect\citeauthoryear{{Fehlberg}}{{Fehlberg}}{1969}]{Fehlberg1969}
{Fehlberg} E.,  1969, NASA Technical Report R-315, Low-Order Classical
  Runge-Kutta Formulas with Step Size Control and Their Application to Some
  Heat Transfer Problems. Washington, USA

\bibitem[\protect\citeauthoryear{{Ferreras}, {La Barbera}, {de La Rosa},
  {Vazdekis}, {de Carvalho}, {Falcon-Barroso} \& {Ricciardelli}}{{Ferreras}
  et~al.}{2013}]{Ferreras_etal2013}
{Ferreras} I.,  {La Barbera} F.,  {de La Rosa} I.~G.,  {Vazdekis} A.,  {de
  Carvalho} R.~R.,  {Falcon-Barroso} J.,    {Ricciardelli} E.,  2013, \mnras,
  429, L15

\bibitem[\protect\citeauthoryear{{Ferreras}, {Saha} \& {Burles}}{{Ferreras}
  et~al.}{2008}]{FerSahBur2008}
{Ferreras} I.,  {Saha} P.,    {Burles} S.,  2008, \mnras, 383, 857

\bibitem[\protect\citeauthoryear{{Ferreras}, {Saha}, {Leier}, {Courbin} \&
  {Falco}}{{Ferreras} et~al.}{2010}]{Ferreras_etal2010}
{Ferreras} I.,  {Saha} P.,  {Leier} D.,  {Courbin} F.,    {Falco} E.~E.,  2010,
  \mnras, 409, L30

\bibitem[\protect\citeauthoryear{{Ferreras}, {Saha} \& {Williams}}{{Ferreras}
  et~al.}{2005}]{FerSahWil2005}
{Ferreras} I.,  {Saha} P.,    {Williams} L. L.~R.,  2005, \apjl, 623, L5

\bibitem[\protect\citeauthoryear{{Ferreras}, {Weidner}, {Vazdekis} \& {La
  Barbera}}{{Ferreras} et~al.}{2015}]{Ferreras_etal2015}
{Ferreras} I.,  {Weidner} C.,  {Vazdekis} A.,    {La Barbera} F.,  2015,
  \mnras, 448, L82

\bibitem[\protect\citeauthoryear{{Gallazzi}, {Charlot}, {Brinchmann} \&
  {White}}{{Gallazzi} et~al.}{2006}]{Gallazzi_etal2006}
{Gallazzi} A.,  {Charlot} S.,  {Brinchmann} J.,    {White} S. D.~M.,  2006,
  \mnras, 370, 1106

\bibitem[\protect\citeauthoryear{{Galliano}, {Galametz} \& {Jones}}{{Galliano}
  et~al.}{2018}]{GalGalJon2018}
{Galliano} F.,  {Galametz} M.,    {Jones} A.~P.,  2018, \araa, 56, 673

\bibitem[\protect\citeauthoryear{{Galliano}, {Nersesian}, {Bianchi}, {De
  Looze}, {Roychowdhury}, {Baes}, {Casasola}, {Cassar{\'a}}, {Dobbels},
  {Fritz}, {Galametz}, {Jones}, {Madden}, {Mosenkov}, {Xilouris} \&
  {Ysard}}{{Galliano} et~al.}{2021}]{Galliano_etal2021}
{Galliano} F.,  {Nersesian} A.,  {Bianchi} S.,  {De Looze} I.,  {Roychowdhury}
  S.,  {Baes} M.,  {Casasola} V.,  {Cassar{\'a}} L.~P.,  {Dobbels} W.,  {Fritz}
  J.,  {Galametz} M.,  {Jones} A.~P.,  {Madden} S.~C.,  {Mosenkov} A.,
  {Xilouris} E.~M.,    {Ysard} N.,  2021, \aap, 649, A18

\bibitem[\protect\citeauthoryear{{Glover} \& {Clark}}{{Glover} \&
  {Clark}}{2012}]{GloCla2012c}
{Glover} S.~C.~O.,  {Clark} P.~C.,  2012, \mnras, 426, 377

\bibitem[\protect\citeauthoryear{{Glover}, {Federrath}, {Mac Low} \&
  {Klessen}}{{Glover} et~al.}{2010}]{Gloveretal2010}
{Glover} S.~C.~O.,  {Federrath} C.,  {Mac Low} M.-M.,    {Klessen} R.~S.,
  2010, \mnras, 404, 2

\bibitem[\protect\citeauthoryear{{Graves}, {Faber} \& {Schiavon}}{{Graves}
  et~al.}{2009a}]{GraFabSch2009a}
{Graves} G.~J.,  {Faber} S.~M.,    {Schiavon} R.~P.,  2009a, \apj, 693, 486

\bibitem[\protect\citeauthoryear{{Graves}, {Faber} \& {Schiavon}}{{Graves}
  et~al.}{2009b}]{GraFabSch2009b}
{Graves} G.~J.,  {Faber} S.~M.,    {Schiavon} R.~P.,  2009b, \apj, 698, 1590

\bibitem[\protect\citeauthoryear{{Greene}, {Janish}, {Ma}, {McConnell},
  {Blakeslee}, {Thomas} \& {Murphy}}{{Greene} et~al.}{2015}]{Greene_etal2015}
{Greene} J.~E.,  {Janish} R.,  {Ma} C.-P.,  {McConnell} N.~J.,  {Blakeslee}
  J.~P.,  {Thomas} J.,    {Murphy} J.~D.,  2015, \apj, 807, 11

\bibitem[\protect\citeauthoryear{{Grudi{\'c}}, {Guszejnov}, {Offner}, {Rosen},
  {Raju}, {Faucher-Gigu{\`e}re} \& {Hopkins}}{{Grudi{\'c}}
  et~al.}{2022}]{Grudic_etal2022}
{Grudi{\'c}} M.~Y.,  {Guszejnov} D.,  {Offner} S. S.~R.,  {Rosen} A.~L.,
  {Raju} A.~N.,  {Faucher-Gigu{\`e}re} C.-A.,    {Hopkins} P.~F.,  2022,
  \mnras, 512, 216

\bibitem[\protect\citeauthoryear{{Grudi{\'c}} \& {Hopkins}}{{Grudi{\'c}} \&
  {Hopkins}}{2023}]{GruHop2023}
{Grudi{\'c}} M.~Y.,  {Hopkins} P.~F.,  2023, arXiv e-prints, p.
  arXiv:2308.16268

\bibitem[\protect\citeauthoryear{{Gunawardhana}, {Hopkins}, {Sharp}, {Brough},
  {Taylor}, {Bland-Hawthorn}, {Maraston}, {Tuffs}, {Popescu} \& {et
  al.}}{{Gunawardhana} et~al.}{2011}]{Gunawardhana_etal2011}
{Gunawardhana} M.~L.~P.,  {Hopkins} A.~M.,  {Sharp} R.~G.,  {Brough} S.,
  {Taylor} E.,  {Bland-Hawthorn} J.,  {Maraston} C.,  {Tuffs} R.~J.,  {Popescu}
  C.~C.,    {et al.} 2011, \mnras, 415, 1647

\bibitem[\protect\citeauthoryear{{Guszejnov}, {Grudi{\'c}}, {Hopkins}, {Offner}
  \& {Faucher-Gigu{\`e}re}}{{Guszejnov} et~al.}{2021}]{Guszejnov_etal2021}
{Guszejnov} D.,  {Grudi{\'c}} M.~Y.,  {Hopkins} P.~F.,  {Offner} S. S.~R.,
  {Faucher-Gigu{\`e}re} C.-A.,  2021, \mnras, 502, 3646

\bibitem[\protect\citeauthoryear{{Guszejnov}, {Grudi{\'c}}, {Offner},
  {Faucher-Gigu{\`e}re}, {Hopkins} \& {Rosen}}{{Guszejnov}
  et~al.}{2022}]{Guszejnov_etal2022}
{Guszejnov} D.,  {Grudi{\'c}} M.~Y.,  {Offner} S. S.~R.,  {Faucher-Gigu{\`e}re}
  C.-A.,  {Hopkins} P.~F.,    {Rosen} A.~L.,  2022, \mnras, 515, 4929

\bibitem[\protect\citeauthoryear{{Guszejnov}, {Hopkins} \& {Graus}}{{Guszejnov}
  et~al.}{2019}]{GusHopGra2019}
{Guszejnov} D.,  {Hopkins} P.~F.,    {Graus} A.~S.,  2019, \mnras, 485, 4852

\bibitem[\protect\citeauthoryear{{Guzman}, {Lucey} \& {Bower}}{{Guzman}
  et~al.}{1993}]{GuzLucBow1993}
{Guzman} R.,  {Lucey} J.~R.,    {Bower} R.~G.,  1993, \mnras, 265, 731

\bibitem[\protect\citeauthoryear{{Hardy} \& {Couture}}{{Hardy} \&
  {Couture}}{1988}]{HarCou1988}
{Hardy} E.,  {Couture} J.,  1988, \apjl, 325, L29

\bibitem[\protect\citeauthoryear{{Hennebelle}, {Lee} \&
  {Chabrier}}{{Hennebelle} et~al.}{2019}]{HenLeeCha2019}
{Hennebelle} P.,  {Lee} Y.-N.,    {Chabrier} G.,  2019, \apj, 883, 140

\bibitem[\protect\citeauthoryear{{Hopkins}}{{Hopkins}}{2013}]{Hopkins2013}
{Hopkins} P.~F.,  2013, \mnras, 433, 170

\bibitem[\protect\citeauthoryear{{Hoyle}}{{Hoyle}}{1953}]{Hoyle1953}
{Hoyle} F.,  1953, \apj, 118, 513

\bibitem[\protect\citeauthoryear{{Hubber}, {Goodwin} \& {Whitworth}}{{Hubber}
  et~al.}{2006}]{HubGooWhi2006}
{Hubber} D.~A.,  {Goodwin} S.~P.,    {Whitworth} A.~P.,  2006, \aap, 450, 881

\bibitem[\protect\citeauthoryear{{Jones} \& {Bate}}{{Jones} \&
  {Bate}}{2018a}]{JonBat2018b}
{Jones} M.~O.,  {Bate} M.~R.,  2018a, \mnras, 480, 2562

\bibitem[\protect\citeauthoryear{{Jones} \& {Bate}}{{Jones} \&
  {Bate}}{2018b}]{JonBat2018a}
{Jones} M.~O.,  {Bate} M.~R.,  2018b, \mnras, 478, 2650

\bibitem[\protect\citeauthoryear{{Jorgensen}, {Franx} \&
  {Kjaergaard}}{{Jorgensen} et~al.}{1996}]{JorFRaKja1996}
{Jorgensen} I.,  {Franx} M.,    {Kjaergaard} P.,  1996, \mnras, 280, 167

\bibitem[\protect\citeauthoryear{{Keto} \& {Caselli}}{{Keto} \&
  {Caselli}}{2008}]{KetCas2008}
{Keto} E.,  {Caselli} P.,  2008, \apj, 683, 238

\bibitem[\protect\citeauthoryear{{Konstantopoulou}, {De Cia}, {Ledoux},
  {Krogager}, {Mattsson}, {Watson}, {Heintz}, {P{\'e}roux}, {Noterdaeme},
  {Andersen}, {Fynbo}, {Jermann} \& {Ramburuth-Hurt}}{{Konstantopoulou}
  et~al.}{2024}]{Konstantopoulou_etal2024}
{Konstantopoulou} C.,  {De Cia} A.,  {Ledoux} C.,  {Krogager} J.-K.,
  {Mattsson} L.,  {Watson} D.,  {Heintz} K.~E.,  {P{\'e}roux} C.,  {Noterdaeme}
  P.,  {Andersen} A.~C.,  {Fynbo} J. P.~U.,  {Jermann} I.,    {Ramburuth-Hurt}
  T.,  2024, \aap, 681, A64

\bibitem[\protect\citeauthoryear{{Kroupa}}{{Kroupa}}{2001}]{Kroupa2001}
{Kroupa} P.,  2001, \mnras, 322, 231

\bibitem[\protect\citeauthoryear{{Krumholz}, {Klein} \& {McKee}}{{Krumholz}
  et~al.}{2012}]{KruKleMcK2012}
{Krumholz} M.~R.,  {Klein} R.~I.,    {McKee} C.~F.,  2012, \apj, 754, 71

\bibitem[\protect\citeauthoryear{{Krumholz}, {Myers}, {Klein} \&
  {McKee}}{{Krumholz} et~al.}{2016}]{Krumholz_etal2016}
{Krumholz} M.~R.,  {Myers} A.~T.,  {Klein} R.~I.,    {McKee} C.~F.,  2016,
  \mnras, 460, 3272

\bibitem[\protect\citeauthoryear{{Kuntschner}, {Emsellem}, {Bacon},
  {Cappellari}, {Davies}, {de Zeeuw}, {Falc{\'o}n-Barroso}, {Krajnovi{\'c}},
  {McDermid}, {Peletier}, {Sarzi}, {Shapiro}, {van den Bosch} \& {van de
  Ven}}{{Kuntschner} et~al.}{2010}]{Kuntschner_etal2010}
{Kuntschner} H.,  {Emsellem} E.,  {Bacon} R.,  {Cappellari} M.,  {Davies}
  R.~L.,  {de Zeeuw} P.~T.,  {Falc{\'o}n-Barroso} J.,  {Krajnovi{\'c}} D.,
  {McDermid} R.~M.,  {Peletier} R.~F.,  {Sarzi} M.,  {Shapiro} K.~L.,  {van den
  Bosch} R. C.~E.,    {van de Ven} G.,  2010, \mnras, 408, 97

\bibitem[\protect\citeauthoryear{{La Barbera}, {Ferreras}, {Vazdekis}, {de la
  Rosa}, {de Carvalho}, {Trevisan}, {Falc{\'o}n-Barroso} \& {Ricciardelli}}{{La
  Barbera} et~al.}{2013}]{LaBarbera_etal2013}
{La Barbera} F.,  {Ferreras} I.,  {Vazdekis} A.,  {de la Rosa} I.~G.,  {de
  Carvalho} R.~R.,  {Trevisan} M.,  {Falc{\'o}n-Barroso} J.,    {Ricciardelli}
  E.,  2013, \mnras, 433, 3017

\bibitem[\protect\citeauthoryear{{La Barbera}, {Vazdekis}, {Ferreras},
  {Pasquali}, {Allende Prieto}, {R{\"o}ck}, {Aguado} \& {Peletier}}{{La
  Barbera} et~al.}{2017}]{LaBarbera_etal2017}
{La Barbera} F.,  {Vazdekis} A.,  {Ferreras} I.,  {Pasquali} A.,  {Allende
  Prieto} C.,  {R{\"o}ck} B.,  {Aguado} D.~S.,    {Peletier} R.~F.,  2017,
  \mnras, 464, 3597

\bibitem[\protect\citeauthoryear{{Labb{\'e}}, {van Dokkum}, {Nelson},
  {Bezanson}, {Suess}, {Leja}, {Brammer}, {Whitaker}, {Mathews}, {Stefanon} \&
  {Wang}}{{Labb{\'e}} et~al.}{2023}]{Labbe_etal2023}
{Labb{\'e}} I.,  {van Dokkum} P.,  {Nelson} E.,  {Bezanson} R.,  {Suess} K.~A.,
   {Leja} J.,  {Brammer} G.,  {Whitaker} K.,  {Mathews} E.,  {Stefanon} M.,
  {Wang} B.,  2023, \nat, 616, 266

\bibitem[\protect\citeauthoryear{{Larson}}{{Larson}}{1969}]{Larson1969}
{Larson} R.~B.,  1969, \mnras, 145, 271

\bibitem[\protect\citeauthoryear{{Larson}}{{Larson}}{1998}]{Larson1998}
{Larson} R.~B.,  1998, \mnras, 301, 569

\bibitem[\protect\citeauthoryear{{Lebreuilly}, {Hennebelle}, {Colman}, {Maury},
  {Tung}, {Testi}, {Klessen}, {Molinari}, {Commer{\c{c}}on}, {Gonz{\'a}lez},
  {Pacetti}, {Somigliana} \& {Rosotti}}{{Lebreuilly}
  et~al.}{2024}]{Lebreuilly_etal2024}
{Lebreuilly} U.,  {Hennebelle} P.,  {Colman} T.,  {Maury} A.,  {Tung} N.~D.,
  {Testi} L.,  {Klessen} R.,  {Molinari} S.,  {Commer{\c{c}}on} B.,
  {Gonz{\'a}lez} M.,  {Pacetti} E.,  {Somigliana} A.,    {Rosotti} G.,  2024,
  \aap, 682, A30

\bibitem[\protect\citeauthoryear{{Lee} \& {Hennebelle}}{{Lee} \&
  {Hennebelle}}{2018}]{LeeHen2018b}
{Lee} Y.-N.,  {Hennebelle} P.,  2018, \aap, 611, A89

\bibitem[\protect\citeauthoryear{{Leier}, {Ferreras}, {Saha}, {Charlot},
  {Bruzual} \& {La Barbera}}{{Leier} et~al.}{2016}]{Leier_etal2016}
{Leier} D.,  {Ferreras} I.,  {Saha} P.,  {Charlot} S.,  {Bruzual} G.,    {La
  Barbera} F.,  2016, \mnras, 459, 3677

\bibitem[\protect\citeauthoryear{{Li}, {Liu}, {Hasegawa} \& {Hirano}}{{Li}
  et~al.}{2017}]{Li_etal2017}
{Li} J. I.-H.,  {Liu} H.~B.,  {Hasegawa} Y.,    {Hirano} N.,  2017, \apj, 840,
  72

\bibitem[\protect\citeauthoryear{{Lomax}, {Whitworth} \& {Hubber}}{{Lomax}
  et~al.}{2015}]{LomWhiHub2015}
{Lomax} O.,  {Whitworth} A.~P.,    {Hubber} D.~A.,  2015, \mnras, 449, 662

\bibitem[\protect\citeauthoryear{{Low} \& {Lynden-Bell}}{{Low} \&
  {Lynden-Bell}}{1976}]{LowLyn1976}
{Low} C.,  {Lynden-Bell} D.,  1976, \mnras, 176, 367

\bibitem[\protect\citeauthoryear{{Mart{\'\i}n-Navarro}, {Vazdekis}, {La
  Barbera}, {Falc{\'o}n-Barroso}, {Lyubenova}, {van de Ven}, {Ferreras},
  {S{\'a}nchez}, {Trager} \& {et al.}}{{Mart{\'\i}n-Navarro}
  et~al.}{2015}]{MartinNavarro_etal2015}
{Mart{\'\i}n-Navarro} I.,  {Vazdekis} A.,  {La Barbera} F.,
  {Falc{\'o}n-Barroso} J.,  {Lyubenova} M.,  {van de Ven} G.,  {Ferreras} I.,
  {S{\'a}nchez} S.~F.,  {Trager} S.~C.,    {et al.} 2015, \apjl, 806, L31

\bibitem[\protect\citeauthoryear{{Maschberger}}{{Maschberger}}{2013}]{Maschberger2013}
{Maschberger} T.,  2013, \mnras, 429, 1725

\bibitem[\protect\citeauthoryear{{Mathew} \& {Federrath}}{{Mathew} \&
  {Federrath}}{2021}]{MatFed2021}
{Mathew} S.~S.,  {Federrath} C.,  2021, \mnras, 507, 2448

\bibitem[\protect\citeauthoryear{{Mathew}, {Federrath} \& {Seta}}{{Mathew}
  et~al.}{2022}]{MatFedSet2022}
{Mathew} S.~S.,  {Federrath} C.,    {Seta} A.,  2022, arXiv e-prints, p.
  arXiv:2208.08802

\bibitem[\protect\citeauthoryear{{Mathew}, {Federrath} \& {Seta}}{{Mathew}
  et~al.}{2023}]{MatFedSet2023}
{Mathew} S.~S.,  {Federrath} C.,    {Seta} A.,  2023, \mnras, 518, 5190

\bibitem[\protect\citeauthoryear{{McDermid}, {Alatalo}, {Blitz}, {Bournaud},
  {Bureau}, {Cappellari}, {Crocker}, {Davies}, {Davis} \& {et al.}}{{McDermid}
  et~al.}{2015}]{McDermid_etal2015}
{McDermid} R.~M.,  {Alatalo} K.,  {Blitz} L.,  {Bournaud} F.,  {Bureau} M.,
  {Cappellari} M.,  {Crocker} A.~F.,  {Davies} R.~L.,  {Davis} T.~A.,    {et
  al.} 2015, \mnras, 448, 3484

\bibitem[\protect\citeauthoryear{{McDermid}, {Cappellari}, {Alatalo}, {Bayet},
  {Blitz}, {Bois}, {Bournaud}, {Bureau}, {Crocker} \& {et al.}}{{McDermid}
  et~al.}{2014}]{McDermid_etal2014}
{McDermid} R.~M.,  {Cappellari} M.,  {Alatalo} K.,  {Bayet} E.,  {Blitz} L.,
  {Bois} M.,  {Bournaud} F.,  {Bureau} M.,  {Crocker} A.~F.,    {et al.} 2014,
  \apjl, 792, L37

\bibitem[\protect\citeauthoryear{{Moe}, {Kratter} \& {Badenes}}{{Moe}
  et~al.}{2019}]{MoeKraBad2019}
{Moe} M.,  {Kratter} K.~M.,    {Badenes} C.,  2019, \apj, 875, 61

\bibitem[\protect\citeauthoryear{{Morris} \& {Monaghan}}{{Morris} \&
  {Monaghan}}{1997}]{MorMon1997}
{Morris} J.~P.,  {Monaghan} J.~J.,  1997, J.\ Comp.\ Phys., 136, 41

\bibitem[\protect\citeauthoryear{{Mowla}, {Iyer}, {Asada}, {Desprez}, {Tan},
  {Martis}, {Sarrouh}, {Strait}, {Abraham}, {Brada{\v{c}}}, {Brammer},
  {Muzzin}, {Pacifici}, {Ravindranath}, {Sawicki}, {Willott},
  {Estrada-Carpenter}, {Jahan}, {Noirot}, {Matharu}, {Rihtar{\v{s}}i{\v{c}}} \&
  {Zabl}}{{Mowla} et~al.}{2024}]{Mowla_etal2024}
{Mowla} L.,  {Iyer} K.,  {Asada} Y.,  {Desprez} G.,  {Tan} V. Y.~Y.,  {Martis}
  N.,  {Sarrouh} G.,  {Strait} V.,  {Abraham} R.,  {Brada{\v{c}}} M.,
  {Brammer} G.,  {Muzzin} A.,  {Pacifici} C.,  {Ravindranath} S.,  {Sawicki}
  M.,  {Willott} C.,  {Estrada-Carpenter} V.,  {Jahan} N.,  {Noirot} G.,
  {Matharu} J.,  {Rihtar{\v{s}}i{\v{c}}} G.,    {Zabl} J.,  2024, arXiv
  e-prints, p. arXiv:2402.08696

\bibitem[\protect\citeauthoryear{{Mowla}, {Iyer}, {Desprez},
  {Estrada-Carpenter}, {Martis}, {Noirot}, {Sarrouh}, {Strait}, {Asada},
  {Abraham}, {Brammer}, {Sawicki}, {Willott}, {Bradac}, {Doyon}, {Muzzin},
  {Pacifici}, {Ravindranath} \& {Zabl}}{{Mowla} et~al.}{2022}]{Mowla_etal2022}
{Mowla} L.,  {Iyer} K.~G.,  {Desprez} G.,  {Estrada-Carpenter} V.,  {Martis}
  N.~S.,  {Noirot} G.,  {Sarrouh} G.~T.,  {Strait} V.,  {Asada} Y.,  {Abraham}
  R.~G.,  {Brammer} G.,  {Sawicki} M.,  {Willott} C.~J.,  {Bradac} M.,  {Doyon}
  R.,  {Muzzin} A.,  {Pacifici} C.,  {Ravindranath} S.,    {Zabl} J.,  2022,
  \apjl, 937, L35

\bibitem[\protect\citeauthoryear{{Myers}, {Klein}, {Krumholz} \&
  {McKee}}{{Myers} et~al.}{2014}]{Myers_etal2014}
{Myers} A.~T.,  {Klein} R.~I.,  {Krumholz} M.~R.,    {McKee} C.~F.,  2014,
  \mnras, 439, 3420

\bibitem[\protect\citeauthoryear{{Myers}, {Krumholz}, {Klein} \&
  {McKee}}{{Myers} et~al.}{2011}]{Myers_etal2011}
{Myers} A.~T.,  {Krumholz} M.~R.,  {Klein} R.~I.,    {McKee} C.~F.,  2011,
  \apj, 735, 49

\bibitem[\protect\citeauthoryear{{Myers}, {McKee}, {Cunningham}, {Klein} \&
  {Krumholz}}{{Myers} et~al.}{2013}]{Myersetal2013}
{Myers} A.~T.,  {McKee} C.~F.,  {Cunningham} A.~J.,  {Klein} R.~I.,
  {Krumholz} M.~R.,  2013, \apj, 766, 97

\bibitem[\protect\citeauthoryear{{Naidu}, {Oesch}, {van Dokkum}, {Nelson},
  {Suess}, {Brammer}, {Whitaker}, {Illingworth}, {Bouwens}, {Tacchella},
  {Matthee}, {Allen}, {Bezanson}, {Conroy}, {Labbe}, {Leja}, {Leonova},
  {Magee}, {Price}, {Setton}, {Strait}, {Stefanon}, {Toft}, {Weaver} \&
  {Weibel}}{{Naidu} et~al.}{2022}]{Naidu_etal2022}
{Naidu} R.~P.,  {Oesch} P.~A.,  {van Dokkum} P.,  {Nelson} E.~J.,  {Suess}
  K.~A.,  {Brammer} G.,  {Whitaker} K.~E.,  {Illingworth} G.,  {Bouwens} R.,
  {Tacchella} S.,  {Matthee} J.,  {Allen} N.,  {Bezanson} R.,  {Conroy} C.,
  {Labbe} I.,  {Leja} J.,  {Leonova} E.,  {Magee} D.,  {Price} S.~H.,  {Setton}
  D.~J.,  {Strait} V.,  {Stefanon} M.,  {Toft} S.,  {Weaver} J.~R.,    {Weibel}
  A.,  2022, \apjl, 940, L14

\bibitem[\protect\citeauthoryear{{Nam}, {Federrath} \& {Krumholz}}{{Nam}
  et~al.}{2021}]{NamFedKru2021}
{Nam} D.~G.,  {Federrath} C.,    {Krumholz} M.~R.,  2021, \mnras, 503, 1138

\bibitem[\protect\citeauthoryear{{Narayanan} \& {Dav{\'e}}}{{Narayanan} \&
  {Dav{\'e}}}{2012}]{NarDav2012}
{Narayanan} D.,  {Dav{\'e}} R.,  2012, \mnras, 423, 3601

\bibitem[\protect\citeauthoryear{{Narayanan} \& {Dav{\'e}}}{{Narayanan} \&
  {Dav{\'e}}}{2013}]{NarDav2013}
{Narayanan} D.,  {Dav{\'e}} R.,  2013, \mnras, 436, 2892

\bibitem[\protect\citeauthoryear{{Naylor}}{{Naylor}}{2009}]{Naylor2009}
{Naylor} T.,  2009, \mnras, 399, 432

\bibitem[\protect\citeauthoryear{{Newman}, {Smith}, {Conroy}, {Villaume} \&
  {van Dokkum}}{{Newman} et~al.}{2017}]{Newman_etal2017}
{Newman} A.~B.,  {Smith} R.~J.,  {Conroy} C.,  {Villaume} A.,    {van Dokkum}
  P.,  2017, \apj, 845, 157

\bibitem[\protect\citeauthoryear{{Ostriker}, {Stone} \& {Gammie}}{{Ostriker}
  et~al.}{2001}]{OstStoGam2001}
{Ostriker} E.~C.,  {Stone} J.~M.,    {Gammie} C.~F.,  2001, \apj, 546, 980

\bibitem[\protect\citeauthoryear{{Posacki}, {Cappellari}, {Treu}, {Pellegrini}
  \& {Ciotti}}{{Posacki} et~al.}{2015}]{Posacki_etal2015}
{Posacki} S.,  {Cappellari} M.,  {Treu} T.,  {Pellegrini} S.,    {Ciotti} L.,
  2015, \mnras, 446, 493

\bibitem[\protect\citeauthoryear{{Price}}{{Price}}{2007}]{Price2007}
{Price} D.~J.,  2007, Publ. Astron. Soc. Australia, 24, 159

\bibitem[\protect\citeauthoryear{{Price}, {Cuello}, {Pinte}, {Mentiplay},
  {Casassus}, {Christiaens}, {Kennedy}, {Cuadra}, {Sebastian Perez}, {Marino},
  {Armitage}, {Zurlo}, {Juhasz}, {Ragusa}, {Laibe} \& {Lodato}}{{Price}
  et~al.}{2018}]{Price_etal2018}
{Price} D.~J.,  {Cuello} N.,  {Pinte} C.,  {Mentiplay} D.,  {Casassus} S.,
  {Christiaens} V.,  {Kennedy} G.~M.,  {Cuadra} J.,  {Sebastian Perez} M.,
  {Marino} S.,  {Armitage} P.~J.,  {Zurlo} A.,  {Juhasz} A.,  {Ragusa} E.,
  {Laibe} G.,    {Lodato} G.,  2018, \mnras, 477, 1270

\bibitem[\protect\citeauthoryear{{Price} \& {Monaghan}}{{Price} \&
  {Monaghan}}{2005}]{PriMon2005}
{Price} D.~J.,  {Monaghan} J.~J.,  2005, \mnras, 364, 384

\bibitem[\protect\citeauthoryear{{Price} \& {Monaghan}}{{Price} \&
  {Monaghan}}{2007}]{PriMon2007}
{Price} D.~J.,  {Monaghan} J.~J.,  2007, \mnras, 374, 1347

\bibitem[\protect\citeauthoryear{{Rees}}{{Rees}}{1976}]{Rees1976}
{Rees} M.~J.,  1976, \mnras, 176, 483

\bibitem[\protect\citeauthoryear{{Rohde}, {Walch}, {Clarke}, {Seifried},
  {Whitworth} \& {Klepitko}}{{Rohde} et~al.}{2021}]{Rohde_etal2021}
{Rohde} P.~F.,  {Walch} S.,  {Clarke} S.~D.,  {Seifried} D.,  {Whitworth}
  A.~P.,    {Klepitko} A.,  2021, \mnras, 500, 3594

\bibitem[\protect\citeauthoryear{{Salmon}, {Coe}, {Bradley}, {Brada{\v{c}}},
  {Strait}, {Paterno-Mahler}, {Huang}, {Oesch}, {Zitrin}, {Acebron}, {Cibirka},
  {Kikuchihara}, {Oguri}, {Brammer}, {Sharon}, {Trenti}, {Avila}, {Ogaz},
  {Andrade-Santos}, {Carrasco}, {Cerny}, {Dawson}, {Frye}, {Hoag}, {Jones},
  {Mainali}, {Ouchi}, {Rodney}, {Stark} \& {Umetsu}}{{Salmon}
  et~al.}{2018}]{Salmon_etal2018}
{Salmon} B.,  {Coe} D.,  {Bradley} L.,  {Brada{\v{c}}} M.,  {Strait} V.,
  {Paterno-Mahler} R.,  {Huang} K.-H.,  {Oesch} P.~A.,  {Zitrin} A.,  {Acebron}
  A.,  {Cibirka} N.,  {Kikuchihara} S.,  {Oguri} M.,  {Brammer} G.~B.,
  {Sharon} K.,  {Trenti} M.,  {Avila} R.~J.,  {Ogaz} S.,  {Andrade-Santos} F.,
  {Carrasco} D.,  {Cerny} C.,  {Dawson} W.,  {Frye} B.~L.,  {Hoag} A.,  {Jones}
  C.,  {Mainali} R.,  {Ouchi} M.,  {Rodney} S.~A.,  {Stark} D.,    {Umetsu} K.,
   2018, \apjl, 864, L22

\bibitem[\protect\citeauthoryear{{Shanahan} \& {Gieles}}{{Shanahan} \&
  {Gieles}}{2015}]{ShaGie2015}
{Shanahan} R.~L.,  {Gieles} M.,  2015, \mnras, 448, L94

\bibitem[\protect\citeauthoryear{{Shetty}, {Cappellari}, {McDermid},
  {Krajnovi{\'c}}, {de Zeeuw}, {Davies} \& {Kobayashi}}{{Shetty}
  et~al.}{2020}]{Shetty_etal2020}
{Shetty} S.,  {Cappellari} M.,  {McDermid} R.~M.,  {Krajnovi{\'c}} D.,  {de
  Zeeuw} P.~T.,  {Davies} R.~L.,    {Kobayashi} C.,  2020, \mnras, 494, 5619

\bibitem[\protect\citeauthoryear{{Smith}}{{Smith}}{2020}]{Smith2020}
{Smith} R.~J.,  2020, \araa, 58, 577

\bibitem[\protect\citeauthoryear{{Smith} \& {Lucey}}{{Smith} \&
  {Lucey}}{2013}]{SmiLuc2013}
{Smith} R.~J.,  {Lucey} J.~R.,  2013, \mnras, 434, 1964

\bibitem[\protect\citeauthoryear{{Smith}, {Lucey} \& {Carter}}{{Smith}
  et~al.}{2012}]{SmiLucCar2012}
{Smith} R.~J.,  {Lucey} J.~R.,    {Carter} D.,  2012, \mnras, 426, 2994

\bibitem[\protect\citeauthoryear{{Smith}, {Lucey} \& {Conroy}}{{Smith}
  et~al.}{2015}]{SmiLucCon2015}
{Smith} R.~J.,  {Lucey} J.~R.,    {Conroy} C.,  2015, \mnras, 449, 3441

\bibitem[\protect\citeauthoryear{{Spiniello}, {Koopmans}, {Trager},
  {Barnab{\`e}}, {Treu}, {Czoske}, {Vegetti} \& {Bolton}}{{Spiniello}
  et~al.}{2015}]{Spiniello_etal2015}
{Spiniello} C.,  {Koopmans} L.~V.~E.,  {Trager} S.~C.,  {Barnab{\`e}} M.,
  {Treu} T.,  {Czoske} O.,  {Vegetti} S.,    {Bolton} A.,  2015, \mnras, 452,
  2434

\bibitem[\protect\citeauthoryear{{Spiniello}, {Trager}, {Koopmans} \&
  {Conroy}}{{Spiniello} et~al.}{2014}]{Spiniello_etal2014}
{Spiniello} C.,  {Trager} S.,  {Koopmans} L. V.~E.,    {Conroy} C.,  2014,
  \mnras, 438, 1483

\bibitem[\protect\citeauthoryear{{Spiniello}, {Trager}, {Koopmans} \&
  {Chen}}{{Spiniello} et~al.}{2012}]{Spiniello_etal2012}
{Spiniello} C.,  {Trager} S.~C.,  {Koopmans} L.~V.~E.,    {Chen} Y.~P.,  2012,
  \apjl, 753, L32

\bibitem[\protect\citeauthoryear{{Spinrad}}{{Spinrad}}{1962}]{Spinrad1962}
{Spinrad} H.,  1962, \apj, 135, 715

\bibitem[\protect\citeauthoryear{{Steinhardt}, {Kokorev}, {Rusakov}, {Garcia}
  \& {Sneppen}}{{Steinhardt} et~al.}{2023}]{Steinhardt_etal2023}
{Steinhardt} C.~L.,  {Kokorev} V.,  {Rusakov} V.,  {Garcia} E.,    {Sneppen}
  A.,  2023, \apjl, 951, L40

\bibitem[\protect\citeauthoryear{{Strader}, {Caldwell} \& {Seth}}{{Strader}
  et~al.}{2011}]{StrCalSet2011}
{Strader} J.,  {Caldwell} N.,    {Seth} A.~C.,  2011, \aj, 142, 8

\bibitem[\protect\citeauthoryear{{Tanvir} \& {Krumholz}}{{Tanvir} \&
  {Krumholz}}{2024}]{TanKru2024}
{Tanvir} T.~S.,  {Krumholz} M.~R.,  2024, \mnras, 527, 7306

\bibitem[\protect\citeauthoryear{{Tanvir}, {Krumholz} \& {Federrath}}{{Tanvir}
  et~al.}{2022}]{TanKruFed2022}
{Tanvir} T.~S.,  {Krumholz} M.~R.,    {Federrath} C.,  2022, \mnras, 516, 5712

\bibitem[\protect\citeauthoryear{{Thomas}, {Saglia}, {Bender}, {Thomas},
  {Gebhardt}, {Magorrian}, {Corsini}, {Wegner} \& {Seitz}}{{Thomas}
  et~al.}{2011}]{Thomas_etal2011}
{Thomas} J.,  {Saglia} R.~P.,  {Bender} R.,  {Thomas} D.,  {Gebhardt} K.,
  {Magorrian} J.,  {Corsini} E.~M.,  {Wegner} G.,    {Seitz} S.,  2011, \mnras,
  415, 545

\bibitem[\protect\citeauthoryear{{Tobin}, {Sheehan}, {Megeath},
  {D{\'i}az-Rodr{\'i}guez}, {Offner}, {Murillo}, {van 't Hoff}, {van Dishoeck},
  {Osorio} \& {et al.}}{{Tobin} et~al.}{2020}]{Tobin_etal2020}
{Tobin} J.~J.,  {Sheehan} P.~D.,  {Megeath} S.~T.,  {D{\'i}az-Rodr{\'i}guez}
  A.~K.,  {Offner} S. S.~R.,  {Murillo} N.~M.,  {van 't Hoff} M. L.~R.,  {van
  Dishoeck} E.~F.,  {Osorio} M.,    {et al.} 2020, \apj, 890, 130

\bibitem[\protect\citeauthoryear{{Tortora}, {Romanowsky} \&
  {Napolitano}}{{Tortora} et~al.}{2013}]{TorRomNap2013}
{Tortora} C.,  {Romanowsky} A.~J.,    {Napolitano} N.~R.,  2013, \apj, 765, 8

\bibitem[\protect\citeauthoryear{{Trager}, {Faber}, {Worthey} \&
  {Gonz{\'a}lez}}{{Trager} et~al.}{2000}]{Trager_etal2000}
{Trager} S.~C.,  {Faber} S.~M.,  {Worthey} G.,    {Gonz{\'a}lez} J.~J.,  2000,
  \aj, 120, 165

\bibitem[\protect\citeauthoryear{{Treu}, {Auger}, {Koopmans}, {Gavazzi},
  {Marshall} \& {Bolton}}{{Treu} et~al.}{2010}]{Treu_etal2010}
{Treu} T.,  {Auger} M.~W.,  {Koopmans} L. V.~E.,  {Gavazzi} R.,  {Marshall}
  P.~J.,    {Bolton} A.~S.,  2010, \apj, 709, 1195

\bibitem[\protect\citeauthoryear{{Truelove}, {Klein}, {McKee}, {Holliman} II,
  {Howell} \& {Greenough}}{{Truelove} et~al.}{1997}]{Trueloveetal1997}
{Truelove} J.~K.,  {Klein} R.~I.,  {McKee} C.~F.,  {Holliman} II J.~H.,
  {Howell} L.~H.,    {Greenough} J.~A.,  1997, \apjl, 489, L179

\bibitem[\protect\citeauthoryear{{Tung}, {Testi}, {Lebreuilly}, {Hennebelle},
  {Maury}, {Klessen}, {Cacciapuoti}, {Gonz{\'a}lez}, {Rosotti} \&
  {Molinari}}{{Tung} et~al.}{2024}]{Tung_etal2024}
{Tung} N.-D.,  {Testi} L.,  {Lebreuilly} U.,  {Hennebelle} P.,  {Maury} A.,
  {Klessen} R.~S.,  {Cacciapuoti} L.,  {Gonz{\'a}lez} M.,  {Rosotti} G.,
  {Molinari} S.,  2024, \aap, 684, A36

\bibitem[\protect\citeauthoryear{{Tychoniec}, {Tobin}, {Karska}, {Chandler},
  {Dunham}, {Harris}, {Kratter}, {Li}, {Looney}, {Melis}, {P{\'e}rez},
  {Sadavoy}, {Segura-Cox} \& {van Dishoeck}}{{Tychoniec}
  et~al.}{2018}]{Tychoniec_etal2018}
{Tychoniec} {\L}.,  {Tobin} J.~J.,  {Karska} A.,  {Chandler} C.,  {Dunham}
  M.~M.,  {Harris} R.~J.,  {Kratter} K.~M.,  {Li} Z.-Y.,  {Looney} L.~W.,
  {Melis} C.,  {P{\'e}rez} L.~M.,  {Sadavoy} S.~I.,  {Segura-Cox} D.,    {van
  Dishoeck} E.~F.,  2018, \apjs, 238, 19

\bibitem[\protect\citeauthoryear{{van Dokkum} \& {Conroy}}{{van Dokkum} \&
  {Conroy}}{2024}]{vanDokCon2024}
{van Dokkum} P.,  {Conroy} C.,  2024, arXiv e-prints, p. arXiv:2407.06281

\bibitem[\protect\citeauthoryear{{van Dokkum}, {Conroy}, {Villaume}, {Brodie}
  \& {Romanowsky}}{{van Dokkum} et~al.}{2017}]{vanDokkum_etal2017}
{van Dokkum} P.,  {Conroy} C.,  {Villaume} A.,  {Brodie} J.,    {Romanowsky}
  A.~J.,  2017, \apj, 841, 68

\bibitem[\protect\citeauthoryear{{van Dokkum} \& {Conroy}}{{van Dokkum} \&
  {Conroy}}{2010}]{vanCon2010}
{van Dokkum} P.~G.,  {Conroy} C.,  2010, \nat, 468, 940

\bibitem[\protect\citeauthoryear{{van Dokkum} \& {Conroy}}{{van Dokkum} \&
  {Conroy}}{2011}]{vanCon2011}
{van Dokkum} P.~G.,  {Conroy} C.,  2011, \apjl, 735, L13

\bibitem[\protect\citeauthoryear{{Vanzella}, {Castellano}, {Bergamini}, {Treu},
  {Mercurio}, {Scarlata}, {Rosati}, {Grillo}, {Acebron}, {Caminha}, {Nonino},
  {Nanayakkara}, {Roberts-Borsani}, {Bradac}, {Wang}, {Brammer}, {Strait},
  {Vulcani}, {Me{\v{s}}tri{\'c}}, {Meneghetti}, {Calura}, {Henry}, {Zanella},
  {Trenti}, {Boyett}, {Morishita}, {Calabr{\`o}}, {Glazebrook}, {Marchesini},
  {Birrer}, {Yang} \& {Jones}}{{Vanzella} et~al.}{2022}]{Vanzella_etal2022}
{Vanzella} E.,  {Castellano} M.,  {Bergamini} P.,  {Treu} T.,  {Mercurio} A.,
  {Scarlata} C.,  {Rosati} P.,  {Grillo} C.,  {Acebron} A.,  {Caminha} G.~B.,
  {Nonino} M.,  {Nanayakkara} T.,  {Roberts-Borsani} G.,  {Bradac} M.,  {Wang}
  X.,  {Brammer} G.,  {Strait} V.,  {Vulcani} B.,  {Me{\v{s}}tri{\'c}} U.,
  {Meneghetti} M.,  {Calura} F.,  {Henry} A.,  {Zanella} A.,  {Trenti} M.,
  {Boyett} K.,  {Morishita} T.,  {Calabr{\`o}} A.,  {Glazebrook} K.,
  {Marchesini} D.,  {Birrer} S.,  {Yang} L.,    {Jones} T.,  2022, \apjl, 940,
  L53

\bibitem[\protect\citeauthoryear{{Vanzella}, {Claeyssens}, {Welch}, {Adamo},
  {Coe}, {Diego}, {Mahler}, {Khullar}, {Kokorev}, {Oguri}, {Ravindranath},
  {Furtak}, {Hsiao}, {Abdurro'uf}, {Mandelker}, {Brammer}, {Bradley},
  {Brada{\v{c}}}, {Conselice}, {Dayal}, {Nonino}, {Andrade-Santos},
  {Windhorst}, {Pirzkal}, {Sharon}, {de Mink}, {Fujimoto}, {Zitrin}, {Eldridge}
  \& {Norman}}{{Vanzella} et~al.}{2023}]{Vanzella_etal2023a}
{Vanzella} E.,  {Claeyssens} A.,  {Welch} B.,  {Adamo} A.,  {Coe} D.,  {Diego}
  J.~M.,  {Mahler} G.,  {Khullar} G.,  {Kokorev} V.,  {Oguri} M.,
  {Ravindranath} S.,  {Furtak} L.~J.,  {Hsiao} T. Y.-Y.,  {Abdurro'uf}
  {Mandelker} N.,  {Brammer} G.,  {Bradley} L.~D.,  {Brada{\v{c}}} M.,
  {Conselice} C.~J.,  {Dayal} P.,  {Nonino} M.,  {Andrade-Santos} F.,
  {Windhorst} R.~A.,  {Pirzkal} N.,  {Sharon} K.,  {de Mink} S.~E.,  {Fujimoto}
  S.,  {Zitrin} A.,  {Eldridge} J.~J.,    {Norman} C.,  2023, \apj, 945, 53

\bibitem[\protect\citeauthoryear{{Vazdekis}, {Casuso}, {Peletier} \&
  {Beckman}}{{Vazdekis} et~al.}{1996}]{Vazdekis_etal1996}
{Vazdekis} A.,  {Casuso} E.,  {Peletier} R.~F.,    {Beckman} J.~E.,  1996,
  \apjs, 106, 307

\bibitem[\protect\citeauthoryear{{Vazdekis}, {Peletier}, {Beckman} \&
  {Casuso}}{{Vazdekis} et~al.}{1997}]{Vazdekis_etal1997}
{Vazdekis} A.,  {Peletier} R.~F.,  {Beckman} J.~E.,    {Casuso} E.,  1997,
  \apjs, 111, 203

\bibitem[\protect\citeauthoryear{{Wang}, {Leja}, {Atek}, {Labb{\'e}}, {Li},
  {Bezanson}, {Brammer}, {Cutler}, {Dayal}, {Furtak}, {Greene}, {Kokorev},
  {Pan}, {Price}, {Suess}, {Weaver}, {Whitaker} \& {Williams}}{{Wang}
  et~al.}{2024}]{Wang_etal2024}
{Wang} B.,  {Leja} J.,  {Atek} H.,  {Labb{\'e}} I.,  {Li} Y.,  {Bezanson} R.,
  {Brammer} G.,  {Cutler} S.~E.,  {Dayal} P.,  {Furtak} L.~J.,  {Greene} J.~E.,
   {Kokorev} V.,  {Pan} R.,  {Price} S.~H.,  {Suess} K.~A.,  {Weaver} J.~R.,
  {Whitaker} K.~E.,    {Williams} C.~C.,  2024, \apj, 963, 74

\bibitem[\protect\citeauthoryear{{Wegner}, {Corsini}, {Thomas}, {Saglia},
  {Bender} \& {Pu}}{{Wegner} et~al.}{2012}]{Wegner_etal2012}
{Wegner} G.~A.,  {Corsini} E.~M.,  {Thomas} J.,  {Saglia} R.~P.,  {Bender} R.,
    {Pu} S.~B.,  2012, \aj, 144, 78

\bibitem[\protect\citeauthoryear{{Weidner}, {Ferreras}, {Vazdekis} \& {La
  Barbera}}{{Weidner} et~al.}{2013}]{Weidner_etal2013}
{Weidner} C.,  {Ferreras} I.,  {Vazdekis} A.,    {La Barbera} F.,  2013,
  \mnras, 435, 2274

\bibitem[\protect\citeauthoryear{{Whitehouse} \& {Bate}}{{Whitehouse} \&
  {Bate}}{2006}]{WhiBat2006}
{Whitehouse} S.~C.,  {Bate} M.~R.,  2006, \mnras, 367, 32

\bibitem[\protect\citeauthoryear{{Whitehouse}, {Bate} \&
  {Monaghan}}{{Whitehouse} et~al.}{2005}]{WhiBatMon2005}
{Whitehouse} S.~C.,  {Bate} M.~R.,    {Monaghan} J.~J.,  2005, \mnras, 364,
  1367

\bibitem[\protect\citeauthoryear{{Whitworth}}{{Whitworth}}{1998}]{Whitworth1998}
{Whitworth} A.~P.,  1998, \mnras, 296, 442

\bibitem[\protect\citeauthoryear{{Whitworth}, {Priestley}, {W{\"u}nsch} \&
  {Palou{\v{s}}}}{{Whitworth} et~al.}{2024}]{Whitworth_etal2024}
{Whitworth} A.~P.,  {Priestley} F.~D.,  {W{\"u}nsch} R.,    {Palou{\v{s}}} J.,
  2024, \mnras, 529, 3712

\bibitem[\protect\citeauthoryear{{Woodrum}, {Rieke}, {Ji}, {Baker},
  {Bhatawdekar}, {Bunker}, {Charlot}, {Curtis-Lake}, {Eisenstein}, {Hainline},
  {Hausen}, {Helton}, {Hviding}, {Johnson}, {Robertson}, {Sun}, {Tacchella},
  {Whitler}, {Williams} \& {Willmer}}{{Woodrum}
  et~al.}{2023}]{Woodrum_etal2023}
{Woodrum} C.,  {Rieke} M.,  {Ji} Z.,  {Baker} W.~M.,  {Bhatawdekar} R.,
  {Bunker} A.~J.,  {Charlot} S.,  {Curtis-Lake} E.,  {Eisenstein} D.~J.,
  {Hainline} K.,  {Hausen} R.,  {Helton} J.~M.,  {Hviding} R.~E.,  {Johnson}
  B.~D.,  {Robertson} B.,  {Sun} F.,  {Tacchella} S.,  {Whitler} L.,
  {Williams} C.~C.,    {Willmer} C. N.~A.,  2023, arXiv e-prints, p.
  arXiv:2310.18464

\bibitem[\protect\citeauthoryear{{Worthey}, {Faber} \& {Gonzalez}}{{Worthey}
  et~al.}{1992}]{WorFabGon1992}
{Worthey} G.,  {Faber} S.~M.,    {Gonzalez} J.~J.,  1992, \apj, 398, 69

\bibitem[\protect\citeauthoryear{{Wurster}, {Bate} \& {Price}}{{Wurster}
  et~al.}{2019}]{WurBatPri2019}
{Wurster} J.,  {Bate} M.~R.,    {Price} D.~J.,  2019, \mnras, 489, 1719

\bibitem[\protect\citeauthoryear{{Xiao}, {Oesch}, {Elbaz}, {Bing}, {Nelson},
  {Weibel}, {Naidu}, {Daddi}, {Bouwens}, {Matthee}, {Wuyts}, {Chisholm},
  {Brammer}, {Dickinson}, {Magnelli}, {Leroy}, {van Dokkum}, {Schaerer},
  {Herard-Demanche}, {Barrufet}, {Endsley}, {Fudamoto}, {G{\'o}mez-Guijarro},
  {Gottumukkala}, {Illingworth}, {Labbe}, {Magee}, {Marchesini}, {Maseda},
  {Qin}, {Reddy}, {Shapley}, {Shivaei}, {Shuntov}, {Stefanon}, {Whitaker} \&
  {Wyithe}}{{Xiao} et~al.}{2023}]{Xiao_etal2023}
{Xiao} M.,  {Oesch} P.,  {Elbaz} D.,  {Bing} L.,  {Nelson} E.,  {Weibel} A.,
  {Naidu} R.,  {Daddi} E.,  {Bouwens} R.,  {Matthee} J.,  {Wuyts} S.,
  {Chisholm} J.,  {Brammer} G.,  {Dickinson} M.,  {Magnelli} B.,  {Leroy} L.,
  {van Dokkum} P.,  {Schaerer} D.,  {Herard-Demanche} T.,  {Barrufet} L.,
  {Endsley} R.,  {Fudamoto} Y.,  {G{\'o}mez-Guijarro} C.,  {Gottumukkala} R.,
  {Illingworth} G.,  {Labbe} I.,  {Magee} D.,  {Marchesini} D.,  {Maseda} M.,
  {Qin} Y.,  {Reddy} N.,  {Shapley} A.,  {Shivaei} I.,  {Shuntov} M.,
  {Stefanon} M.,  {Whitaker} K.,    {Wyithe} J.~S.,  2023, arXiv e-prints, p.
  arXiv:2309.02492

\bibitem[\protect\citeauthoryear{{Yasui}, {Izumi}, {Saito}, {Lau}, {Kobayashi}
  \& {Ressler}}{{Yasui} et~al.}{2024}]{Yasui_etal2024}
{Yasui} C.,  {Izumi} N.,  {Saito} M.,  {Lau} R.~M.,  {Kobayashi} N.,
  {Ressler} M.~E.,  2024, arXiv e-prints, p. arXiv:2408.15440

\bibitem[\protect\citeauthoryear{{Yasui}, {Kobayashi}, {Saito}, {Izumi} \&
  {Ikeda}}{{Yasui} et~al.}{2023}]{Yasui_etal2023}
{Yasui} C.,  {Kobayashi} N.,  {Saito} M.,  {Izumi} N.,    {Ikeda} Y.,  2023,
  \apj, 943, 137

\bibitem[\protect\citeauthoryear{{Zhang}, {Romano}, {Ivison}, {Papadopoulos} \&
  {Matteucci}}{{Zhang} et~al.}{2018}]{Zhang_etal2018}
{Zhang} Z.-Y.,  {Romano} D.,  {Ivison} R.~J.,  {Papadopoulos} P.~P.,
  {Matteucci} F.,  2018, \nat, 558, 260

\bibitem[\protect\citeauthoryear{{Zucconi}, {Walmsley} \& {Galli}}{{Zucconi}
  et~al.}{2001}]{ZucWalGal2001}
{Zucconi} A.,  {Walmsley} C.~M.,    {Galli} D.,  2001, \aap, 376, 650

\end{thebibliography}

\end{document}